\documentclass[a4paper,11pt]{article}
\usepackage{jheppub} 
\usepackage{lineno}

\usepackage{enumitem}

\usepackage[T1]{fontenc} 

\usepackage{tikz}
\usepackage[compat=1.1.0]{tikz-feynman}

\graphicspath{{Figures/}}

\usepackage{amsmath,amsfonts,amssymb,amsbsy,mathrsfs,units,pifont,slashed,bbm,bbding,textcomp, array}
\usepackage{array,booktabs,multirow,colortbl,hhline}
\usepackage{tabu}
\usepackage[thicklines]{cancel}
\usepackage{graphicx,feynmp,feynmp}
\usepackage{pstricks,epsf,tikz,pst-plot}

\usepackage{dingbat}
 
\usepackage{mathabx}

\usepackage{setspace}
\usepackage{caption}
\usepackage[hang,nooneline]{subfigure}

\renewcommand{\geq}{\geqslant}

\renewcommand{\dag}{\dagger}
\newcommand{\bmat}{\begin{pmatrix}}
\newcommand{\emat}{\end{pmatrix}}
\renewcommand{\to}{\rightarrow}

\renewcommand{\a}{\alpha}
\renewcommand{\b}{\beta}

\newcommand{\e}{\varepsilon}
\newcommand{\g}{\gamma}
\newcommand{\s}{\sigma}

\newcommand{\BBu}{B^{\mu\nu}}
\newcommand{\BBd}{B_{\mu\nu}}

\newcommand{\WWd}{W_{\mu\nu}}

\makeatletter

\newcommand\midtilde@raisedtilde[1][.5]{\raisebox{#1ex}{\texttildelow}}
\def\midtilde@normaltilde{\texttildelow}

\newcommand\midtilde
{
  \expandafter\in@\expandafter{\f@family}
    {cmr,cmss,cmtt,cmm,cmsy,cmx,
    lmr,lmss,lmtt,lmm,lmsy,lmx,
    pxr,pxss,pxm,pxsy,pxx,
    txr,txss,txm,txsy,txx}
  \ifin@
    \midtilde@raisedtilde
  \else
    \expandafter\in@\expandafter{\f@family}
    {pxtt,txtt}
    \ifin@
      \midtilde@raisedtilde[.35]
    \else
      \midtilde@normaltilde
    \fi
  \fi
}

\usepackage{adjustbox}
\newcolumntype{M}[1]{>{\centering\arraybackslash}m{#1}}

\usepackage{pifont}

\arxivnumber{2303.18215} 

\title{\boldmath A sensitivity study of triboson production processes to dimension-6 EFT operators at the LHC}

\author[a,b]{R.~Bellan,}
\author[c]{S.~Bhattacharya,}
\author[d,e]{G.~Boldrini,}
\author[d,e]{F.~Cetorelli,}
\author[d,e]{P.~Govoni,}
\author[e]{A.~Massironi,}
\author[a,b]{A.~Mecca,}
\author[a,b]{C.~Tarricone,}
\author[a,b,1]{A.~Vagnerini\note{Now at the University of Nebraska-Lincoln, Lincoln, Nebraska, USA}}

\affiliation[a]{Dipartimento di Fisica, Università degli Studi di Torino, \\Via Pietro Giuria 1, 10125 Torino, Italy}
\affiliation[b]{INFN, sezione di Torino, \\Via Pietro Giuria 1, 10125 Torino, Italy}
\affiliation[c]{Northwestern University, \\2145 Sheridan Road, Tech F165, Evanston, 60208-3112 Illinois, USA}
\affiliation[d]{Dipartimento di Fisica, Università degli Studi di Milano-Bicocca,\\Piazza della Scienza 3, 20126 Milano, Italy}
\affiliation[e]{INFN, sezione di Milano - Bicocca,\\Piazza della Scienza 3, 20126 Milano, Italy}
\emailAdd{riccardo.bellan@unito.it}
\emailAdd{saptaparna.bhattacharya@cern.ch}
\emailAdd{giacomo.boldrini@cern.ch}
\emailAdd{flavia.cetorelli@cern.ch}
\emailAdd{pietro.govoni@unimib.it}
\emailAdd{andrea.massironi@mib.infn.it}
\emailAdd{alberto.mecca@cern.ch}
\emailAdd{cristiano.tarricone@cern.ch}
\emailAdd{antonio.vagnerini@cern.ch}

\abstract{We present the first parton-level study of anomalous effects in triboson production in both fully and semi-leptonic channels in proton-proton collisions at 13~TeV at the Large Hadron Collider (LHC). 
The sensitivity to anomalies induced by a minimal set of bosonic dimension-6 operators from the Warsaw basis is evaluated with specific analyses for each final state. 
A likelihood-based strategy is employed to  assess the most sensitive kinematic observables per channel, where the contribution of Effective Field Theory operators is parameterized at either the linear or quadratic level.  
The impact of the mutual interference terms of pairs of operators on the sensitivity is also examined.
This benchmark study explores the complementarity and overlap in sensitivity between different triboson measurements and paves the way for future analyses at the LHC experiments. 
The statistical combination of the considered final states allows setting stringent bounds on five bosonic Wilson coefficients.}

\begin{document} 
\maketitle
\flushbottom

\section{Introduction}
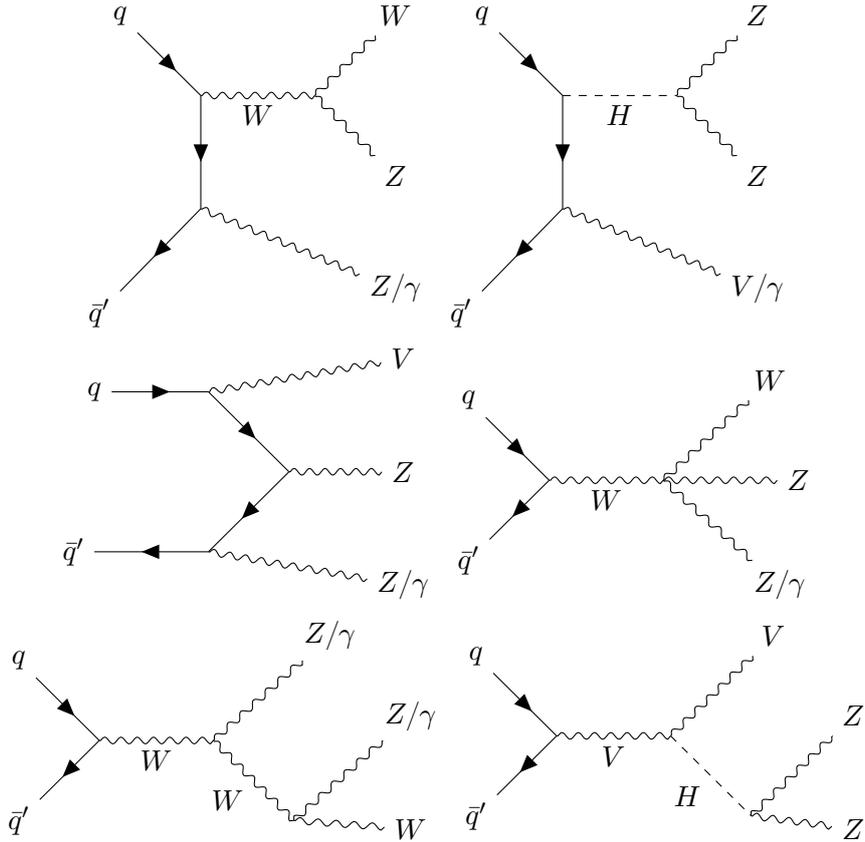
\begin{figure}[t!]%
    	\centering
%
    	\begin{tikzpicture}
    	\begin{feynman}
    	\vertex (a) {\(q\)};
    	\vertex [below right of= a] (1);
    	\vertex [right of= 1] (i);
    	\vertex [below of= 1] (2);
    	\vertex [above right of= i] (c) {\(W\)};
    	\vertex [below right of= i] (d) {\(Z\)};
    	\vertex [below of= d] (e) {\(Z/\g\)};
    	\vertex [below left =of 2] (b) {\(\bar q'\)};
    	\diagram* {
    		(a) -- [fermion] (1) -- [fermion] (2) -- [fermion] (b),
    		(1) -- [boson, edge label'=\(W\)] (i),
    		(i) -- [boson] (c),
    		(i) -- [boson] (d),
    		(2) -- [boson] (e)
    	};
    	\end{feynman}
    	\end{tikzpicture}
    	\begin{tikzpicture}
    	\begin{feynman}
    	\vertex (a) {\(q\)};
    	\vertex [below right of= a] (1);
    	\vertex [right of= 1] (H);
    	\vertex [below of= 1] (2);
    	\vertex [above right of= H] (c) {\(Z\)};
    	\vertex [below right of= H] (d) {\(Z\)};
    	\vertex [below of= d] (e) {\(V/\g\)};
    	\vertex [below left =of 2] (b) {\(\bar q'\)};
    	\diagram* {
    		(a) -- [fermion] (1) -- [fermion] (2) -- [fermion] (b),
    		(1) -- [scalar, edge label'=\(H\)] (H),
    		(H) -- [boson] (c),
    		(H) -- [boson] (d),
    		(2) -- [boson] (e)
    	};
    	\end{feynman}
    	\end{tikzpicture}
\\
    	\begin{tikzpicture}
    	\begin{feynman}
    	\vertex (a) {\(q\)};
    	\vertex [right of= a] (1);
    	\vertex [below right of= 1] (2);
    	\vertex [below left of= 2] (3);
    	\vertex [left=of 3] (b){\(\bar q'\)};
    	\vertex [right of= 2] (d) {\(Z\)};
    	\vertex [above of= d] (c) {\(V\)};
    	\vertex [below of= d] (e) {\(Z/\g\)};
    	\diagram* {
    		(a) -- [fermion] (1) -- [fermion] (2) -- [fermion] (3) -- [fermion] (b),
    		(1) -- [boson] (c),
    		(2) -- [boson] (d),
    		(3) -- [boson] (e)
    	};
    	\end{feynman}
    	\end{tikzpicture}
    	\begin{tikzpicture}
    	\begin{feynman}
    	\vertex (a) {\(q\)};
    	\vertex [below right of= a] (1);
    	\vertex [below left  of= 1] (b) {\(\bar q'\)};
    	\vertex [right=of 1] (2);
    	\vertex [above right=of 2] (c) {\(W\)};
    	\vertex [right=of 2] (d) {\(Z\)};
    	\vertex [below right=of 2] (e) {\(Z/\g\)};
    	\diagram* {
    		(a) -- [fermion] (1) -- [fermion] (b),
    		(1) -- [boson, edge label'=\(W\)] (2),
    		(2) -- [boson] (c),
    		(2) -- [boson] (d),
    		(2) -- [boson] (e)
    	};
    	\end{feynman}
    	\end{tikzpicture}
\\
    	\begin{tikzpicture}
    	\begin{feynman}
    	\vertex (a) {\(q\)};
    	\vertex [below right of= a] (1);
    	\vertex [below left  of= 1] (b) {\(\bar q'\)};
    	\vertex [right=of 1] (2);
    	\vertex [above right=of 2] (c) {\(Z/\g\)};
    	\vertex [below right=of 2] (3) ;
        \vertex [above right=of 3] (d) {\(Z/\g\)};
        \vertex [below =of d] (e) {\(W\)};
    	\diagram* {
    		(a) -- [fermion] (1) -- [fermion] (b),
    		(1) -- [boson, edge label'=\(W\)] (2),
    		(2) -- [boson] (c),
    		(2)  -- [boson, edge label'=\(W\)] (3),
    		(d) -- [boson] (3) -- [boson] (e)
    	};
    	\end{feynman}

        \end{tikzpicture}
    	\begin{tikzpicture}
    	\begin{feynman}
    	\vertex (a) {\(q\)};
    	\vertex [below right of= a] (1);
    	\vertex [below left  of= 1] (b) {\(\bar q'\)};
    	\vertex [right=of 1] (2);
    	\vertex [above right=of 2] (c) {\(V\)};
    	\vertex [below right=of 2] (3) ;
        \vertex [above right=of 3] (d) {\(Z\)};
        \vertex [below =of d] (e) {\(Z\)};
    	\diagram* {
    		(a) -- [fermion] (1) -- [fermion] (b),
    		(1) -- [boson, edge label'=\(V\)] (2),
    		(2) -- [boson] (c),
    		(2)  -- [scalar, edge label'=\(H\)] (3),
    		(d) -- [boson] (3) -- [boson] (e)
    	};
    	\end{feynman}

        \end{tikzpicture}

\caption{ Representative Standard Model Feynman diagrams at the tree level for
              the electroweak VZ$\gamma$ and VZZ triboson production.
\label{fig:EWK_Feynm}}
\end{figure}

The experiments at the Large Hadron Collider (LHC) have probed the Standard Model (SM) up to the TeV scale, and so far no evidence for significant anomalies has been found. 
The electroweak sector remains partially unexplored since several extremely rare processes have not yet been observed. 
The electroweak production of three gauge bosons is of paramount importance as it provides tree-level access to triple and quartic gauge couplings, as well as to Higgs-gauge boson couplings. These processes have a complementary topology with respect to the vector boson scattering (VBS), and play a fundamental role in the study of the gauge interactions. In particular, the search for triboson production provides an essential test of the non-Abelian gauge structure of the electroweak group at the accessible LHC energies. In addition, they complement the precision measurements of the Higgs boson at an energy scale that can differ significantly from its observed mass.
Any deviation from the SM predictions in the high-energy tails of the observables relevant to these processes would be a direct indication of the existence of new physics beyond the Standard Model (BSM). The Standard Model Effective Field Theory (SM-EFT) 
is a theoretical framework developed in the last decade that describes the effects of BSM physics in terms of higher-dimensional operators (see a review in \cite{Brivio:2017vri}). These operators are constructed from the fields and symmetries of the SM and are suppressed by a scale $\Lambda$, which is assumed to be the energy scale of the new physics phenomena. 
The SM-EFT Lagrangian is generically defined as an expansion of the SM Lagrangian, 
\begin{equation}\label{EFTlagr}
    \mathcal L_\text{SM-EFT}=
    \mathcal L_\text{SM}^{(4)}
    +\sum_{i,n}\frac{c_i^{(n)}}{\Lambda^{n-4}} Q_i^{(n)} ,
\end{equation}
where the Wilson coefficients, $c_i^{(n)}$, are the dimensionless parameters that determine the magnitude of the contribution of the corresponding operators. The SM gauge symmetry constraints restrict the dimension-5 set to a single operator~\cite{weinberg1979baryon} which violates lepton-number conservation. Since the conservation of the total lepton number is one of the underlying assumptions, this operator is outside the scope of this study. The dimension-6 SM-EFT operators can be organized in the Warsaw basis~\cite{grzadkowski2010dimension}. Considering only the effect of EFT operators with this dimension, the amplitude $\mathcal{A}$ of the scattering matrix of a certain process receives a correction:
\begin{equation}
\label{AmplitudeQi}
    \mathcal{A} = \mathcal{A}_{\text{SM}} + \sum_i\frac{c_{i}}{\Lambda^{2}} \cdot \mathcal{A}_{Q_{i}}\ .
\end{equation}
The squared amplitude is proportional to the total event yield $N$, which is the sum of a SM contribution, a linear interference term $N_\a^{int}$ between the SM diagrams and a single operator, the pure BSM contribution $N_{\a}^{quad}$, which scales quadratically with the Wilson coefficient, and the mixed interference term $N_{\a,\b}^{mix}$ between pairs of dimension-6 operators:  
\begin{equation}
\label{eq:NEvents}
       N =N_ {SM} + \sum_\a \left[\frac{c_\a}{\Lambda^2}N_\a^{int} +\frac{c_\a^2}{\Lambda^4}N_{\a}^{quad}\right]
    +
    \sum_{\a\neq\beta}\frac{c_\a c_\beta}{\Lambda^4}N_{\a,\b}^{mix}\,. 
\end{equation}
The present work follows the same strategy as the diboson analysis in Ref.~\cite{bellan2021sensitivity}, but in the case of triboson production processes. Specifically, we perform a parton-level study of the anomalous effects of dimension-6 operators on the fully leptonic channels WZ$\g$ and ZZ$\g$, and the semi-leptonic ones VZ$\g$ and VZZ, with the V=\{W,Z\} boson decaying hadronically. 
The triboson processes under study, exemplified in figure~\ref{fig:EWK_Feynm}, are associated with very low yields in the SM. In fact, only the inclusive production of three massive gauge bosons VVV~\cite{Sirunyan_VVV,Aad_2019} and of one massive gauge boson and two photons V$\gamma\gamma$ \cite{tumasyan2021measurements} have been measured at LHC at 13 TeV. 
Upper bounds on the production cross section of VV$\gamma$ processes with one photon in the final state have been set in LHC searches at 8 TeV~\cite{cms2013measurement,atlas2017study,chatrchyan2014search}. 
The aforementioned studies focus on the anomalies induced by dimension-8 operators on quartic couplings, while neglecting the potential presence of anomalous triple gauge couplings. 
Unlike at dimension-8, triple and quartic gauge couplings are intrinsically related at dimension-6 \cite{Covarelli:2021gyz}. 
In this work, we show the potential contribution to anomalies in triboson production induced by a representative set of bosonic dimension-6 SM-EFT operators from the Warsaw basis. The selected operators affect the triple and quartic gauge vertices involved in these processes as well as the corresponding Higgs-mediated diagrams. 
The fermionic operators are outside the scope of this study. Previous works, such as \cite{falkowski2021light}, address the contribution of fermionic dimension-6 operators to triboson production processes. 
In the current study, we perform the combination of the final states to assess how the orthogonality and the interplay between different analyses enhance the sensitivity reached to anomalous effects.   

A sensitivity study of the fully leptonic WWW channel in a specific tri-lepton final state to anomalous effects induced by dimension-6 operators is presented in Section~\ref{subsec:WWW}. 
Unlike the triboson channels discussed above, evidence for the WWW production has been found at the LHC at a center-of-mass-energy of 13~TeV~\cite{Sirunyan_VVV,Aad_2019} and the observed cross-section is in agreement with the SM expectation. 
For this channel, we provide a benchmark result on the sensitivity to anomalous gauge couplings and a comparison with the other triboson channels. 

\section{Processes and operators studied}
We study triboson production processes in proton-proton collisions at 13~TeV, outlining a scenario corresponding to an integrated luminosity of 300~fb$^{-1}$. 
The channels targeted in this work are WZZ, ZZZ, WZ$\g$, ZZ$\g$ 
in the final states listed in table~\ref{tab:VVVprocesses}. 
\begin{table}[h!]
\centering
\captionsetup{skip=7pt}
\renewcommand{\arraystretch}{1.5}
\begin{tabular}{c|c|c}
\hline
    channels & final states & decays \\
\hline
    WZ$\g$          & $\mu^\pm\overset{\scriptscriptstyle(-)}{\nu_\mu}\, $e$^+$e$^-\, \g$& W$^{\pm}\rightarrow \mu^\pm\overset{\scriptscriptstyle(-)}{\nu_\mu},\,~$Z$\rightarrow$e$^+$e$^-$\\
    ZZ$\g$          & $\mu^+\mu^-\, $e$^+$e$^-\, \g$& Z~$\rightarrow~\mu^+\mu^-/$e$^+$e$^-$\\
    VZ$\g$          & 2 jets$\,+~$l$^+$l$^-\, \g$&
    V=\{W,Z\}$\rightarrow$ q$\Bar{\text{q'}}$,\,~Z~$\rightarrow~\mu^+\mu^-/$e$^+$e$^-$\\
    VZZ          & 2 jets$\,+~\mu^+\mu^-\, $e$^+$e$^-$& V=\{W,Z\}$\rightarrow$ q$\Bar{\text{q'}}$,\,~Z~$\rightarrow~\mu^+\mu^-/$e$^+$e$^-$
    \\\hline
\end{tabular}
  \caption{ Triboson processes in this work, with the respective final states studied.
  Jets in the final state are the product of hadronizing quarks (or gluons, in the case of QCD-induced backgrounds). 
        \label{tab:VVVprocesses}
    }
\end{table}
\\
In the case of WZZ and ZZZ, 
the semi-leptonic final state VZZ$\,\rightarrow\,$~2j~2$\mu$~2e is examined. 
We study also the semi-leptonic final state the process VZ$\g\,\rightarrow\,$2j~2l~$\g$, where the Z and V bosons decay leptonically and hadronically, respectively. 
The channels WZ$\g$ and ZZ$\g$ are also investigated in the fully leptonic decays. 
At this level, we do not constrain any intermediate triboson states in the generation, hence the background processes with the same final state are included. 
They can contribute to the deviations from the SM predictions, depending on the operators considered. 
The kinematic selection, in particular the constraints on the invariant mass of the vector bosons,  
mitigates the contribution of processes other than triboson production. 

All processes are modeled inclusively as full $2\to 6$ fermions and $2\to 4$ fermions + $\g$ scatterings with the inclusion of non-resonant diagrams.
Vector boson scattering (VBS) and fusion (VBF) contributions are included for the semi-leptonic final states. However, they are heavily suppressed by imposing the requirement of central jets ($|\eta_j|$ < 2.5) pairs with an invariant mass close to the nominal W/Z-boson peak (50 < $m_{jj}$ < 120 GeV). 
The event generation is performed at the leading order using the amplitude decomposition technique with \textsc{MadGraph5\_aMC@NLO} 2.6.5~\cite{Alwall:2014hca} interfaced with the \textsc{SMEFTsim} 3.0 package~\cite{Brivio:2017btx,Brivio:2020onw}. In addition, for the semi-leptonic channels, the generation is reproduced with the re-weighting technique to ensure the stability of the results in the profiled likelihood analysis. Both the amplitude decomposition and the re-weighting techniques are described in Ref.~\cite{bellan2021sensitivity}. 

This analysis is performed under the assumption of CP symmetry conservation, U(3)$^5$ flavor symmetry, and \{$m_W,m_Z,G_F$\} as the input parameter scheme~\cite{brivio2021electroweak}. No unitarization procedure or clipping of the high-energy distribution bins is applied to avoid introducing additional assumptions into the model prediction.

The present study focuses on the anomalies induced in the electroweak sector by dimension-6 CP-even operators in triple and quartic gauge couplings, as well as couplings between the Higgs boson and the electroweak mediators. 
Therefore, the operators considered in this work are the bosonic operators, classified as in Ref.~\cite{grzadkowski2010dimension}, depending on the number of vector gauge fields ($X^n$), Higgs doublets ($\phi^n$), and covariant derivatives ($D^n$). The classes of dimension-6 operators considered are the following: $X^3$, $X^2\phi^2$, $\phi^4 D^2$, $\phi^2 D^4$, and $\phi^6$. 
The only $\phi^6$-operator is Q$_H=(\phi^\dagger\phi)^3$ which affects the Higgs self-couplings, but not the couplings with the vector bosons, hence it can be safely disregarded. 
The bosonic operators inducing the gluon self-couplings and the anomalous couplings of gluons with the Higgs boson, namely Q$_G$ and Q$_{HG}$, are also excluded since they do not affect the electroweak sector. The resulting set of CP-even operators considered in this work is listed in table~\ref{tab:operators}. 
\begin{table}[h!]
\centering
\renewcommand{\arraystretch}{1}
\begin{tabular}{l|c||l|c||l|c}
\hline
\multicolumn{2}{c}{$X^3$}&
\multicolumn{2}{c}{$X^2\phi^2$}&
\multicolumn{2}{c}{$\phi^4 D^2$}\\
\hline
    &&&&&\\
    &&$Q_{HB}$          & $(\phi ^\dag \phi ) \BBd B^{\mu\nu}$    
    &&
    \\
    &&&&$Q_{HD}$          & $(\phi^\dag D_\mu \phi ) (\phi ^\dag D^\mu \phi )$
    \\
    $Q_{W}$ & $\e^{IJK} W^{I\nu}_{\mu}W^{J\rho}_{\nu} W^{K\mu}_\rho$ 
    &$Q_{HW}$          & $(\phi ^\dag \phi ) \WWd^I W^{I\mu\nu}$
    &&\\
    &&&&$Q_{H\Box}$  &$(\phi^\dag \phi)\Box(\phi^\dag \phi)$\\
    &&
    $Q_{HWB}$         & $(\phi ^\dag \s^I \phi ) \WWd^I\BBu$
    &&
    \\
    &&&&&\\\hline
\end{tabular}
  \caption{ Subset of dimension-6 operators extracted from the Warsaw basis, considered for this work. \hspace{1.0cm}
        \label{tab:operators}
    }
\end{table}

This subset of operators enables efficient study of the overlap between different triboson analyses in constraining EFT parameters, while avoiding the use of an excessively complex parameter space in the global fit. 
They induce anomalous effects in the majority of the triboson channels under scrutiny. The main exceptions are respectively the operator Q$_W$, which does not affect the ZZ$\g$ production process, and the operator Q$_{H\Box}$ affecting solely the VZZ channel.
In the latter, the anomalous effects induced by Q$_{H\Box}$ and Q$_{HB}$ are not sufficient to constrain the corresponding Wilson coefficients, hence they are excluded from the results.  

The \textsc{SMEFTsim} convention is chosen for the generation of linear and quadratic EFT components, namely the interference with the SM and the purely BSM term, respectively. It is also used for the generation of mixed interference among two EFT operators. 
All the BSM generations are performed with the non-zero Wilson coefficients set to unity and $\Lambda=1$ TeV. 

The contribution of the single operators and their interference are generated separately. 
The SM term is generated with every BSM dynamics set to zero. 
In the context of the one-dimensional study 
only the single operator contributions are considered, while for the two-dimensional case, the contribution of the interference between an operator pair is also taken into account. 
The interference terms between all the possible couples of the operators in the given subset are computed to extract the profiled constraints. 

In the study of the semi-leptonic decay channels of VZZ and VZ$\g$, we also generate at leading order the processes with jet pairs induced by QCD vertices, denoted as \emph{QCD-ZZjj} and \emph{QCD-Z$\g$jj}, respectively. 
The processes 
with QCD-induced jets in the final states constitute the dominant background source from processes with the same final particles as the signal. Their production cross-sections are up to two orders of magnitude larger than the electroweak processes with the same final states.
However, some of the considered operators also affect the processes with QCD-induced jets, contributing to deviations from the SM predictions. 
Table~\ref{tab:ProcVsOps} provides a  list of the benchmark EFT operators highlighting the ones inducing diagrams for each process. 
\renewcommand{\arraystretch}{1.5}
\begin{table}[t!]
\begin{center}
{\small
\begin{tabular}{ ||M{37mm}|M{10mm}|M{10mm}|M{10mm}|M{10mm}|M{10mm}|M{10mm}|| } 
 \hline
    Operators $\rightarrow$ \newline \rotatebox[origin=c]{90}{$\Lsh$} Processes   & $Q_{W}$ & $Q_{HB}$ & $Q_{HW}$ &  $Q_{HWB}$ & $Q_{HD}$ & $Q_{H\Box}$ \\
   
   \hline
   \textbf{WZ$\g$} & \ding{51} & \ding{51} & \ding{51} & \ding{51} & \ding{51} & \\
   \hline
   \textbf{ZZ$\g$} & & \ding{51} & \ding{51} & \ding{51} & \ding{51} & \\
   \hline
   \textbf{VZ$\g$} & \ding{51} & \ding{51} & \ding{51} & \ding{51} & \ding{51} & \\
   \textbf{QCD-Z$\g$jj} & & & & \ding{51} & \ding{51} & \\
   \hline
   \textbf{VZZ} & \ding{51} & (\ding{51}) & \ding{51} & \ding{51} & \ding{51} & (\ding{51}) \\
   \textbf{QCD-ZZjj} & & & & \ding{51} & \ding{51} & \\
   \hline
   
\end{tabular}
}
\end{center}
\caption{ Dependency of the studied processes on the bosonic EFT operators. Empty cells indicate the absence of diagrams affected by a particular operator for that channel. 
The brackets indicate operators that introduce anomalies in the studied channels, but these effects are strongly suppressed by the kinematic selection criteria. Hence, for the purpose of this study, these operators are neglected for the corresponding channels. 
%
\\}
  \label{tab:ProcVsOps}
\vspace{-0.5cm}
\end{table}
%
\section{Kinematic selection}
The event selection applied is based on the phase space regions outlined in table~\ref{tab:kinSel}.
We study the differential distributions of the number of events as a function of the variables listed. 
For the channels with two Z bosons, VZZ and ZZ$\g$, the leptonically decaying Z-candidates are ranked according to the difference in mass from the nominal Z mass peak (i.e. $|m_{Z_1}-m_{Z}|<|m_{Z_2}-m_{Z}|$ with $m_Z = (91.1876\pm 0.0021)$~GeV~\cite{pdgMZ}).
\renewcommand{\arraystretch}{2.0}
\begin{table}[!htbp]
 \centering
  \begin{tabular}{||p{30mm} p{55mm} p{50mm} p{0mm}||} 
   \hline
   \textbf{Process} & \textbf{Variables of interest} & \textbf{Selections}   &\\
   \hline
   
   $\mathbf{WZ\g}$\newline 
   ($pp \rightarrow \mu^\pm\overset{\scriptscriptstyle(-)}{\nu_\mu}~2e~\g$) \newline \newline \newline \newline 
   Expected \newline events: (EW) 50& 
   $\cancel{\it{E}}_{T}$, $m_{ee}$, $m_{T,W}$, $p_{T}^{Z}$, $p_{T}^{W}$, $p_{T}^{\g}$, $p_{T}^{SFOS-ll}$, $p_{T}^{l_{i}}$, 
   $p_{T}^{\g}$, 
   $p_{T}^{e^+\mu^+}$, $\eta_{l_i}$, 
   $\eta^{\g}$, $\phi^{\g}$,
   $p_{T(Z\g)}^{l_i}$, 
   $p_{T(Z)}^{l_i}$, 
   $p_{T(WZ)}^{l_i}$, 
   $p_{T(W)}^{l_i}$, 
   $p_{T(Z)}^{\g}$, $p_{T(Z)}^{W}$, $p_{T(\g)}^{W}$, 
   $p_{T(WZ)}^{\g}$,
   $H_\ell^T(ee)$, $H_\ell^T(3l\nu\g)$
   
   & $50 \hspace{-1mm}<\hspace{-1mm}m_{\mu\nu}\hspace{-1mm}<\hspace{-1mm}110$ GeV \newline 
   $60 \hspace{-1mm}<\hspace{-1mm}m_{ee}\hspace{-1mm}<\hspace{-1mm}\ 120$ GeV \newline 
   $p_{T,l^{1}}$ > 20 GeV \ \ \newline 
   $p_{T,l^{2}}$ > 10 GeV \newline 
   $p_{T,l^{i}}$ > 5 GeV \ \ \ 
   |$\eta_{l^i}$| < 2.5 \newline
   $p_{T,\g}$ > 20 GeV \ \ 
   |$\eta_\g$| < 2.5 \newline 
   $\cancel{\it{E}}_{T}$ > 30 GeV \  
   $\Delta R(l^{i},\g)$ > 0.4 
    & 
   \\
   \hline
   
   $\mathbf{ZZ\g}$\newline 
   ($pp \rightarrow 2e~2\mu~\g$) \newline \newline \newline \newline 
   Expected \newline events: (EW) 22& 
   $m_{SFOS-ll}$, $m_{4l}$, 
   $p_{T}^{Z_i}$, 
   $p_{T}^{l_{i}}$, 
   $p_{T}^{SFOS-ll}$,  
   $p_{T}^{\g}$, $p_{T}^{e^\pm\mu^\pm}$, 
   $\eta_{l_i}$, 
   $\eta^{\g}$, $\phi^{\g}$,
   $p_{T(Z\g)}^{l_i}$, 
   $p_{T(Z_i)}^{l_k}$, 
   $p_{T(ZZ)}^{l_i}$, 
   $p_{T(\g)}^{l_i}$, 
   $p_{T(Z_i)}^{\g}$, 
   $p_{T(ZZ)}^{\g}$, 
   $H_\ell^T(SFOS-ll)$, $H_\ell^T(4l\g)$

   & $60 \hspace{-1mm}<\hspace{-1mm}m_{SFOS-ll}\hspace{-1mm}<\hspace{-1mm}120$ GeV \newline 
   $p_{T,l^{1}}$ > 20 GeV \ \ \newline 
   $p_{T,l^{2}}$ > 10 GeV \newline 
   $p_{T,l^{i}}$ > 5 GeV \ \ \ 
   |$\eta_{l^i}$| < 2.5 \newline
   $p_{T,\g}$ > 20 GeV \ \ 
   |$\eta_\g$| < 2.5 \newline 
   $\Delta R(l^{i},\g)$ > 0.4 \ \ 
    & 
   \\
   \hline    
   
   $\mathbf{VZ\g}$\newline 
   $\mathbf{QCD~-Z\g jj}$\newline 
   ($pp \rightarrow 2j~2l~\g$)\newline \newline \newline 
   Expected \newline events: \newline (EW) 620 \newline (QCD) 31385  
   
   &$m_{ll}$, $m_{jj}$, $p_{T}^{Z}$, 
   $p_{T}^{V}$, $p_{T}^{\g}$, 
   $p_{T}^{l_{i}}$, 
   $p_{T}^{j_{i}}$, 
   $\Delta\eta_{jj}$, $\Delta\phi_{jj}$, $\eta_{j_i}$, 
   $\eta_{l_i}$, 
   $\phi_{j_{i}}$, 
   $\eta^{\g}$, $\phi^{\g}$,
   $p_{T(Z\g)}^{l_i}$, 
   $p_{T(Z)}^{l_i}$, 
   $p_{T(VZ)}^{l_i}$, 
   $p_{T(V)}^{l_i}$, 
   $p_{T(\g)}^{l_i}$, 
   $p_{T(\g)}^{j_i}$, 
   $p_{T(Z)}^{\g}$, $p_{T(Z)}^{V}$, 
   $p_{T(\g)}^{V}$, 
   $p_{T(Z\g)}^{V}$, 
   $p_{T(VZ)}^{\g}$,
   $H_\ell^T(jj)$, 
   $H_\ell^T(ll)$, $H_\ell^T(2l~2j\g)$
   
   & $50 \hspace{-1mm}<\hspace{-1mm}m_{jj}\hspace{-1mm}<\hspace{-1mm}120$ GeV \newline 
   $60 \hspace{-1mm}<\hspace{-1mm}m_{ll}\hspace{-1mm}<\hspace{-1mm}\ 120$ GeV \newline 
   $p_{T,l^{1}}$ > 20 GeV \newline 
   $p_{T,l^{2}}$ > 10 GeV \newline 
   |$\eta_{l^i}$| < 2.5 \newline 
   $p_{T,\g}$ > 20 GeV \ \ \ \ \ 
   |$\eta_\g$| < 2.5 \newline 
    $p_{T,j^{1,2}}$ > 30 $\rm GeV$  \ \ 
   |$\eta_{j^{i}}$| < 2.5 \newline 
   $\Delta R(l^{i},\g)$ > 0.4 \ \ \newline
   $\Delta R(l^{i},j^{k})$ > 0.4 \ \newline$\Delta R(\g,j^{k})$ > 0.4
& 
\\
   \hline    
    $\mathbf{VZZ}$\newline 
    $\mathbf{QCD~-ZZjj}$\newline 
    ($pp\rightarrow2j~2e~2\mu$)\newline \newline \newline 
   Expected \newline events: \newline (EW) 4 \newline (QCD) 95

   &$m_{SFOS-ll}$, $m_{jj}$, $m_{4l}$, $m_{4ljj}$, $p_{T}^{Z}$, $p_{T}^{SFOS-ll}$, $p_{T}^{j_{i}}$, 
   $p_{T}^{l_{i}}$, 
   $p_{T}^{V}$, $p_{T}^{e^\pm\mu^\pm}$, $\Delta\eta_{jj}$, $\Delta\phi_{jj}$, 
   $\eta_{j_i}$, 
   $\eta_{l_i}$, 
   $\phi_{j_{i}}$, 
   $p_{T(ZZ)}^{l_i}$, 
   $p_{T(ZZ)}^{j_i}$, 
   $p_{T(Z_i)}^{l_i}$, 
   $p_{T(Z_i)}^{j_i}$, 
   $p_{T(Z_2)}^{Z_1}$, $p_{T(Z_i)}^{V}$, 
   $p_{T(ZZ)}^{V}$, $H_\ell^T(jj)$, $H_\ell^T(\text{SFOS-}ll)$, $H_\ell^T(4ljj)$
   
   & $50 \hspace{-1mm}<\hspace{-1mm}m_{jj}\hspace{-1mm}<\hspace{-1mm}120$ GeV \newline 
   $60 \hspace{-1mm}<\hspace{-1mm}m_{SFOS-ll}\hspace{-1mm}<\hspace{-1.2mm}\ 120$ GeV \newline 
   $p_{T,l^{1}}$ > 20 GeV \ \ \newline 
   $p_{T,l^{2}}$ > 10 GeV \newline 
   $p_{T,l^{i}}$ > 5 GeV \ \ \newline 
   $p_{T,j^{1,2}}$ > 30 $\rm GeV$  \newline 
   |$\eta_{j^{i}}$| < 2.5 \ \ 
   |$\eta_{l^i}$| < 2.5 \newline 
   $\Delta R(l^{i},j^{k})$ > 0.4
   &
   \\

   \hline
  \end{tabular}
  \caption{ Summary of the analyzed processes. The left column shows the process name along with the definition of the final state and the number of events predicted by the SM (at leading order) corresponding to an integrated luminosity of 300~fb$^{-1}$. 
  The middle column lists the observables studied for each channel. 
  The sorting of the charged leptons and jets is based on the transverse momentum (e.g. $p_T^{l1}>p_T^{l2}$). 
  The Z-boson candidates that are leptonically decaying are ranked according to the difference in mass from the nominal Z boson mass. 
  The term SFOS-ll denotes the Same-Flavor-Opposite-Sign (SFOS) charged lepton pairs. 
  The last column summarizes the phase space selections at parton level used in this analysis.
  }
  \label{tab:kinSel}
\end{table}
In addition to the standard kinematic observables, a set of variables sensitive to the correlations between the properties of the final state particles are used. 
Transverse momentum variables are defined in the plane orthogonal to the direction of the incoming partons and are denoted as $p_{T(\text{any dir.})}$. 
Orthogonal momentum projections are also computed for other planes such as the one identified by the direction of the 4-lepton system in the ZZ$\rightarrow$4l process. 
These transverse momentum observables are labeled with a suffix identifying the orthogonal plane. 
Other variables of interest are the Fox-Wolfram moments (FWMs~\cite{fox1979event, bernaciak2013fox}), which constitute a rotationally invariant set of observables.
These event shape variables describe the geometric correlations among the final-state particles originating from the hard process. 
Using the transverse parameterization, the FWMs are defined as 
\begin{equation}\label{FWMdef}
 H_\ell^T=\sum_{i,j=\text{particle}} \frac{p_{T,i} p_{T,j}}{\left(\sum_{k=i,j} p_{T,k}\right)^2}P_\ell\left(\text{cos }\Omega_{i,j}\right)\ ,
\end{equation}
where $i$ and $j$ denote the single particles/jets in a subset of the final state. The integer index $\ell\geq0$ is the order of the moment $H_\ell^T$ and corresponds to the degree of the Legendre polynomial. The argument $\text{cos }\Omega_{i,j}$ is defined as 
\begin{equation}\label{cosOmega}
    \text{cos }\Omega_{i,j}=\text{cos }\theta_{i}\ 
    \text{cos }\theta_{j}+ 
    \text{sin }\theta_i\ 
    \text{sin }\theta_j\ 
    \text{cos}\left(\varphi_i-\varphi_j\right)\ ,
\end{equation}
where $\theta$ and $\varphi$ are the polar and azimuthal angles given by the directions of the i- and j-th
particles' momenta.  
\section{The analyses}
The channels studied for this work are VZZ, VZ$\g$, WZ$\g$, and ZZ$\g$ in the  final states described in table\ref{tab:VVVprocesses}. 
A separate paragraph is dedicated to the benchmark study of the fully leptonic WWW channel, see section~\ref{subsec:WWW}. 
For the semi-leptonic channels, a study focused on the purely electroweak processes is performed to characterize the effect of dimension-6 bosonic operators, 
excluding the QCD-induced SM background and the respective EFT contribution. 
The latter are separately investigated in the phase space region of interest to evaluate the sensitivity of the inclusive electroweak-QCD  channel. 

\paragraph{Semi-leptonic VZZ}
The process of triboson VZZ production is studied with the experimental semi-leptonic signature 2jets+2e+2$\mu$, 
where V indicates the hadronically decaying vector boson W or Z, and the other two Z bosons decay into two charged lepton pairs of opposite flavor. The main background process is the QCD-induced ZZjj production, 
which is generated separately from the electroweak component. 
The main Feynman tree-level diagrams include the triple and quartic gauge couplings as well as the Higgs-gauge boson couplings (see figure~\ref{fig:EWK_Feynm}). 
Both ZZZ and WZZ triboson intermediate states depend on the WWZ coupling at the tree level, while only WZZ depends on the quadrilinear WWZZ coupling. 
\begin{figure}[p!]
  \centering 
	\subfigure[\label{subfig:VZZnoQCD_m4l_cW}]{%
        \includegraphics[width=.48\textwidth]{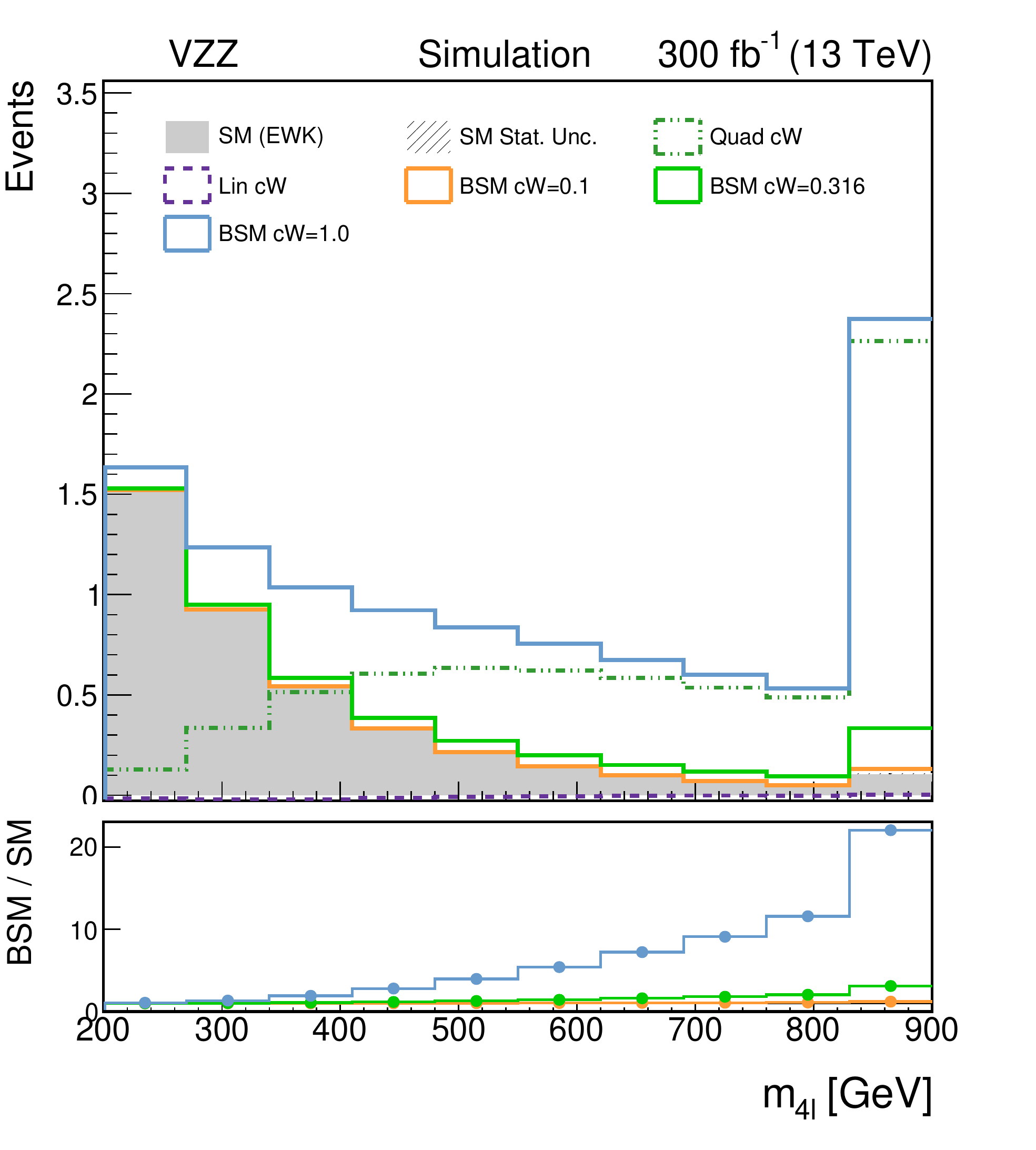}
	}
	\subfigure[\label{subfig:VZZnoQCD_FWM_cW}]{%
	    \includegraphics[width=.48\textwidth]{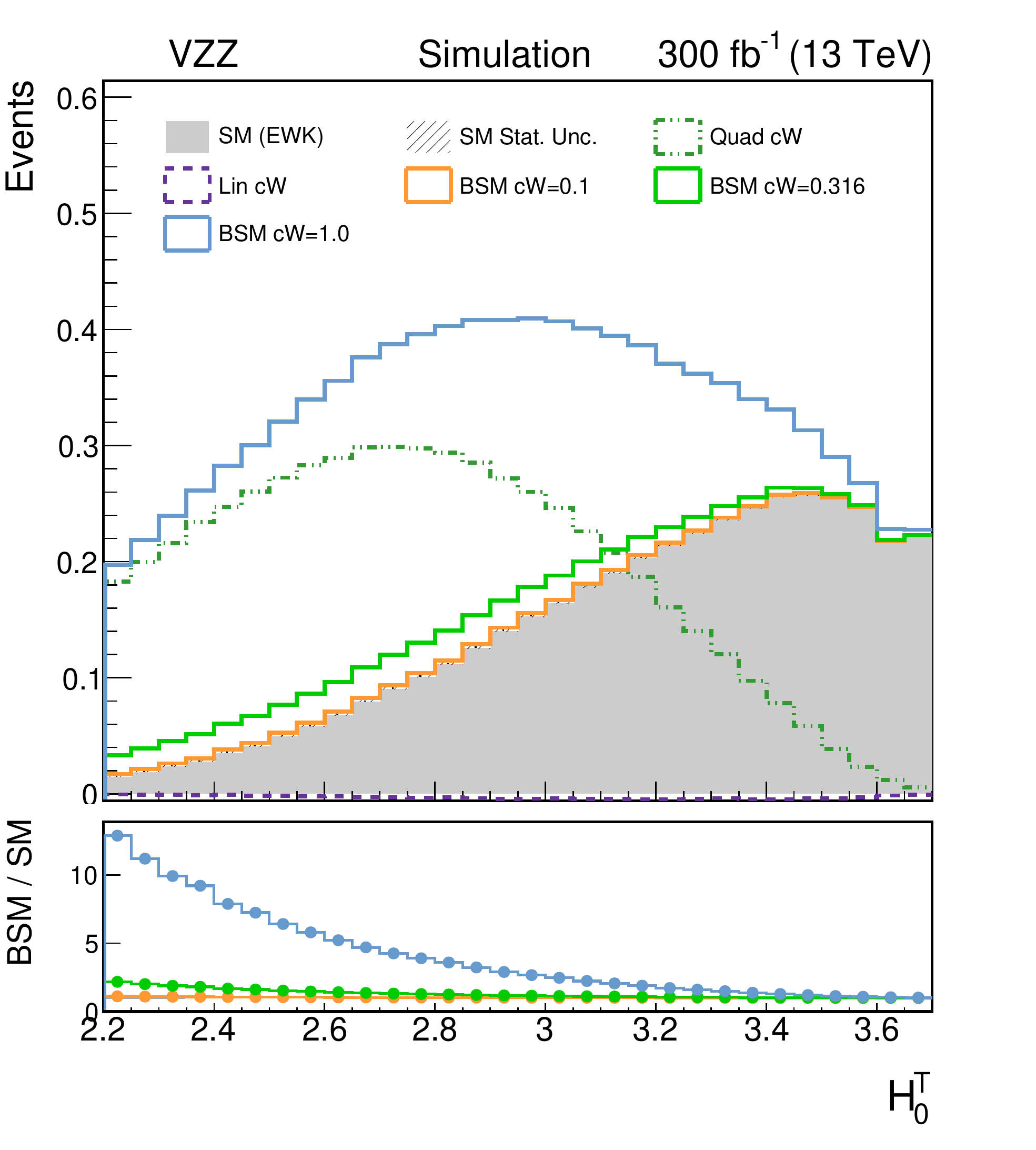}
	}
	\subfigure[\label{subfig:VZZnoQCD_ptV_cHW}]{%
	    \includegraphics[width=.48\textwidth]{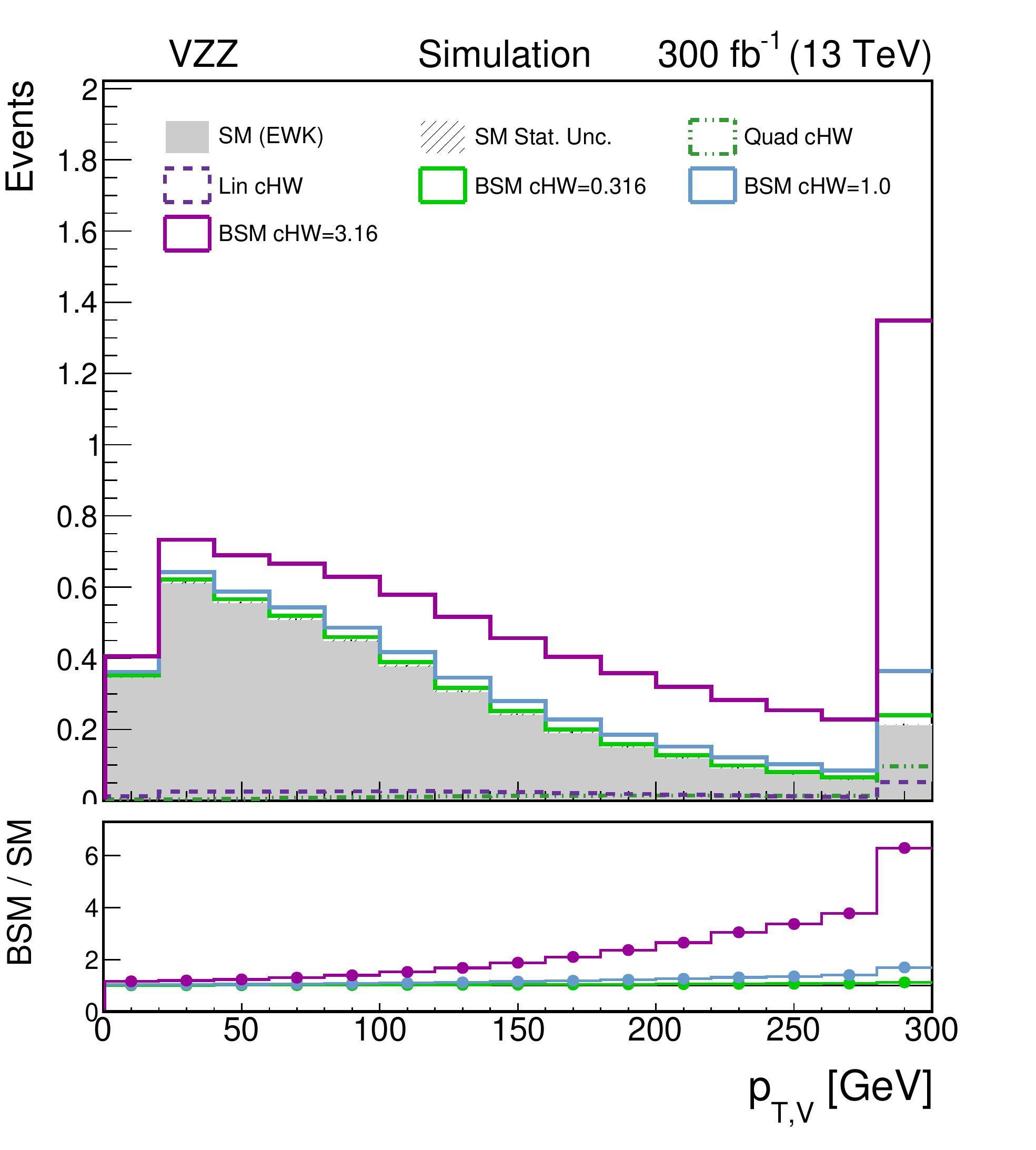}
	}
	\subfigure[\label{subfig:VZZnoQCD_ptemuplus_cHWB}]{%
	    \includegraphics[width=.48\textwidth]{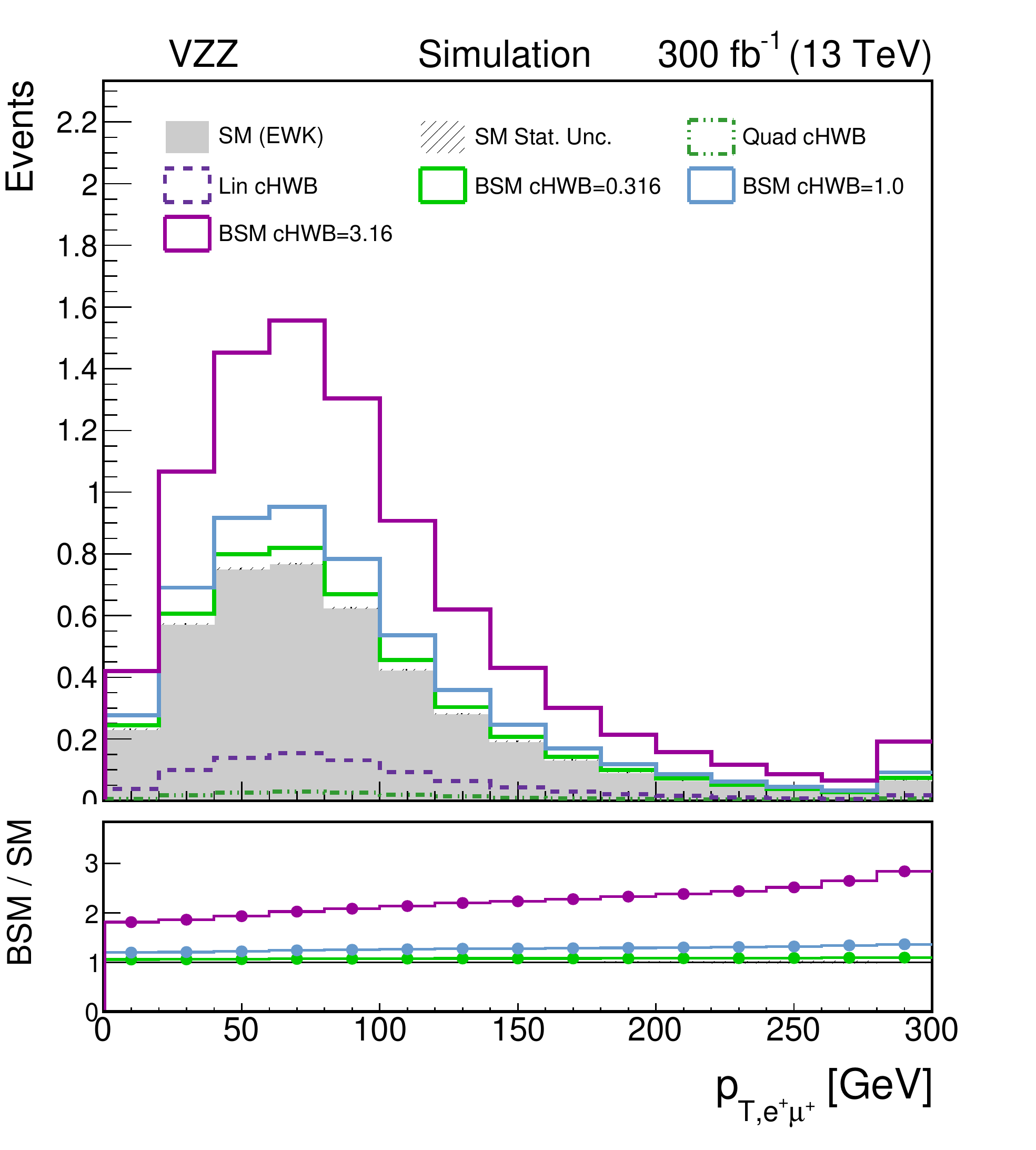}
	}
    
\caption{ Comparison of SM (filled gray histograms) and BSM (lines) expected yield distributions after the selection of table~\ref{tab:kinSel} for the VZZ production process, for an integrated luminosity of 300~fb$^{-1}$. 
Events with QCD-induced vertices (QCD-ZZjj production) are excluded.  
The dashed lines show the distributions of the quadratic component (dashed green) and the linear interference term (dashed violet).  
The solid lines show the behavior of a theory where a single operator Q$_{i}$ is added to the SM Lagrangian, with $\Lambda$~=~1~TeV and the Wilson coefficients set to benchmark values (c$_{i} = 10^{-1},\,10^{-\frac{1}{2}},\,1,\,10^{\frac{1}{2}}$). 
For all the distributions, the last bin contains all the overflow events. 
The ratio of BSM to SM events is shown for each bin. 
The plots show the effects induced by Q$_{W}$ on the distributions of $m_{4l}$ (a) and the FWM $H_0^T$ (b), by Q$_{HW}$ on the transverse momentum of the jet pair (c), and by Q$_{HWB}$ on the transverse momentum of the positively charged lepton pair (d).
\label{fig:Distributions_VZZ_noQCD}  }  
\end{figure}
\begin{figure}[p!]
  \centering 
	\subfigure[\label{subfig:VZZ_ptl1_cW}]{%
        \includegraphics[width=.48\textwidth]{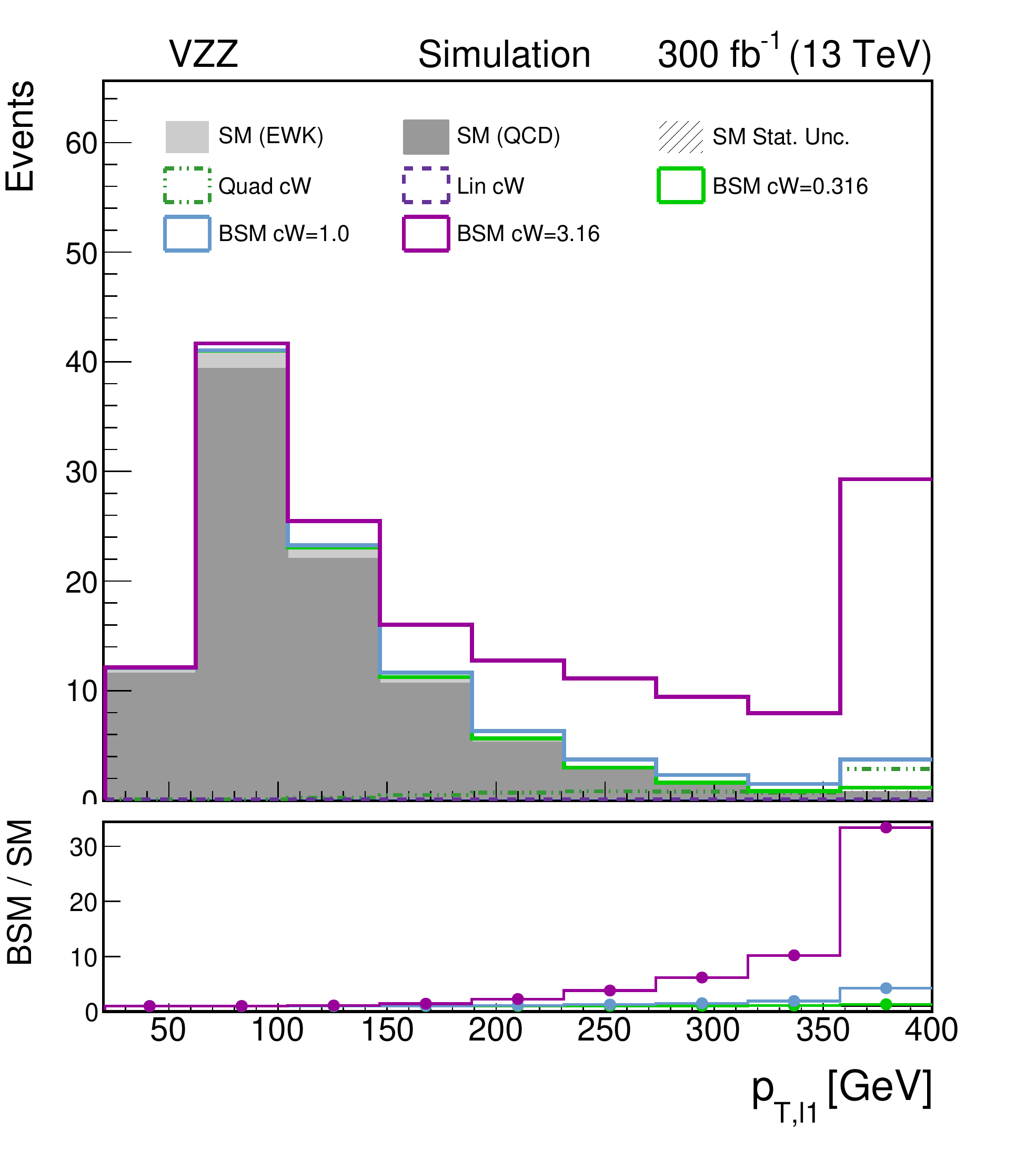}
	}
	\subfigure[\label{subfig:VZZ_mjj_cW}]{%
	    \includegraphics[width=.48\textwidth]{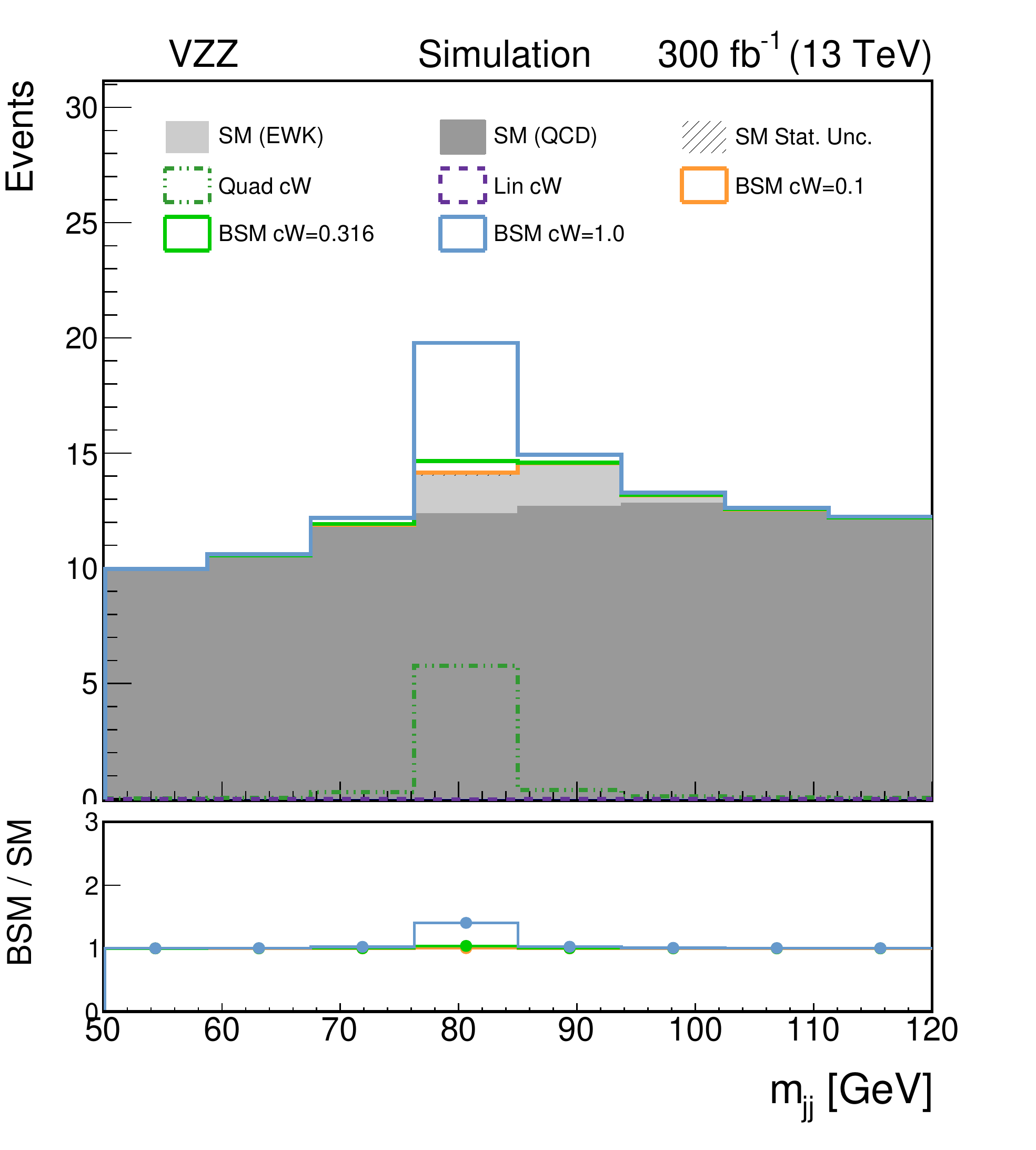}
	}
	\subfigure[\label{subfig:VZZ_ptV_cHW}]{%
	    \includegraphics[width=.48\textwidth]{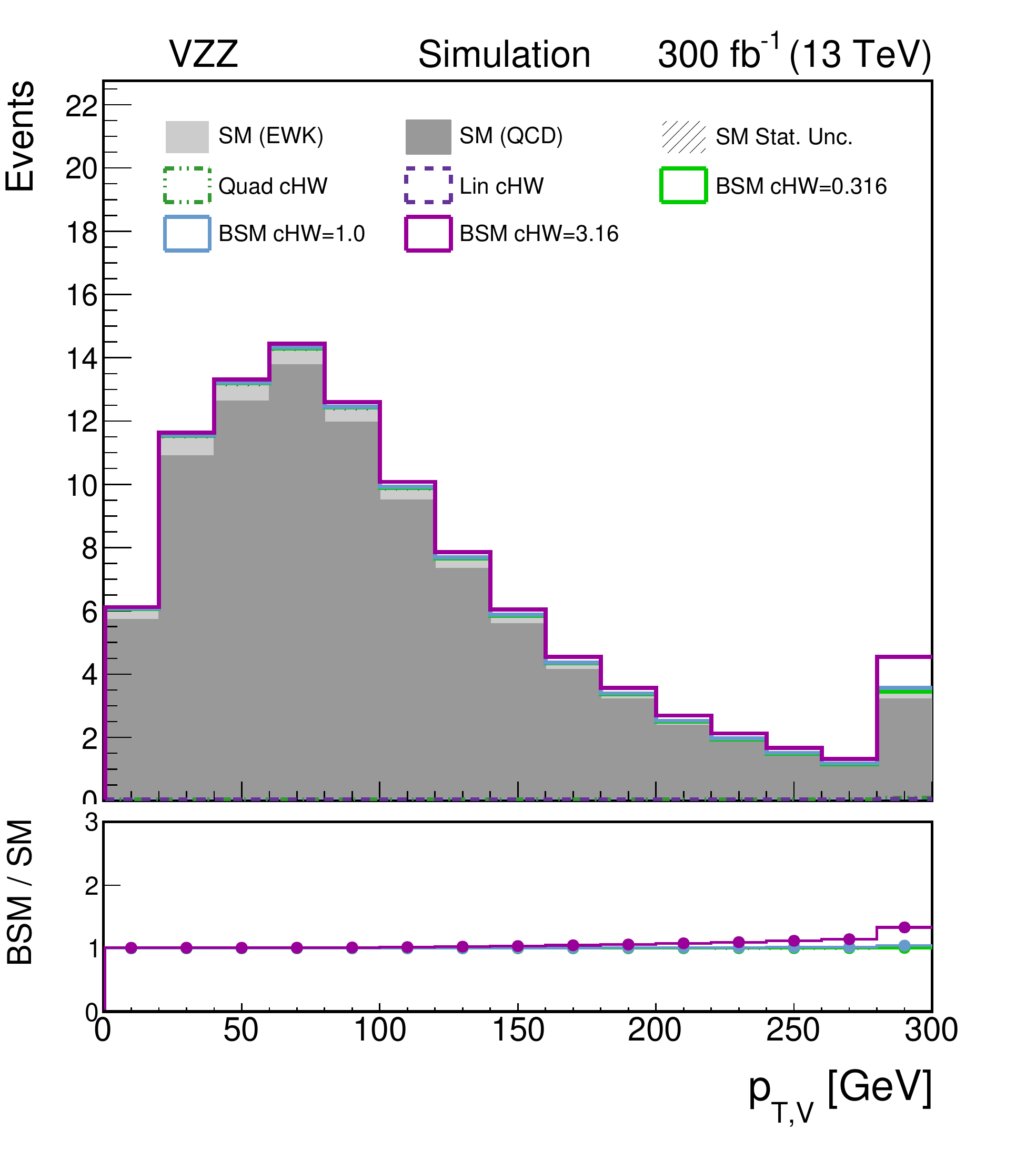}
	}
	\subfigure[\label{subfig:VZZ_ptVRelZ1_cHWB}]{%
	    \includegraphics[width=.48\textwidth]{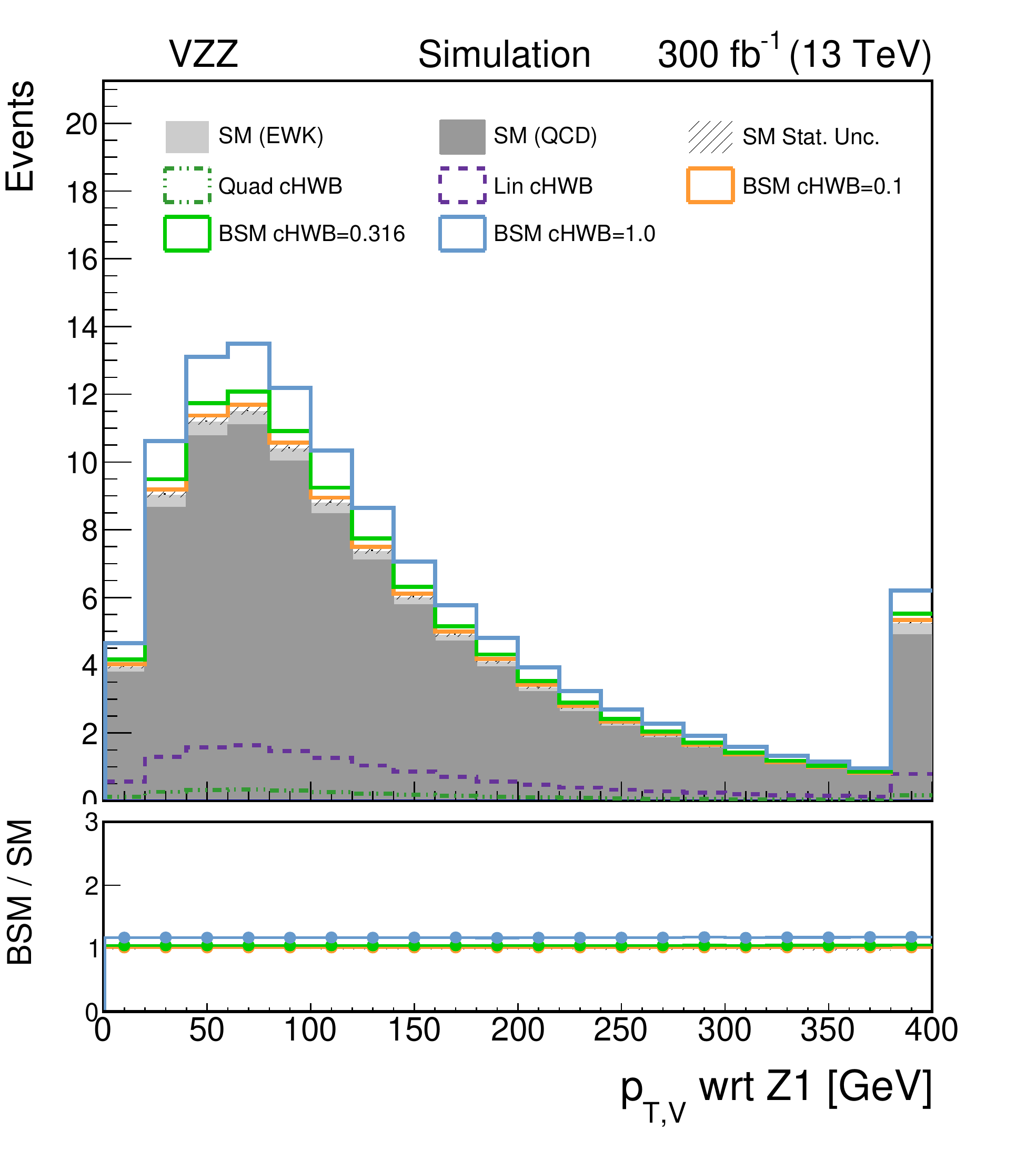}
	}
    
\caption{ Comparison of SM (filled histograms) and BSM (lines)
  expected yield distributions 
  after the selection of table~\ref{tab:kinSel} for the VZZ process, 
  for an integrated luminosity of 300~fb$^{-1}$. 
  The SM distribution is shown as a stacked histogram, summing electroweak VZZ (light gray) and QCD-ZZjj (dark gray) components, while all BSM distributions are superimposed.
  The dashed lines show the distributions of the quadratic component (dashed green) and the linear interference term (dashed violet).  
The solid lines show the behavior of a theory where a single operator Q$_{i}$ is added to the SM Lagrangian, with $\Lambda$~=~1~TeV and the Wilson coefficients set to benchmark values (c$_{i} = 10^{-1},\,10^{-\frac{1}{2}},\,1,\,10^{\frac{1}{2}}$). 
For all the distributions, the last bin contains all the overflow events. 
The bottom plot shows the ratio of BSM to SM events for each bin. 
  The plots show the effects induced by Q$_{W}$ on the distribution of $p_{T}^{l_1}$ (a) and $m_{jj}$ (b), by Q$_{HW}$ on the transverse momentum of the hadron jet pair $p_{T}^{V}$, and by Q$_{HWB}$ on the $p_{T}^{V}$ relative to the longitudinal direction identified by the leading leptonically-decaying Z candidate.
  \label{fig:Distributions_VZZ}  }  
\end{figure}

Figure~\ref{fig:Distributions_VZZ_noQCD} shows several distributions of the variables of interest for a subset of the operators. 
The SM predictions are compared with the EFT components and the overall BSM distribution obtained by setting the Wilson coefficients to non-zero benchmark values ($c_i=10^{-1},~10^{-\frac{1}{2}},~1,~10^{\frac{1}{2}}$).
The distribution of figure~\ref{subfig:VZZnoQCD_m4l_cW} 
is the invariant mass $m_{4l}$ of the charged leptons in the final state. 
The quadratic EFT component associated with the operator Q$_W$ leads to a large deviation in the tail of $m_{4l}$. Similarly, in the case of the zero-th order FWM $H^T_0$ for the 4ljj system, 
the quadratic component is well resolved with respect to the SM background (figure~\ref{subfig:VZZnoQCD_FWM_cW}). 
Figure~\ref{subfig:VZZnoQCD_ptV_cHW} shows the expected distributions for the Q$_{HW}$ operator as a function of the transverse momentum of the best vector boson candidate V$\rightarrow$jj. In this case, the BSM distribution exhibits a small excess with respect to the SM, which is more pronounced at high values of~$p_T^V$. 
Figure~\ref{subfig:VZZnoQCD_ptemuplus_cHWB} shows the the expected distributions for the Q$_{HWB}$ operator as a function of the total transverse momentum of the positively charged leptons in the final states. In contrast to the previous case, the linear interference term Q$_{HWB}$-SM induces a significant deviation. 

Figure~\ref{fig:Distributions_VZZ} refers to the study that considers the contribution of the QCD-ZZjj background processes.
Figure~\ref{subfig:VZZ_ptl1_cW} shows the deviation induced by the Q$_W$ operator on the $p_T^{l1}$ spectrum. The presence of the QCD-ZZjj background suppresses the deviations from SM, except in the high-$p_T^{l1}$ region. Figure~\ref{subfig:VZZ_mjj_cW} shows how the shape of the QCD-ZZjj background is nearly uniform over the probed range of the dijet invariant mass distribution. The electroweak component has a broad peak around the nominal W and Z masses, whereas the quadratic component of Q$_{W}$ resonates in correspondence with the nominal W mass. This can be explained by the unique sensitivity of the WZZ channel to the anomalous triple WWZ and quartic WWZZ gauge couplings induced by Q$_W$, which in turn do not affect the ZZZ channel. 
Figures~\ref{subfig:VZZ_ptV_cHW}-\ref{subfig:VZZ_ptVRelZ1_cHWB} 
illustrate the impact of the QCD-ZZjj background on the Q$_{HW}$ and Q$_{HWB}$ operators for the variables $p_T^V$ and $p_{T(Z_1)}^V$. In both cases, the sensitivity to SM deviations is suppressed by the overwhelming QCD-induced background.  

\begin{figure}[p!]
  \centering 

 	\subfigure[\label{subfig:WZG_cHB}]{%
		\includegraphics[width=0.48\textwidth]{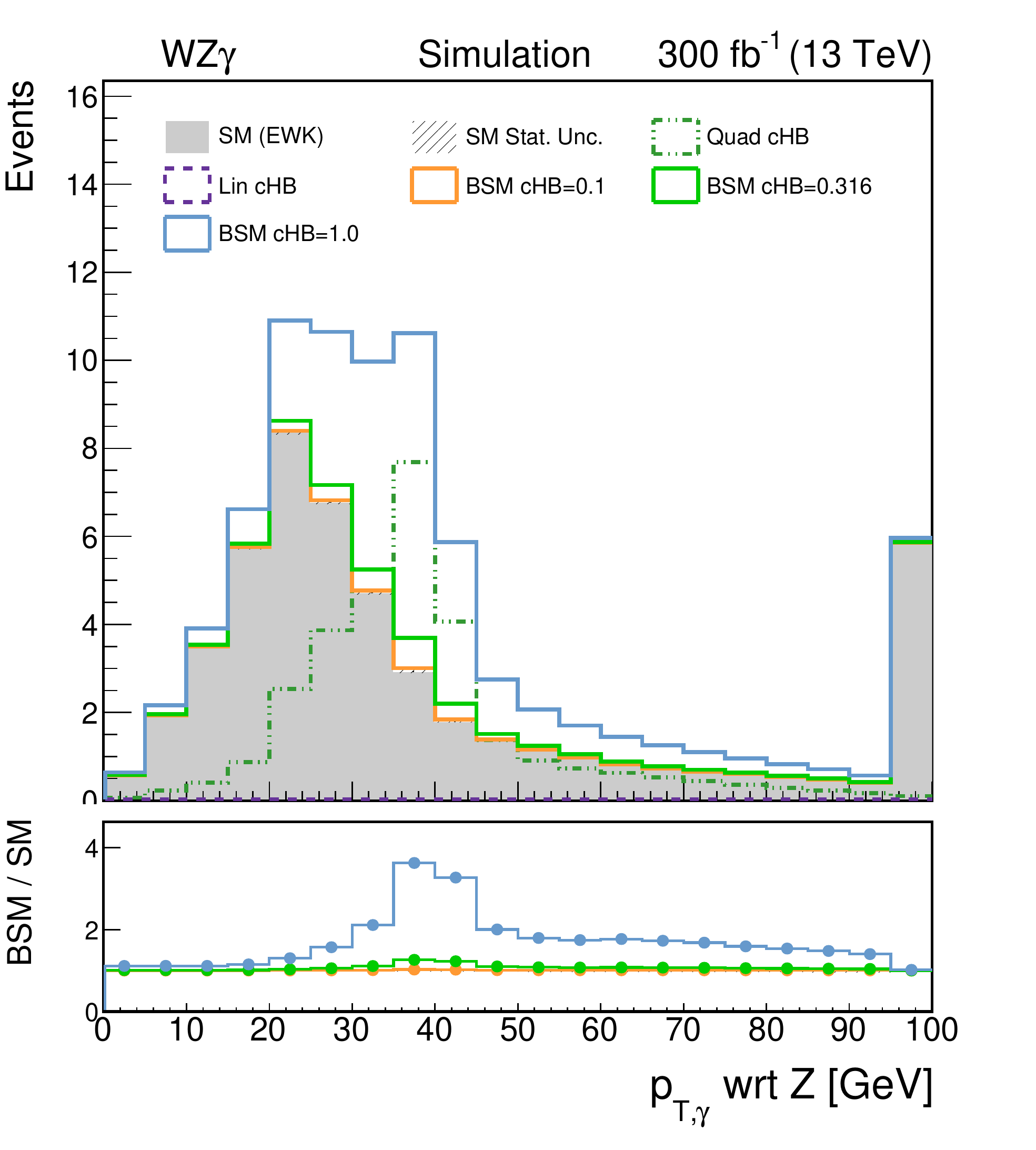}%
	}
	\subfigure[\label{subfig:WZG_cHW}]{%
		\includegraphics[width=0.48\textwidth]{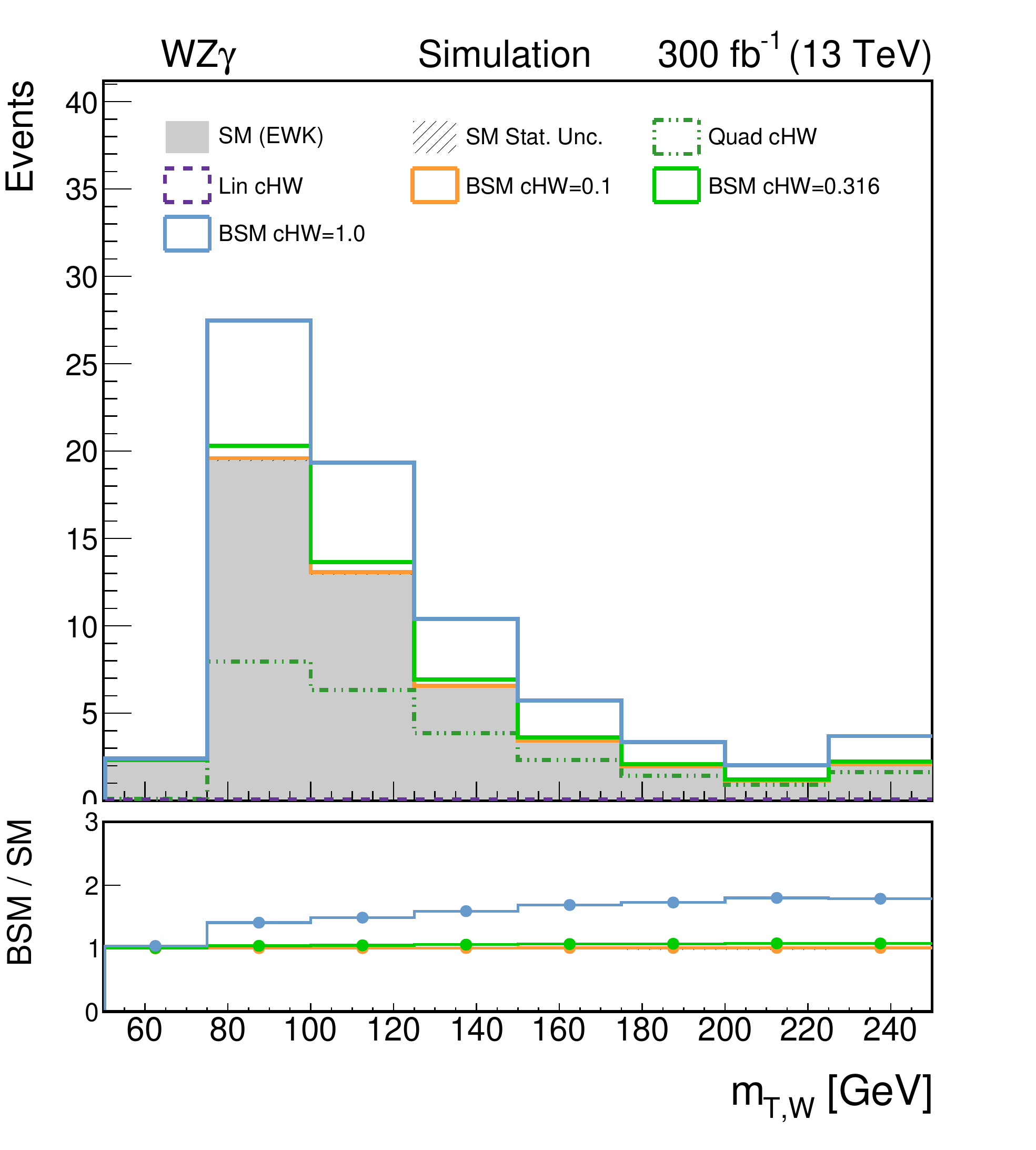}%
	}
    \subfigure[\label{subfig:WZG_cW}]{%
		\includegraphics[width=0.48\textwidth]{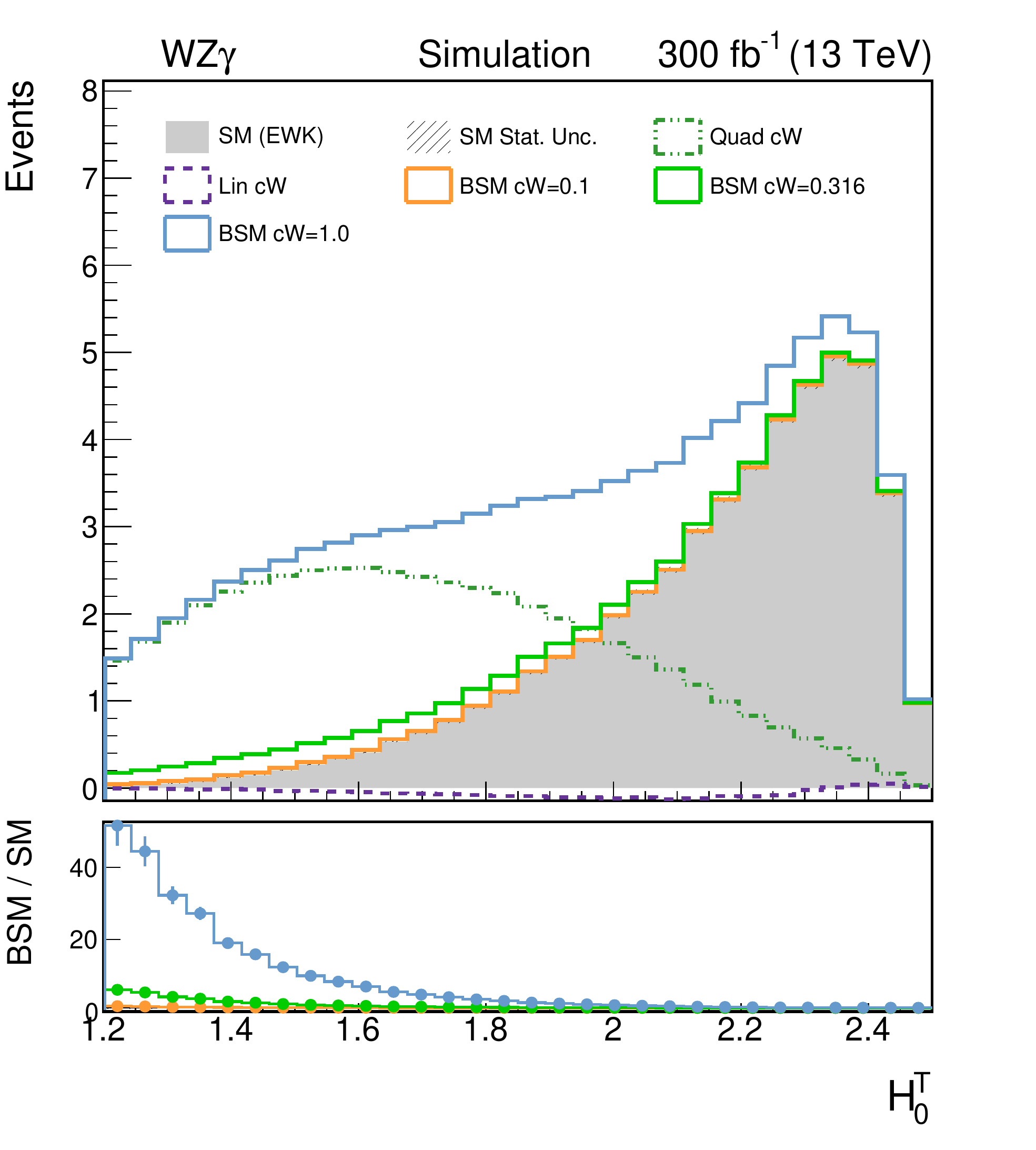}%
	}
	\subfigure[\label{subfig:WZG_cHDD}]{%
		\includegraphics[width=0.48\textwidth]{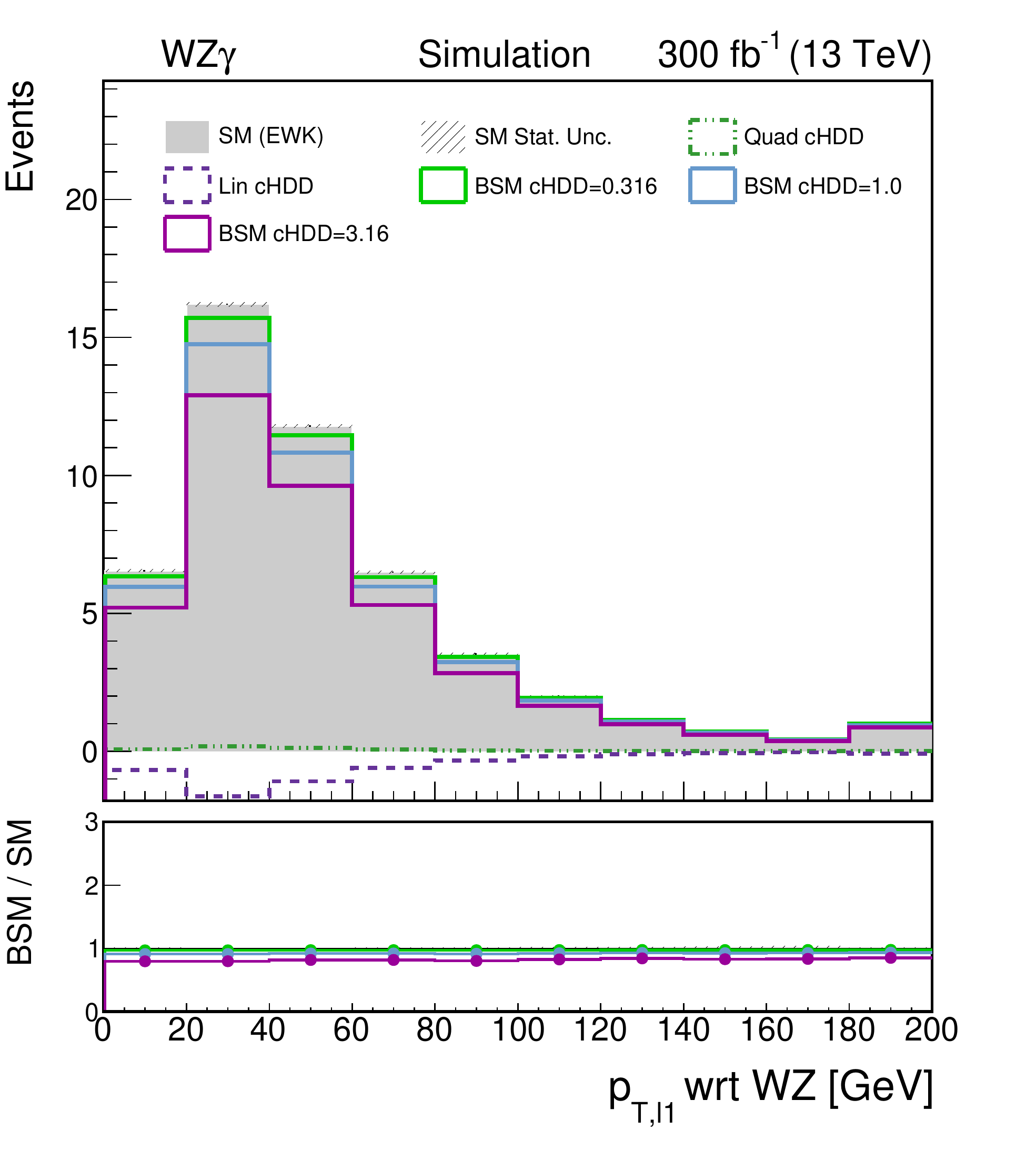}%
	}
\caption{ Comparison of SM (filled gray histograms) and BSM (lines) expected yield distributions 
  after the selection of table~\ref{tab:kinSel} for the WZ$\g$ process,
  for an integrated luminosity of 300 fb$^{-1}$. 
   The dashed lines show the distributions of the quadratic component (dashed green) and the linear interference term (dashed violet).  
The solid lines show the behavior of a theory where a single operator Q$_{i}$ is added to the SM Lagrangian, with $\Lambda$~=~1~TeV and the Wilson coefficients set to benchmark values (c$_{i} = 10^{-1},\,10^{-\frac{1}{2}},\,1,\,10^{\frac{1}{2}}$). 
For all the distributions, the last bin contains all the overflow events. 
The bottom plot shows the ratio of BSM to SM events for each bin. 
  The plots show the effects induced by Q$_{HB}$ on the distribution of the transverse momentum of $\g$ relative to a longitudinal direction defined by the Z candidate 4-momentum (a), by Q$_{HW}$ on the transverse mass of the W boson candidate (b), by Q${_W}$ on the FWM $H_0^T$ (c), and by  Q$_{HD}$ on the transverse momentum of the leading lepton relative to a longitudinal direction defined by the total 4-momentum of the W and Z bosons candidates. 
  \label{fig:Distributions_WZG}}  
\end{figure}

\paragraph{Fully leptonic WZ$\g$}
The process of triboson WZ$\g$ production was studied in the final state $\mu^\pm\overset{\scriptscriptstyle(-)}{\nu_\mu}$ + 2e + $\g$.  
It corresponds to the process pp$\rightarrow $W$^+$Z$\g\rightarrow\mu^+\nu_\mu$e$^+$e$^-\g$, 
where the Z boson decays into an electron-positron pair, 
and the W boson decays into a muon and the corresponding neutrino W$^\pm\rightarrow\mu^\pm\overset{\scriptscriptstyle(-)}{\nu_\mu}$. 
Similarly to the VZZ production, this process also depends on a quartic gauge coupling at the tree level in the SM, which in this case is WWZ$\g$. 
Figure~\ref{subfig:WZG_cHB} shows 
the $p_T^{\g}$ distribution relative to the Z-boson direction for the Q$_{HB}$ operator. 
This kinematic variable is very sensitive to the anomalies induced by the individual quadratic term of the Q$_{HB}$ operator, 
with a large deviation in the bulk of the $p_{T}$ spectrum. 
Since the anomalous HZ$\g$ coupling induced by Q$_{HB}$ is forbidden in the SM at the tree level, the linear interference EFT term is negligible. 
For the same theoretical consideration, the Q$_{HW}$ quadratic component dominates over the linear interference with the SM, as shown for transverse mass spectrum of the W boson in figure~\ref{subfig:WZG_cHW}. 
Figure~\ref{subfig:WZG_cW} shows 
the distribution obtained for the Q$_W$ operator as a function of the zero-th order FWM $H^T_0$ of the $3l\nu\g$ system:
\begin{equation}
    H^T_0=\sum_{a,b\ \in{\{e^\pm, \mu^+, \g\}}}\frac{p_T^ap_T^b}{p_T^\text{Tot.}}
    \text{~with~} p_T^{\nu} \equiv \cancel{\it{E}}_{T}
\end{equation}
where $\cancel{\it{E}}_{T}$ denotes the missing transverse energy per event. 
For the electroweak WZ$\g$ production, the shape of the EFT quadratic component is well resolved with respect to the SM one. 
The Q$_{HD}$ operator, shown in figure~\ref{subfig:WZG_cHDD} exhibits a completely different phenomenological behavior. The transverse momentum of the leading lepton is obtained relative to the longitudinal direction of the total momentum of the WZ diboson system. 
In this case, the linear component accounting for the destructive interference of the Q$_{HD}$-induced diagrams with the SM ones dominates over the quadratic term, leading to a nearly uniform decrease in the expected yield.  

\paragraph{Fully leptonic ZZ$\g$} 
We study the process of triboson ZZ$\g$ production with the experimental fully leptonic signature 2e+2$\mu$+$\g$. 
Unlike the other triboson processes considered, it does not depend at the tree level on gauge couplings, causing the absence of Q$_W$-induced diagrams. Nevertheless, the Higgs-gauge boson couplings affect the electroweak vertices present in the diagrams of the ZZ$\g$ process.

\begin{figure}[p!]
  \centering 
 
	\subfigure[\label{subfig:ZZG_cHB}]{%
		    \includegraphics[width=.48\textwidth]{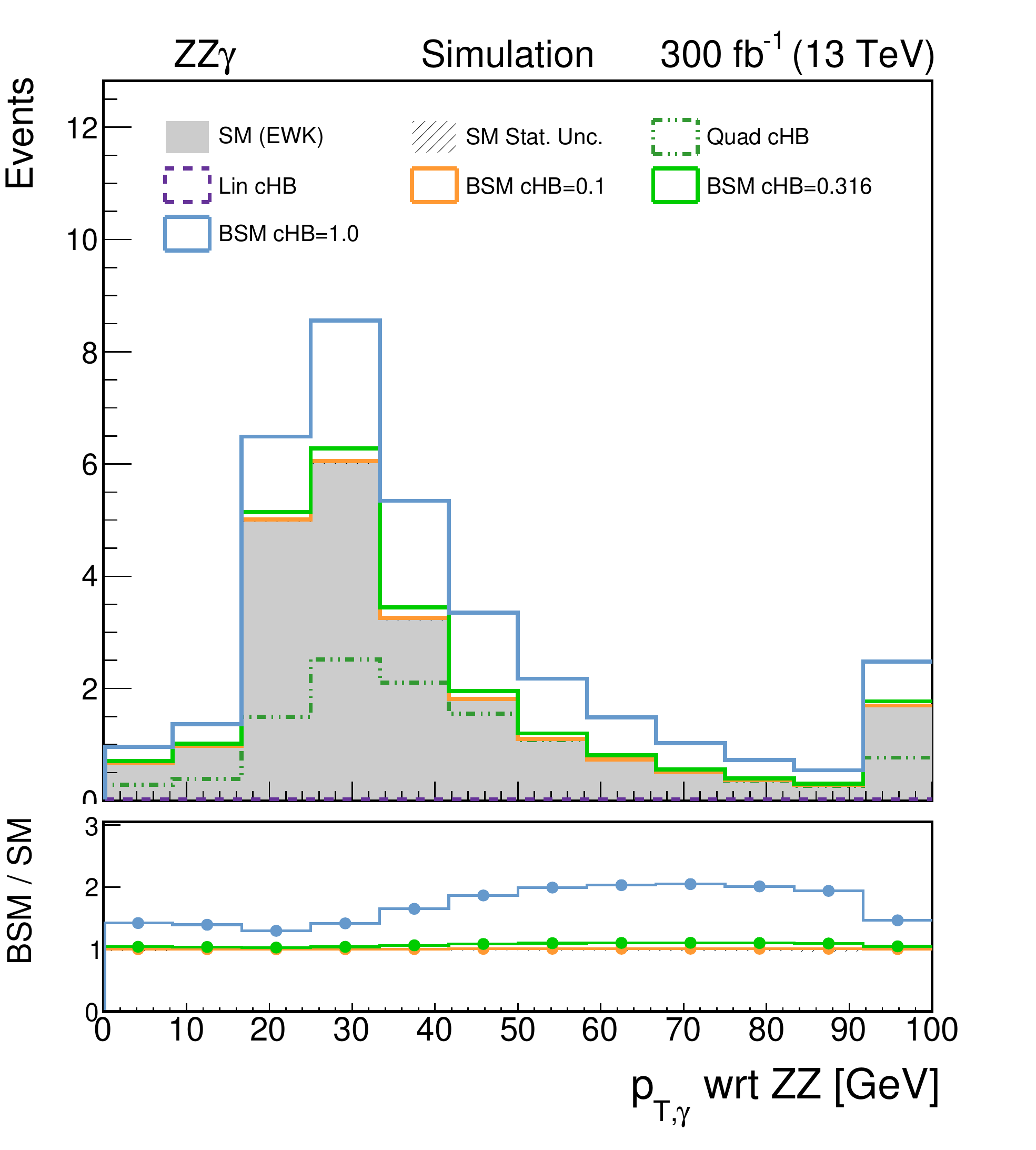}
	}
	\subfigure[\label{subfig:ZZG_cHW}]{%
		    \includegraphics[width=.48\textwidth]{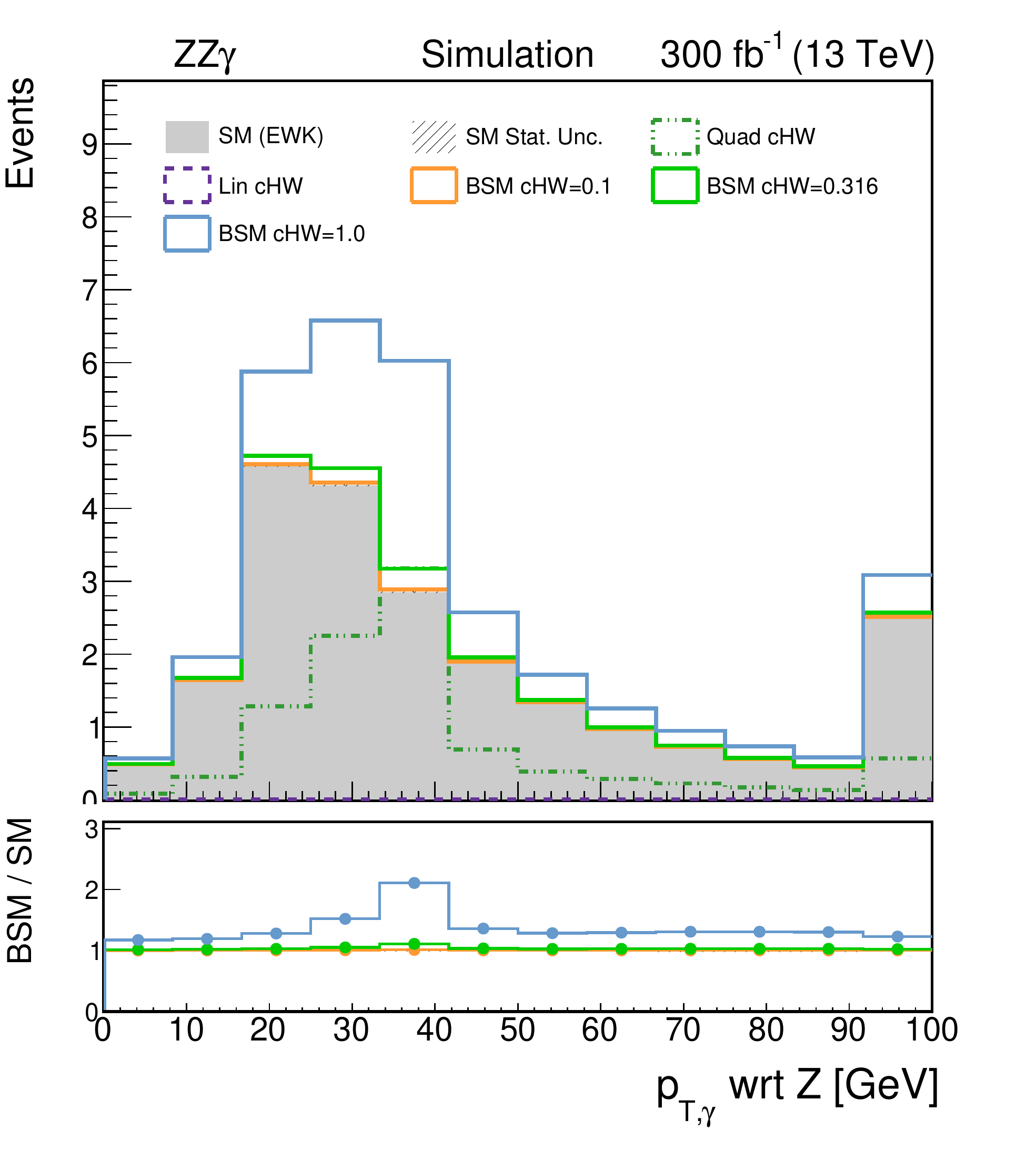}
	}
	\subfigure[\label{subfig:ZZG_cHDD}]{%
		    \includegraphics[width=.48\textwidth]{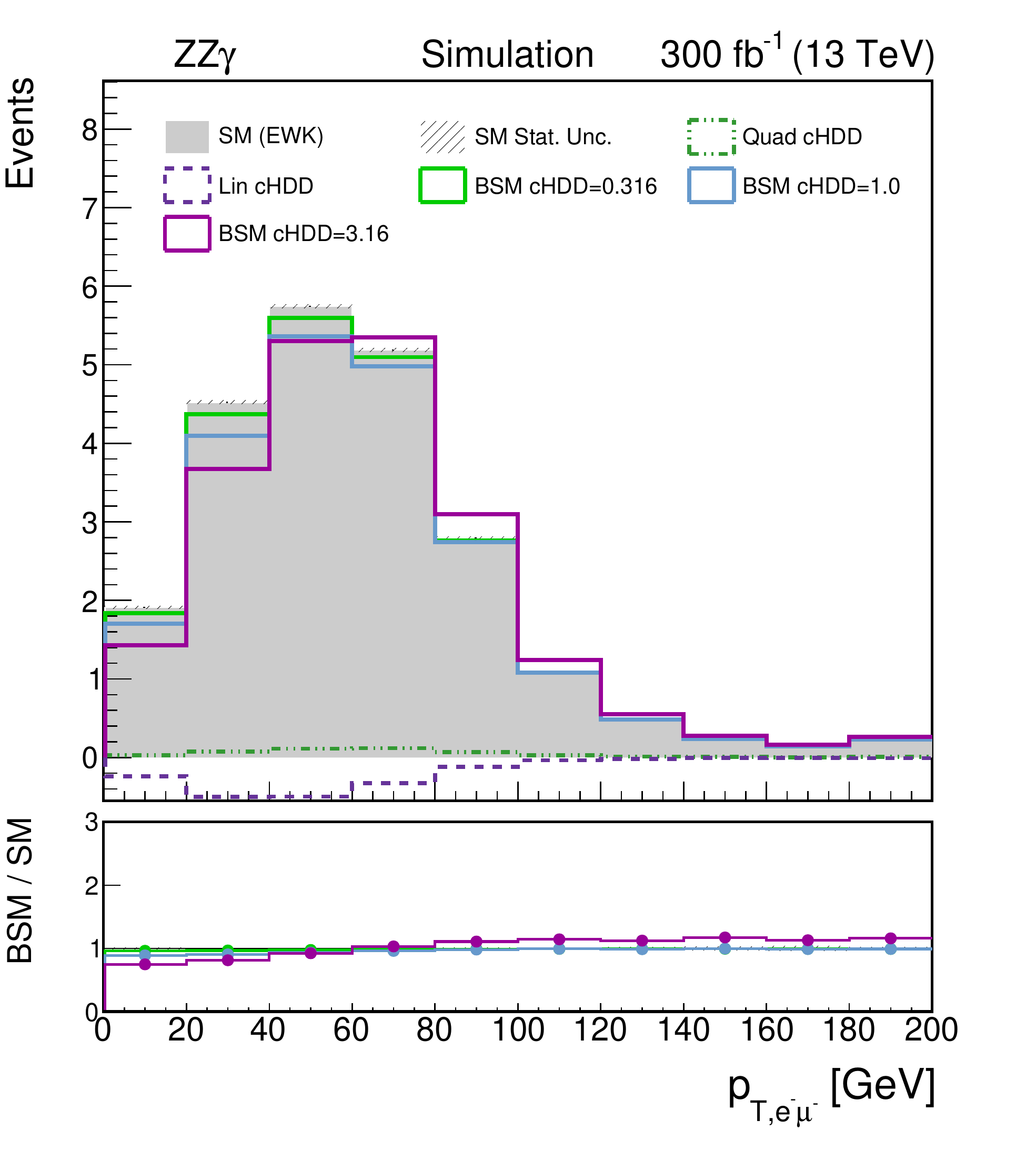}
	}
	\subfigure[\label{subfig:ZZG_cHWB}]{%
		    \includegraphics[width=.48\textwidth]{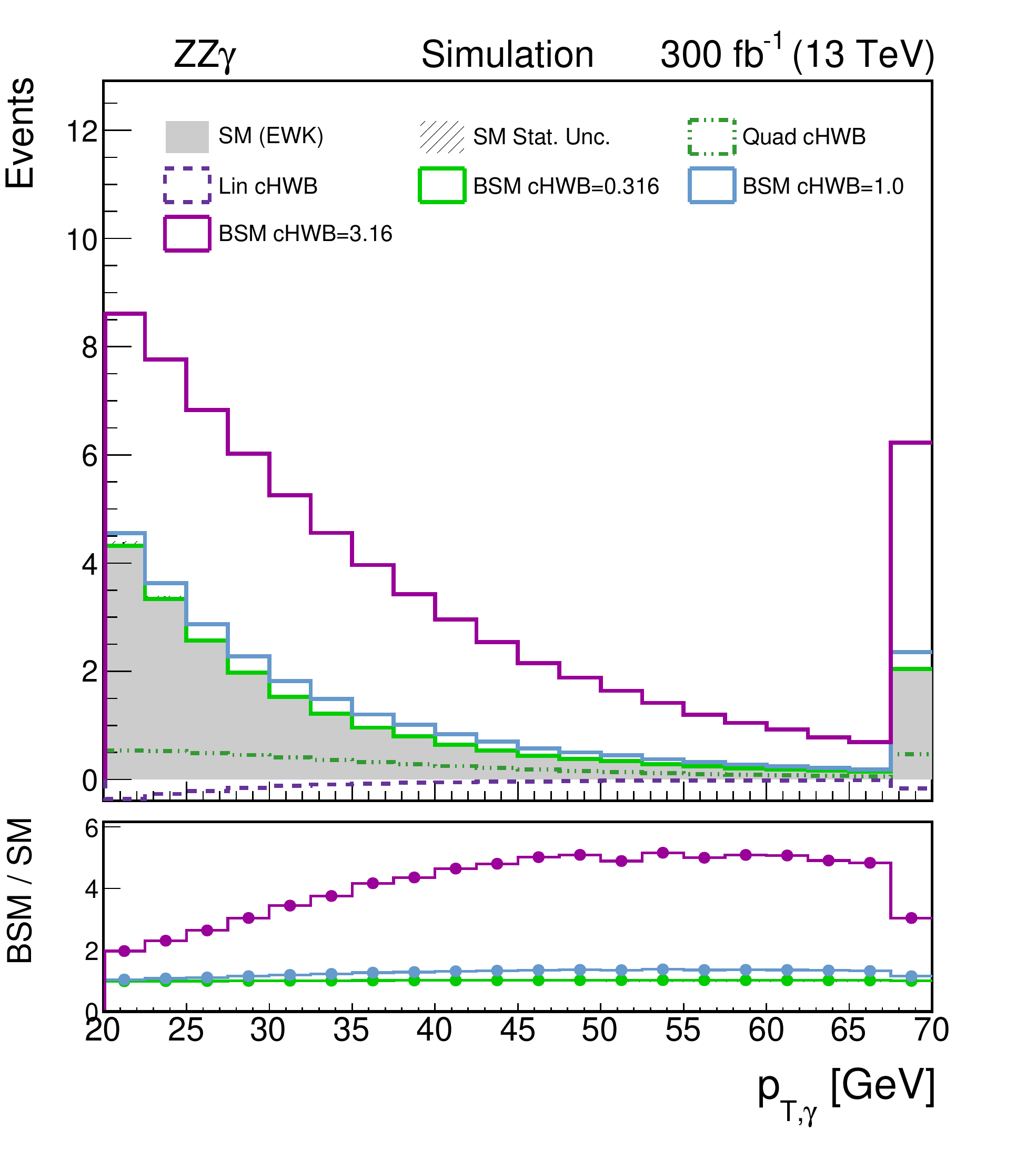}
	}
    
\caption{ Comparison of SM (filled gray histograms) and BSM (lines)
  expected yield distributions 
  after the selection of table~\ref{tab:kinSel} for the ZZ$\g$ process,
  for an integrated luminosity of 300 fb$^{-1}$. 
 The dashed lines show the distributions of the quadratic component (dashed green) and the linear interference term (dashed violet).  
The solid lines show the behavior of a theory where a single operator Q$_{i}$ is added to the SM Lagrangian, with $\Lambda$~=~1~TeV and the Wilson coefficients set to benchmark values (c$_{i} = 10^{-1},\,10^{-\frac{1}{2}},\,1,\,10^{\frac{1}{2}}$). 
For all the distributions, the last bin contains all the overflow events. 
The bottom plot shows the ratio of BSM to SM events for each bin. 
  The distributions show the effects induced by Q$_{HB}$ on the photon transverse momentum relative to the direction defined by the total 4-momentum of the Z boson pair, $p_{T(ZZ)}^\g$, (a), the effects induced by Q$_{HW}$ on the $p_{T(Z1)}^\g$, relative to the leading Z boson candidate (b), by Q$_{HD}$ on the total transverse momentum of the negatively charged leptons (c), and by Q$_{HWB}$ on the standard $p_T^\g$ (d). 
\label{fig:Distributions_ZZG}}  
\end{figure}
\begin{figure}[p!]
  \centering 
 
	\subfigure[\label{subfig:VZGnoQCD_cW}]{%
	    \includegraphics[width=.48\textwidth]{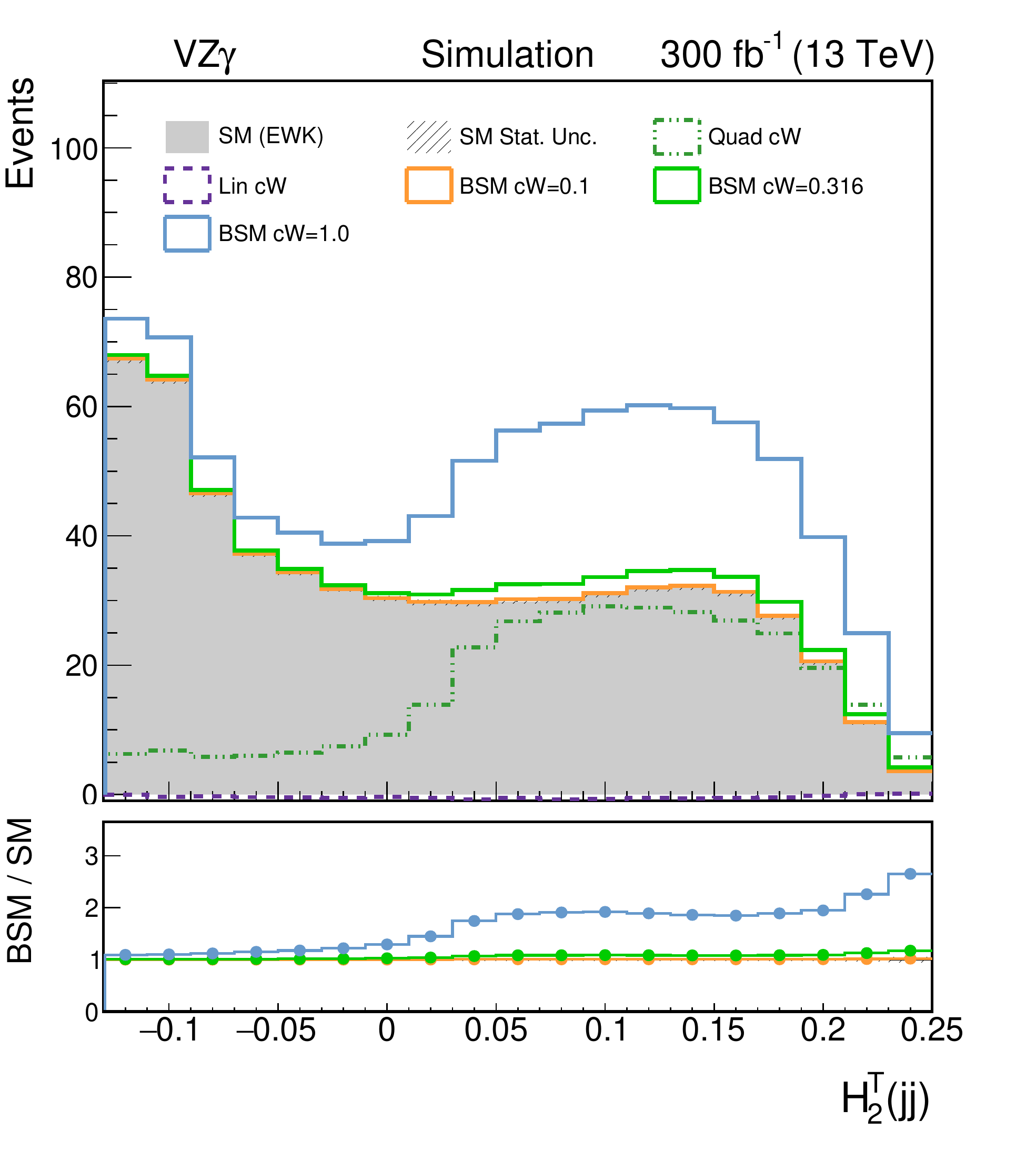}
	}
	\subfigure[\label{subfig:VZGnoQCD_cHWB}]{%
	    \includegraphics[width=.48\textwidth]{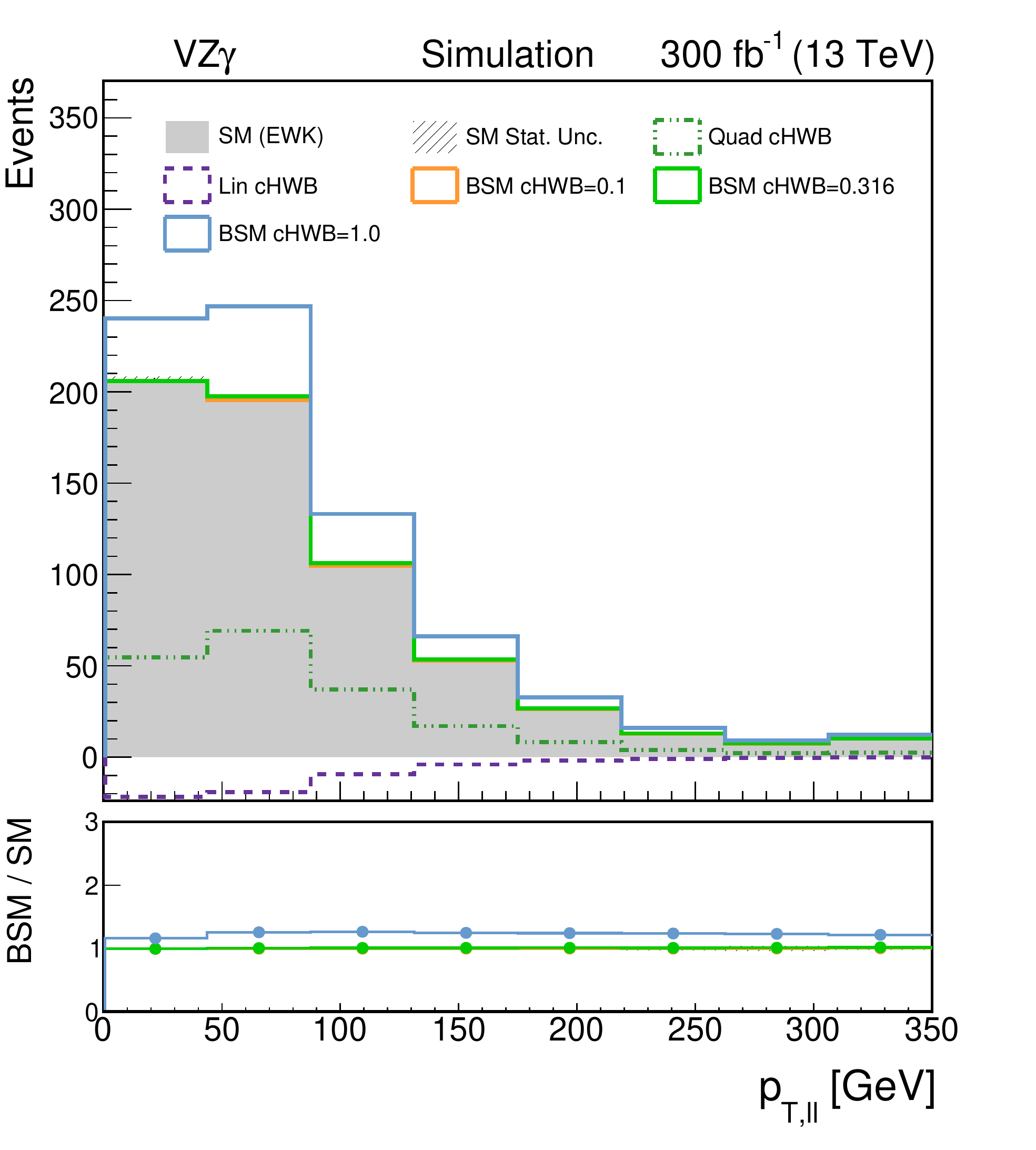}
	}
	\subfigure[\label{subfig:VZGnoQCD_cHW}]{%
	    \includegraphics[width=.48\textwidth]{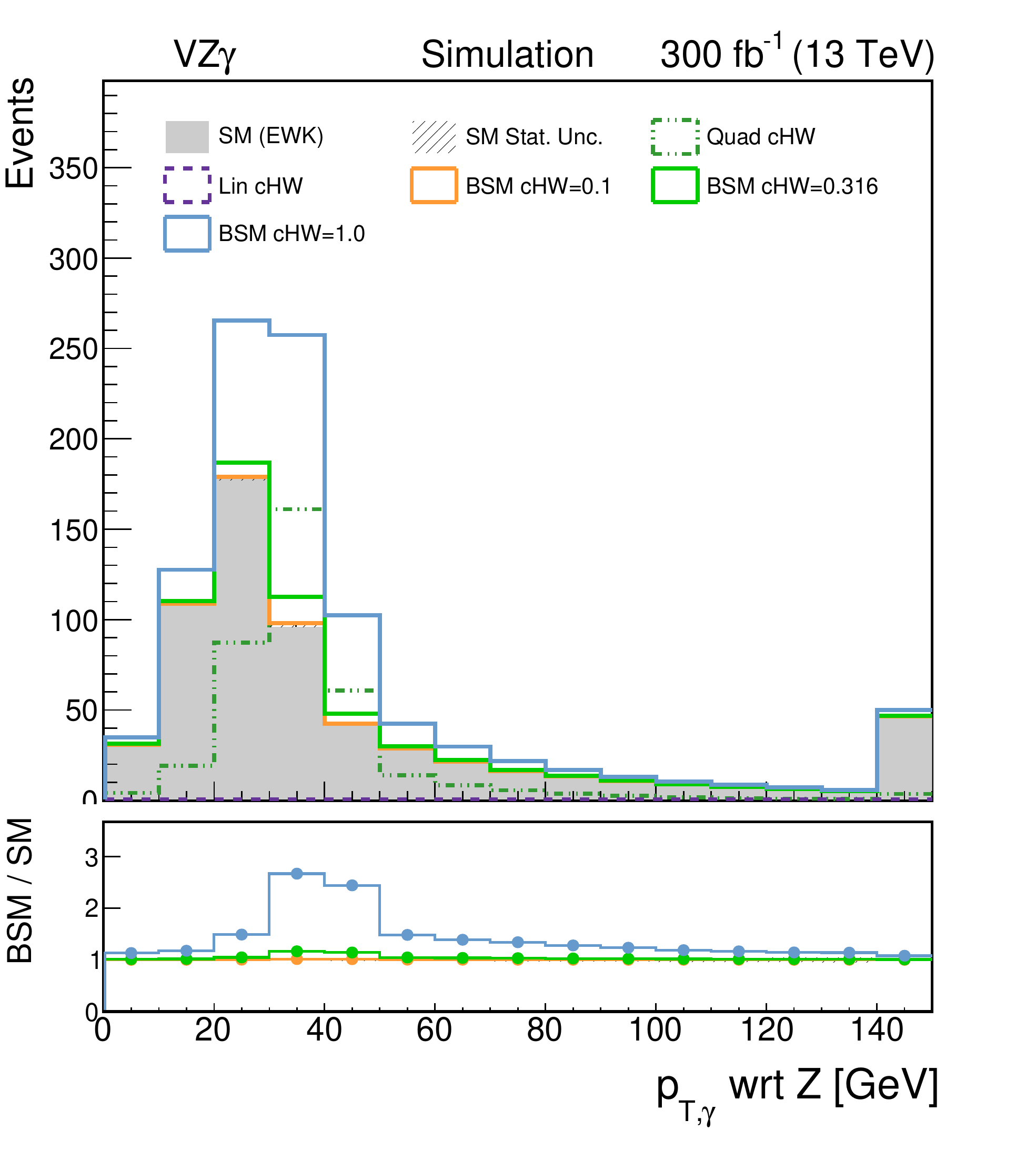}
	}	
	\subfigure[\label{subfig:VZGnoQCD_cHB}]{%
	    \includegraphics[width=.48\textwidth]{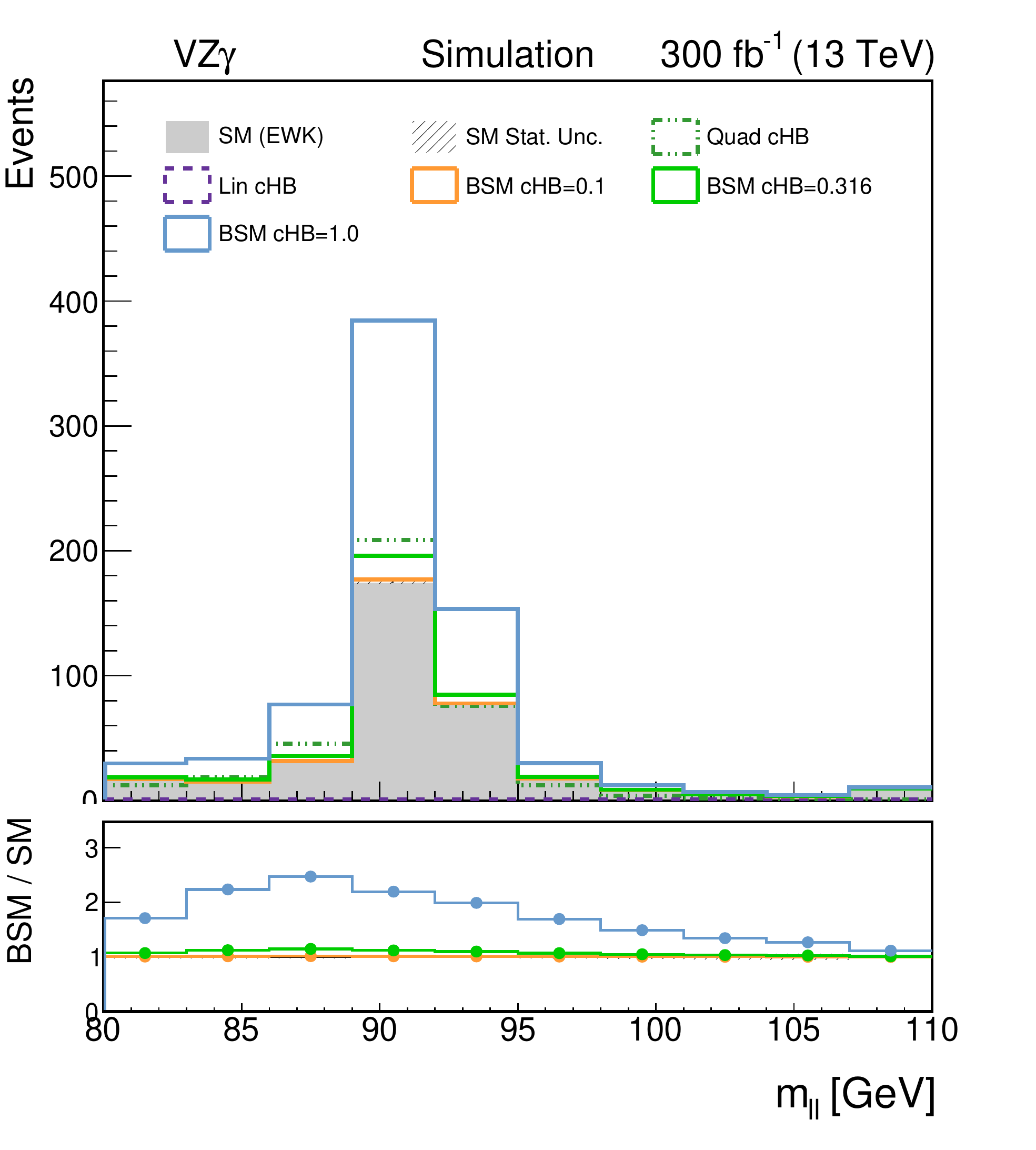}
	}
    
  \caption{ Comparison of SM (filled gray histograms) and BSM (lines)
      expected yield distributions 
      after the selection of table~\ref{tab:kinSel} for the VZ$\g$ process,
      for an integrated luminosity of 300 fb$^{-1}$. 
      Events corresponding to QCD-Z$\g$jj production are excluded. 
 The dashed lines show the distributions of the quadratic component (dashed green) and the linear interference term (dashed violet).  
The solid lines show the behavior of a theory where a single operator Q$_{i}$ is added to the SM Lagrangian, with $\Lambda$~=~1~TeV and the Wilson coefficients set to benchmark values (c$_{i} = 10^{-1},\,10^{-\frac{1}{2}},\,1,\,10^{\frac{1}{2}}$). 
For all the distributions, the last bin contains all the overflow events. 
The bottom plot shows the ratio of BSM to SM events for each bin. 
In the picture are shown the effects induced by Q$_{W}$ on the jet pair FWM $H_2^T$ (a), by Q$_{HWB}$ on the total transverse momentum of the charged lepton pair (b), by Q$_{HW}$ on the photon transverse momentum relative to the direction defined the 4-momentum of the best leptonically decaying Z candidate (c), and by Q$_{HB}$ on the invariant mass of the charged lepton pair (d).
      \label{fig:Distributions_VZG_noQCD}  }  
\end{figure}

\begin{figure}[p!]
  \centering 
 
	\subfigure[\label{subfig:VZG_cW}]{%
        \includegraphics[width=.48\textwidth]{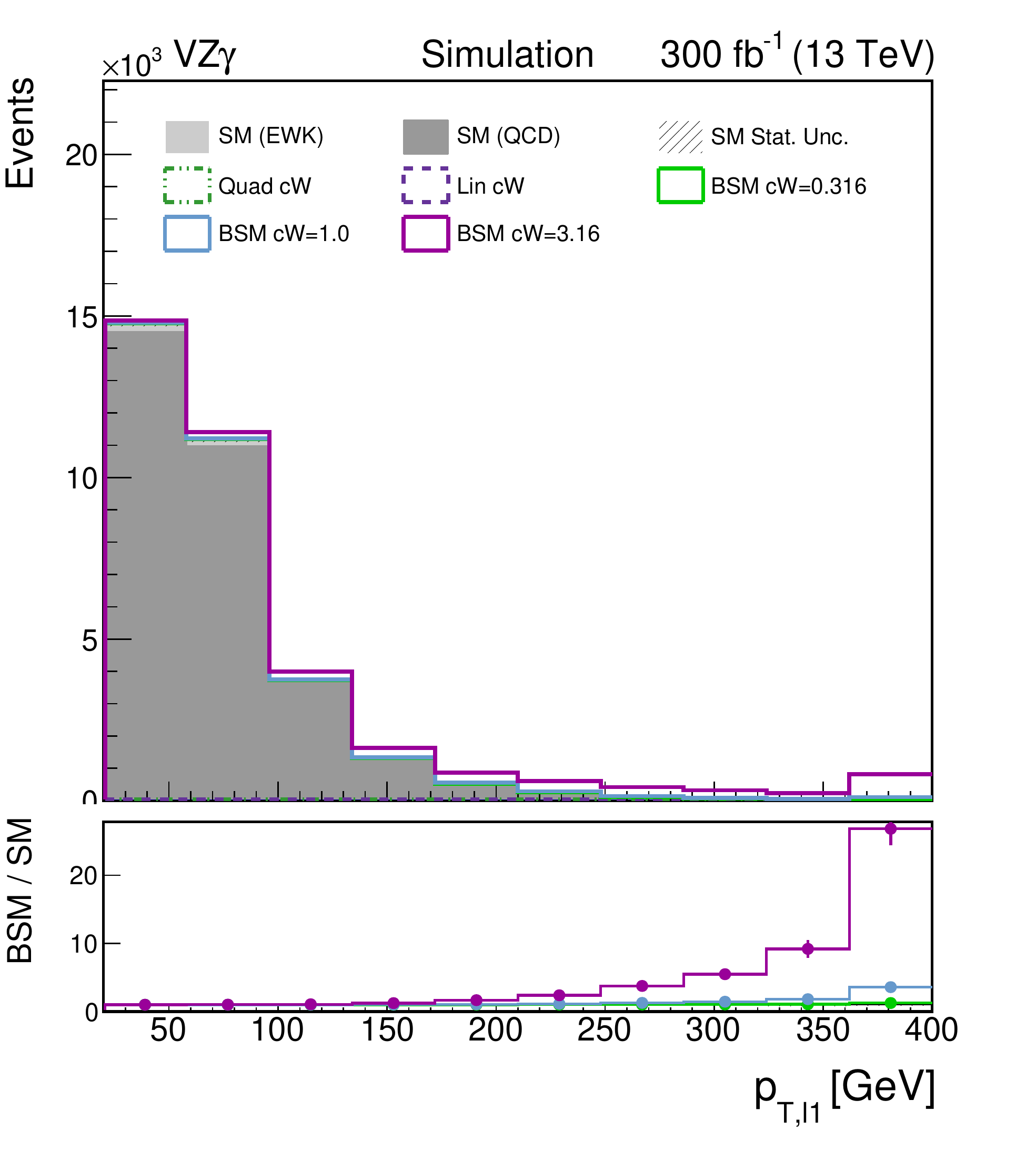}
	}
	\subfigure[\label{subfig:VZG_cHWB}]{%
	    \includegraphics[width=.48\textwidth]{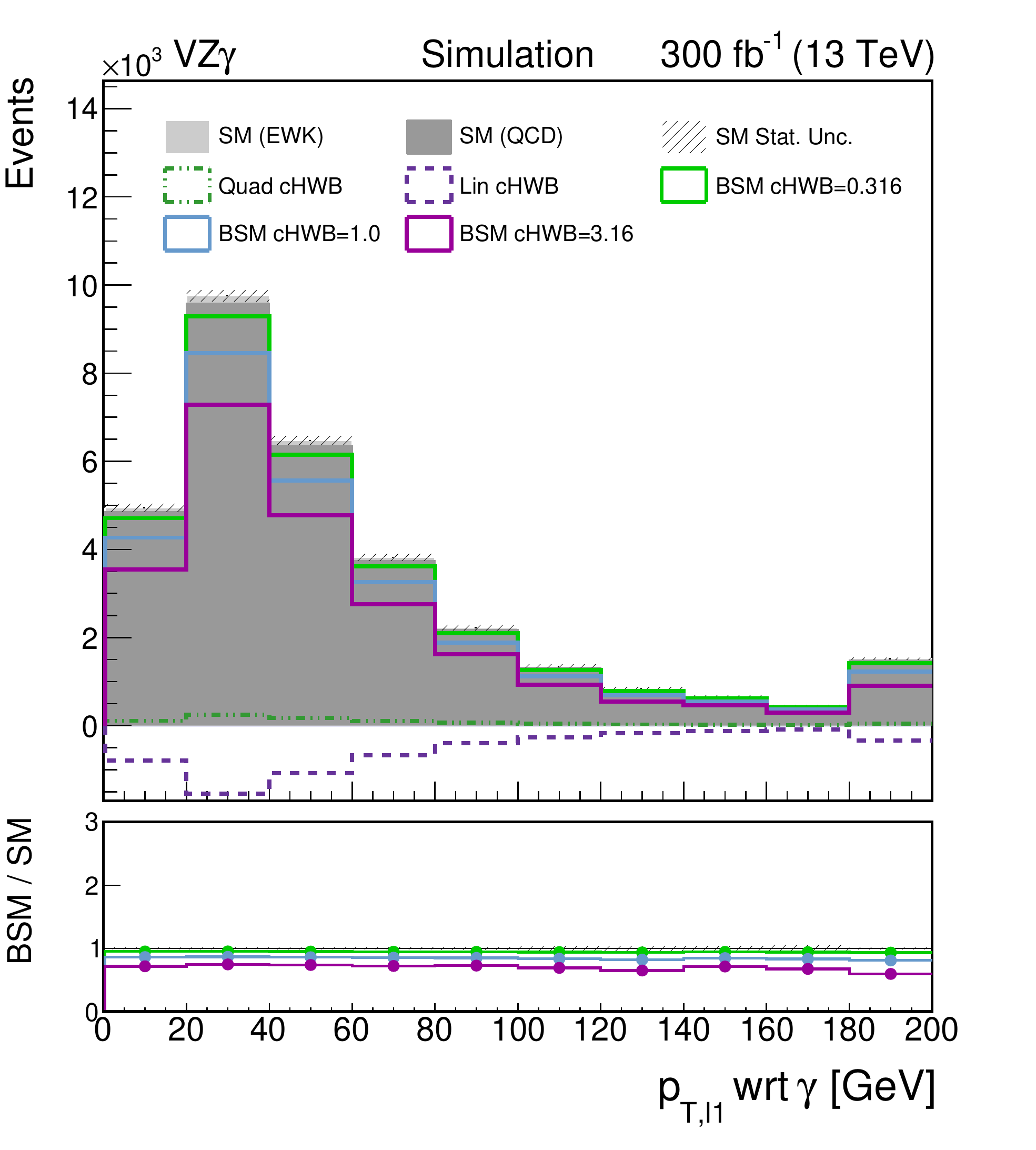}
	}
	\subfigure[\label{subfig:VZG_cHW}]{%
	    \includegraphics[width=.48\textwidth]{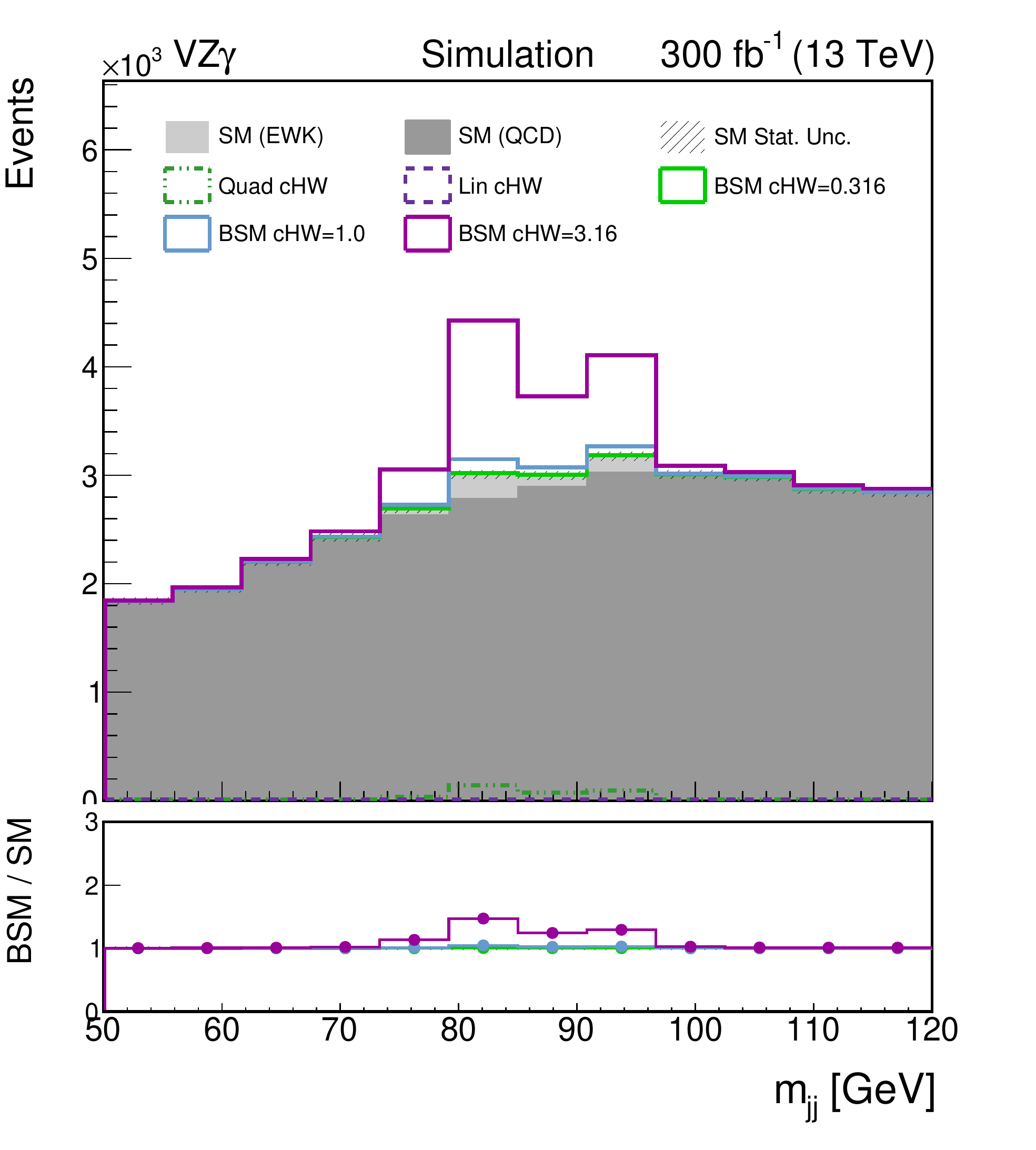}
	}
	\subfigure[\label{subfig:VZG_cHB}]{%
	    \includegraphics[width=.48\textwidth]{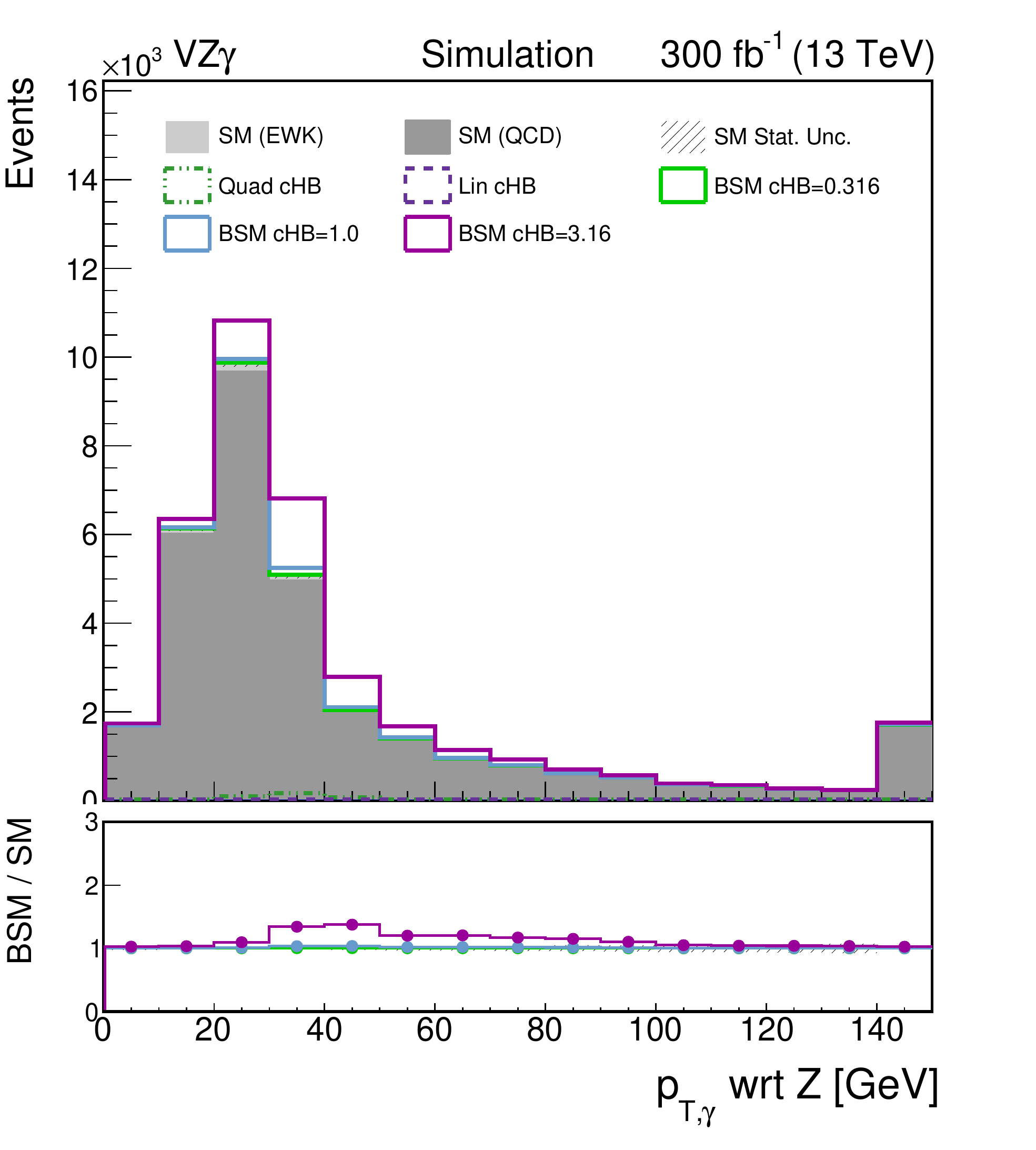}
	}
    
  \caption{ 
      Comparison of SM (filled histograms) and BSM (lines)
      expected yield distributions 
      after the selection of table~\ref{tab:kinSel} for the VZ$\g$ process,
      for an integrated luminosity of 300 fb$^{-1}$. 
      The SM distribution is represented as a stacked histogram, summing electroweak VZ$\g$ (light gray) and QCD-Z$\g$jj (dark gray) components, while all BSM distributions are superimposed.
    The dashed lines show the distributions of the quadratic component (dashed green) and the linear interference term (dashed violet).  
The solid lines show the behavior of a theory where a single operator Q$_{i}$ is added to the SM Lagrangian, with $\Lambda$~=~1~TeV and the Wilson coefficients set to benchmark values (c$_{i} = 10^{-1},\,10^{-\frac{1}{2}},\,1,\,10^{\frac{1}{2}}$). 
For all the distributions, the last bin contains all the overflow events. 
The bottom plot represents the ratio of BSM to SM events for each bin. 
The distribution shows the effects induced by Q$_{W}$ on $p_{T}^{l_1}$ (a), by Q$_{HWB}$ on the $p_{T}^{l_1}$ relative to the photon direction (b), by Q$_{HW}$ on $m_{jj}$ (c), and by Q$_{HB}$ on the $p_{T}^{\g}$ relative to the leptonically decaying Z boson candidate (d).
      \label{fig:Distributions_VZG}  }  
\end{figure}

Figure~\ref{fig:Distributions_ZZG} illustrates the EFT contribution to the photon transverse momentum relative to a non-standard longitudinal direction, the 4-momentum of the Z diboson system for the Q$_{HB}$ operator (figure~\ref{subfig:ZZG_cHB}) and of the best Z-boson candidate (figure~\ref{subfig:ZZG_cHW}) for the Q$_{HW}$ case. 
The linear interference terms associated with Q$_{HW}$ and Q$_{HB}$ are negligible since both operators induce anomalous diagrams involving the Higgs-gauge couplings (HZ$\g$ and H$\g\g$), which are disallowed in the SM at the tree level. 
On the contrary, the quadratic component induced by the Q$_{HB}$ operator dominates in both the low and high momentum tails. 
In the case of the Q$_{HW}$ operator (figure~\ref{subfig:ZZG_cHW}),  the $p_{T}$ spectrum is slightly harder than the SM distribution, as observed in the corresponding plots for the WZ$\g$ production. 
Figures~\ref{subfig:ZZG_cHDD}-\ref{subfig:ZZG_cHWB} show the impact on the transverse momentum $p_{T}^{e^-\mu^-}$ of the same-sign charged leptons of the EFT components induced by the operator Q$_{HD}$. Similarly to figure~\ref{subfig:WZG_cHDD}, the Q$_{HD}$-induced diagrams interfere destructively with the SM component, reducing the overall yield. This effect is partially mitigated by the presence of the quadratic component. A similar compensation between the quadratic and the linear components is visible in the photon $p_T$ spectrum for the Q$_{HWB}$ operator (see figure~\ref{subfig:ZZG_cHWB}). In the low-momentum region, the destructive interference leads to an almost exact cancellation of the quadratic term, resulting in a tiny residual deviation from SM. 

\paragraph{Semi-leptonic VZ$\g$}
The inclusive triboson VZ$\g$ production process is studied with the experimental semi-leptonic signature 2jets+2l+$\g$ using a combination of WZ$\g$ and ZZ$\g$ processes. The vector boson V decays hadronically to produce the jet pair, while the Z-boson decays leptonically via  Z$\rightarrow$e$^+$e$^-/\mu^+\mu^-$. 
The contribution of background processes with the same final state is partially suppressed by the kinematic selection. 
As in the case of VZZ, the dominant background is the QCD-induced Z$\g$jj production (denoted as QCD-Z$\g$jj), which is generated separately. Since the WZ$\g$ diagrams depend on the quartic gauge coupling WWZ$\g$, the inclusive channel VZ$\g$ is sensitive to the operator Q$_W$, unlike the ZZ$\g$ process. 

Figure ~\ref{fig:Distributions_VZG_noQCD} refers to the study of the anomalies induced by the EFT operators on the pure electroweak processes, excluding the QCD-Z$\g$jj contribution. 
Figure~\ref{subfig:VZGnoQCD_cW} illustrates the second order FWM of the jet pair, $H_2^T(jj)$, for the operator Q$_W$. The linear term associated with Q$_W$ is negligible with respect to the quadratic one. 
The latter identifies a high-momentum region with a pronounced deviation from the shape of the SM background, making this observable a viable candidate for targeting potential anomalies. 
The other plots in figure~\ref{fig:Distributions_VZG_noQCD} illustrate the $p_T$ spectrum of the photon relative to the longitudinal direction defined by the dilepton system. 
Figure~\ref{subfig:VZGnoQCD_cHWB} shows the deviation induced by the operator Q$_{HWB}$, whose diagrams interfere destructively with the SM ones. 
The resulting negative contribution of the interference term is not exactly symmetric to the quadratic component so that the cancellation results partially mitigated. 
Figures~\ref{subfig:VZGnoQCD_cHW}-\ref{subfig:VZGnoQCD_cHB} show the $p_{T(Z)}^\g$ and $m_{ll}$ distributions for the operators  Q$_{HW}$ and Q$_{HB}$, respectively. 
Unlike the Q$_{HWB}$ operator, the interference terms have no significant effect, and the quadratic terms are dominant in each case. 
 
Figure~\ref{fig:Distributions_VZG} illustrates the impact of the QCD-Z$\g$jj background on the sensitivity analysis. 
The presence of the dominant background increases the total expected yield by two orders of magnitude. 
Figure~\ref{subfig:VZG_cW} shows the distribution of the leading lepton $p_T$, i.e. the most Q$_W$-sensitive variable. 
The deviation from SM is only relevant at very high transverse momentum. Figure~\ref{subfig:VZG_cW} shows the observable most sensitive to anomalies induced by the Q$_{HWB}$ operator, i.e. the leading lepton $p_T$ relative to the photon direction. 
The destructive interference SM-Q$_{HWB}$ is very prominent, while the quadratic contribution is negligible. 
Interestingly, the Q$_{HWB}$ operator also affects the QCD-Z$\g$jj process, hence the EFT linear component receives a contribution from the Q$_{HWB}$-induced diagrams involving both QCD and electroweak processes.
Similarly, the Q$_{HD}$ operator also yields corrections to the QCD-Z$\g$jj diagrams. 
Figures~\ref{subfig:VZG_cHW}-\ref{subfig:VZG_cHB} display the SM deviations induced by the operators Q$_{HW}$ and Q$_{HB}$ on the dijet invariant mass spectrum and the photon $p_T$ relative to the leptonically decaying Z boson candidate, respectively. 
In both cases, the dominant EFT term is the quadratic one, although much lower than the SM background, leading to a limited sensitivity of VZ$\g$ to Q$_{HB}$ and Q$_{HW}$. 

%
\section{Results}
In this chapter, the bounds on the Wilson coefficients associated with the operators under scrutiny are evaluated at the generator level for an integrated luminosity of 300~fb$^{-1}$. 
We estimate the sensitivity to SM deviations resulting from the inclusion of EFT operators in the Lagrangian by computing the confidence intervals for the considered Wilson coefficients. The latter are evaluated neglecting theoretical and experimental systematic uncertainties,
with the exception of the luminosity uncertainty of $2\%$, which affects the statistical precision of the results. 
The confidence intervals provide an estimate of the constraints on the corresponding Wilson coefficient. 
The more stringent the constraints, the higher is the sensitivity to the kinematic anomalies induced by the corresponding EFT operator. 
The same analysis procedure is performed in the two-dimensional case, by inserting the operators pairwise into the fit, and in a multi-dimensional fit constraining the individual coefficients while profiling all the others. 
The global analysis of the sensitivity estimate is obtained by combining all the triboson production channels for the common operators.

\paragraph{One-dimensional constraints}\label{Individual_constraints}
The likelihood function $\mathcal L(\textbf c)$ is generically defined as: 
\begin{equation}\label{LL}
    \mathcal L(\textbf{c})=\prod_k\frac{\left(N_k(\textbf{c})\right)^{n_k}}{n_k!}\cdot e^{-N_k(\textbf{c})}\ ,
\end{equation}
where $n_k=N_k(\textbf{0})$ is the number of expected SM events in the $k$-th bin. The vector $\textbf{c}$ belongs to the Wilson coefficient space, i.e. the set of values associated with each c$_i$.
\\
To obtain the likelihood shape $\mathcal L(\text c_i)$, a scan is performed over the values of the Wilson coefficient c$_i$ in a plausible range. This procedure is computed for each variable of interest (see table~\ref{tab:kinSel}) using the \textsc{Combine Tool} package~\cite{khachatryan2015precise}. 
\begin{figure}[p!]
  \centering 
  \includegraphics[width=.45\textwidth]{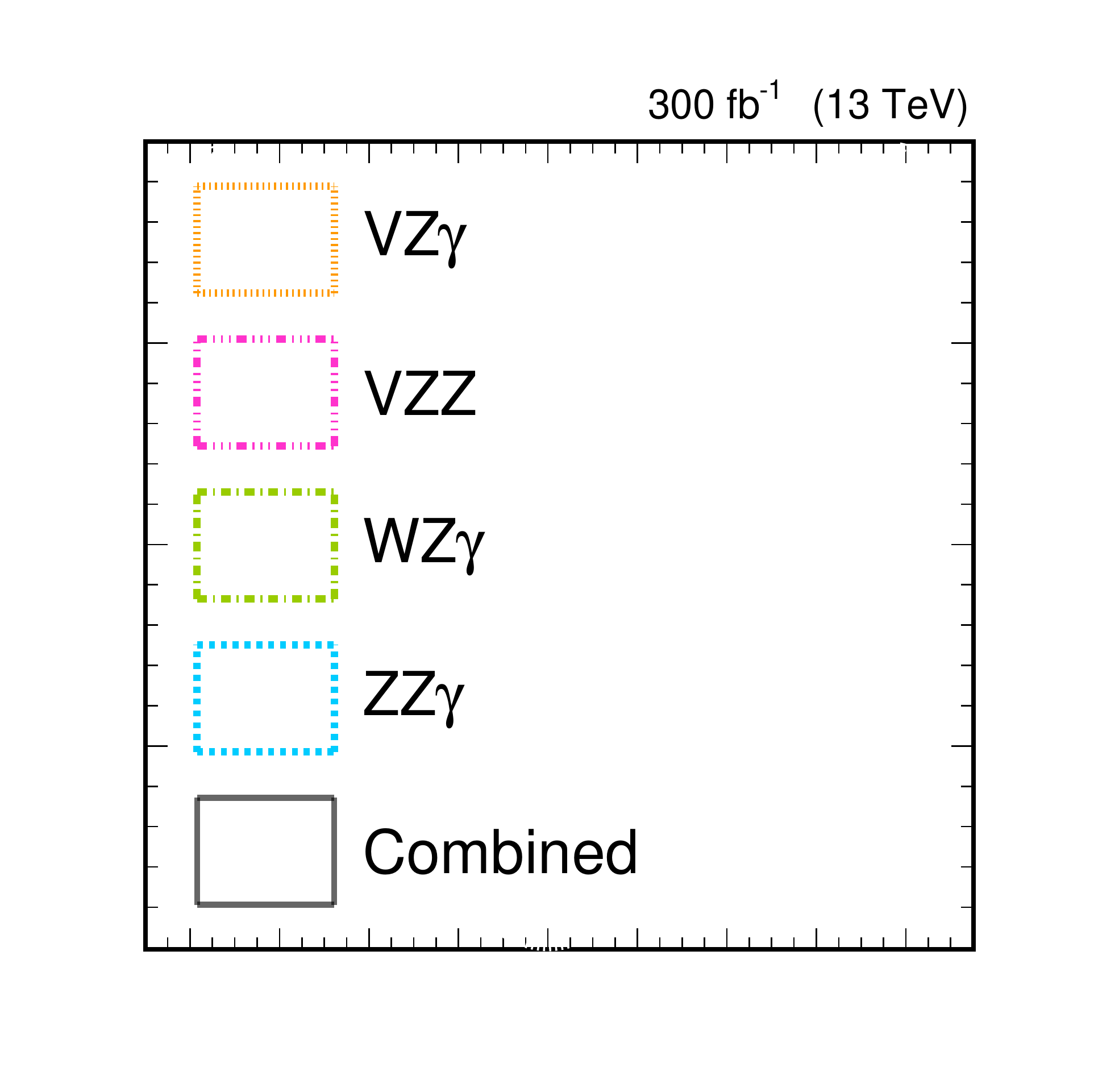}
  \includegraphics[width=.45\textwidth]{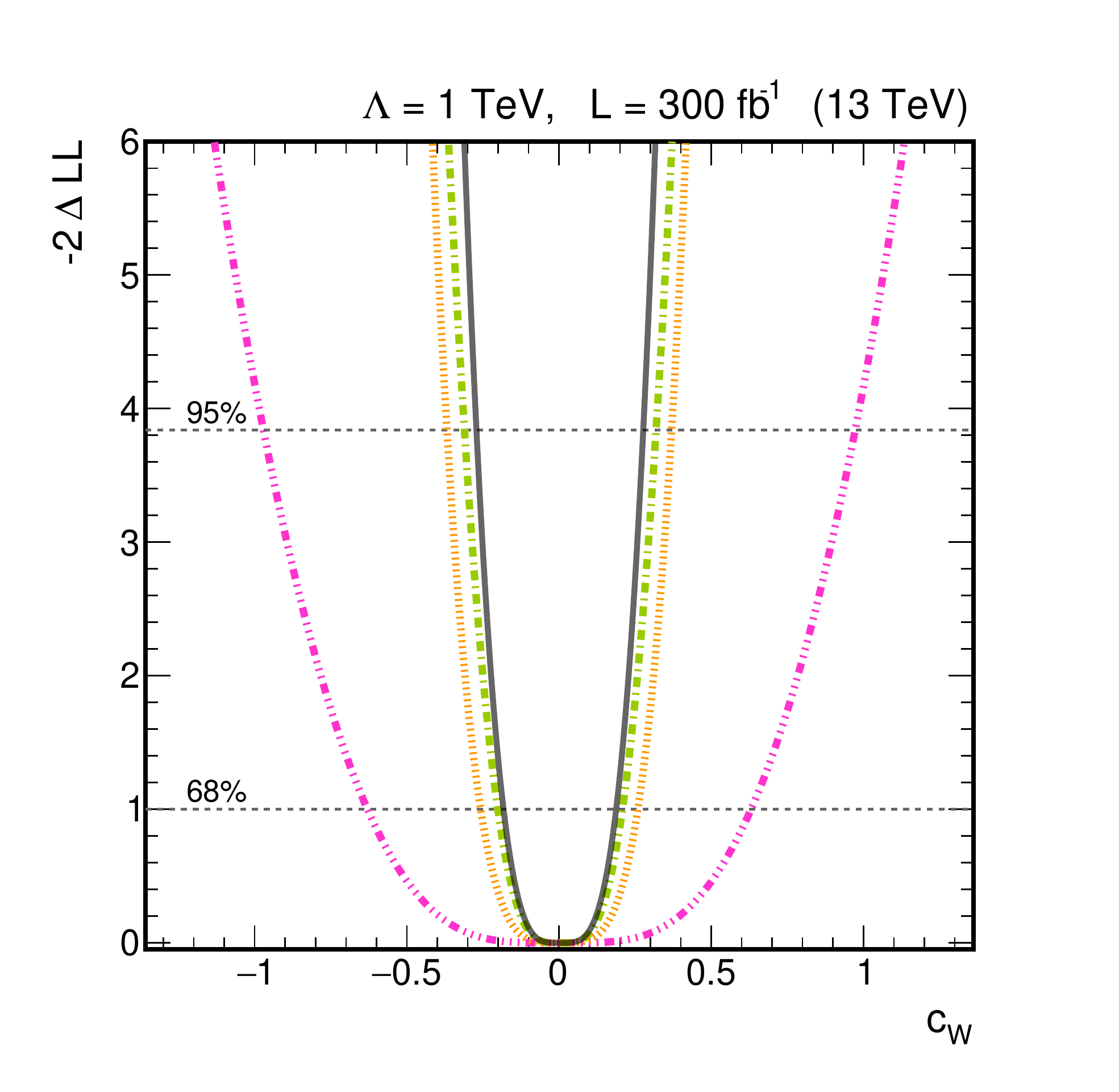}  
  \includegraphics[width=.45\textwidth]{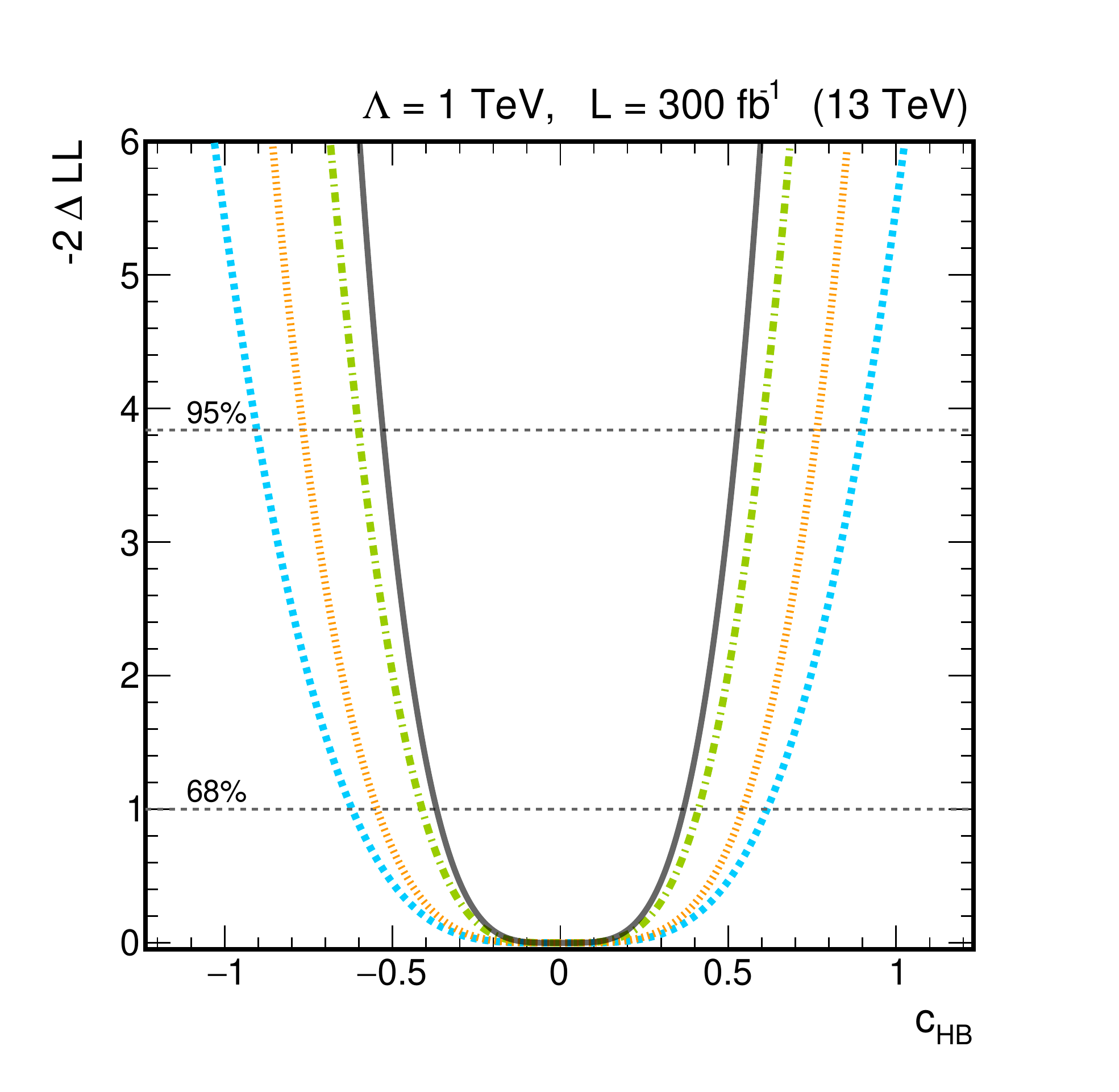}
  \includegraphics[width=.45\textwidth]{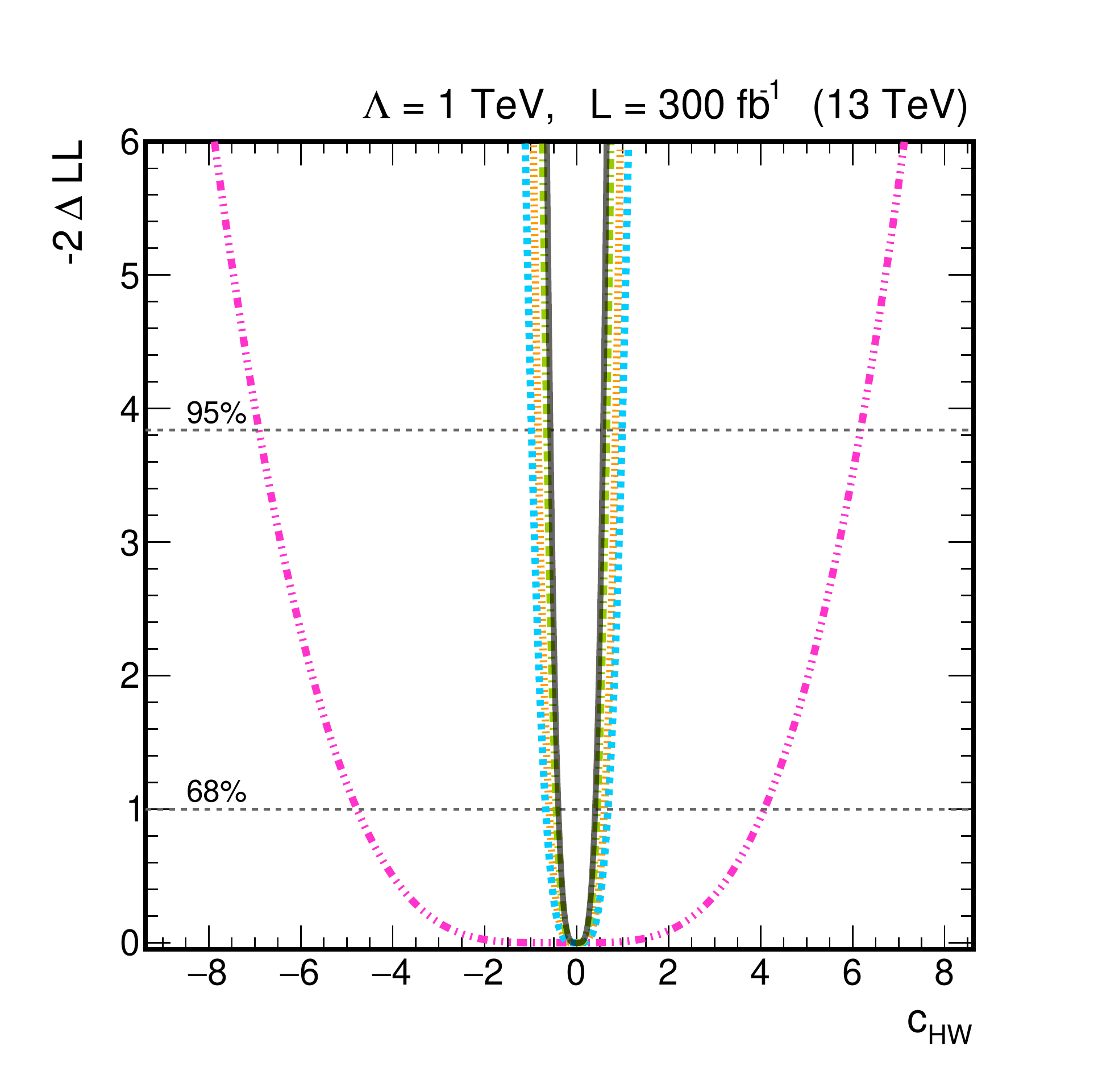}
  \includegraphics[width=.45\textwidth]{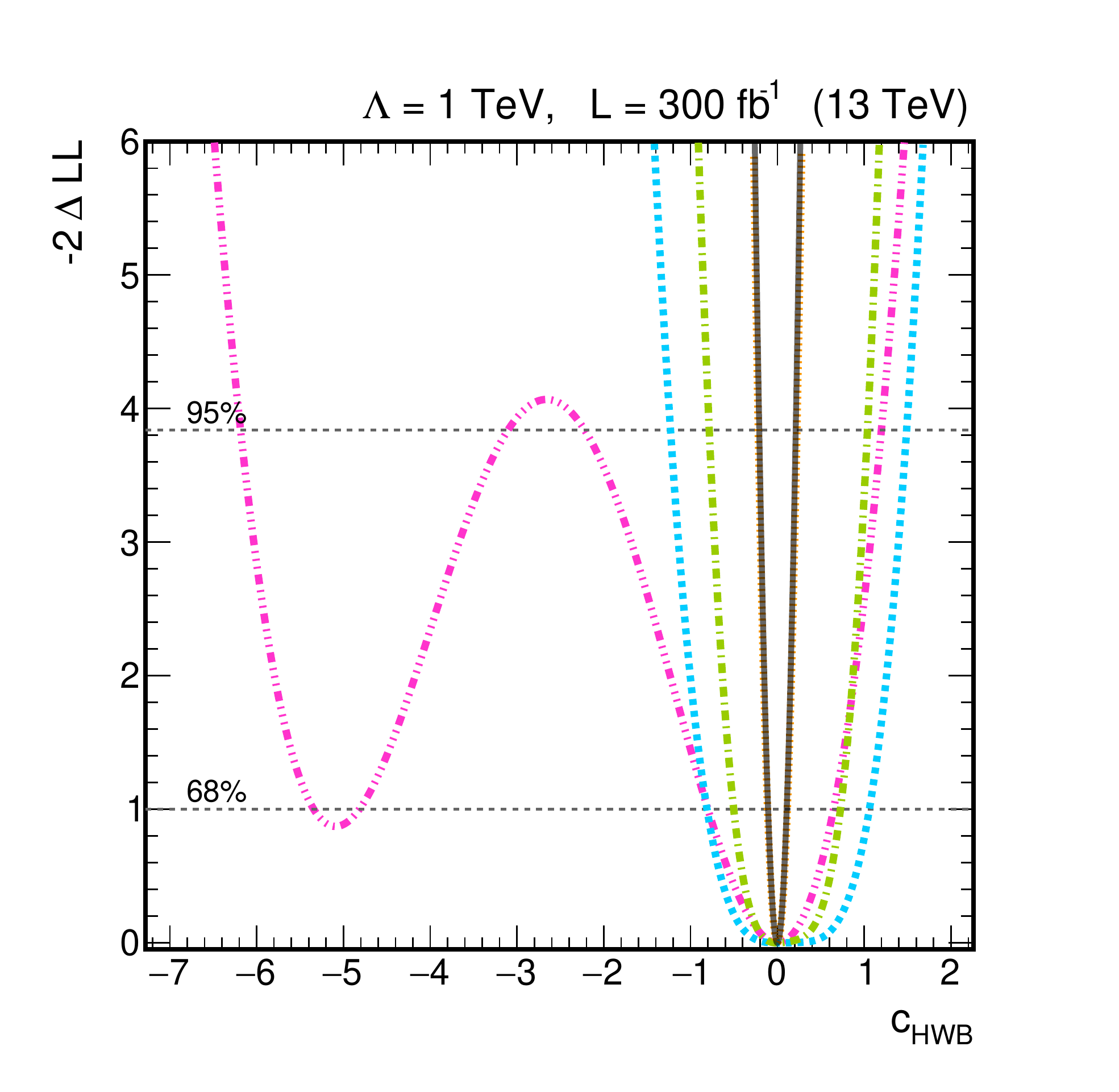}
  \includegraphics[width=.45\textwidth]{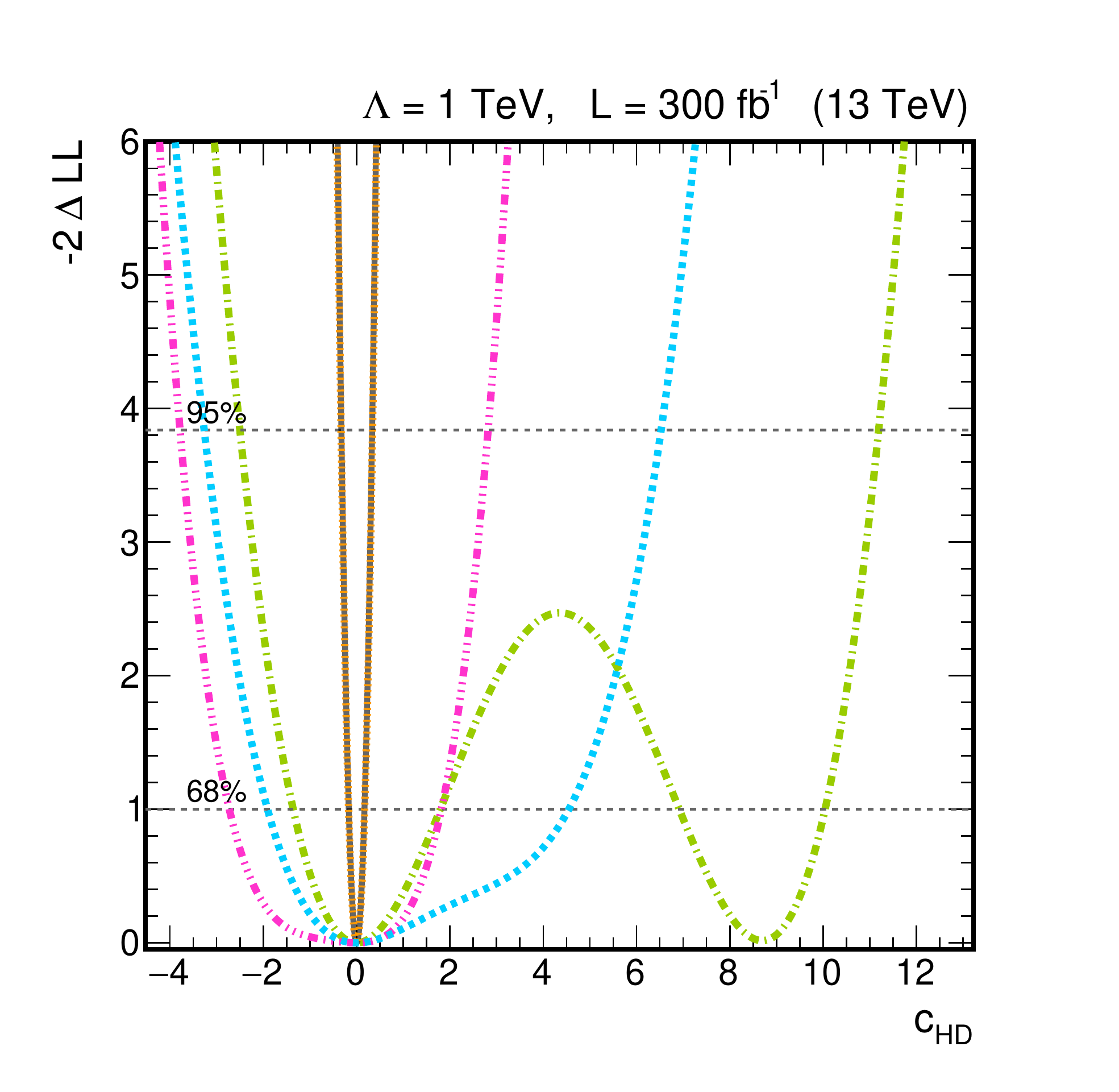}
  
  \caption{ Shapes of $-2\Delta\log \mathcal{L}$ reported for the single channels (colored lines) and their combination (solid transparent black line) as a function of the Wilson coefficients.  \label{fig:Combined_LL_Profiles_1D}}  
\end{figure}

From the likelihood scans, the 68\% and 95\% confidence level intervals in the c$_i$ estimates are extracted by requiring $-2\Delta\text{log}\mathcal L<1$ and $-2\Delta\text{log}\mathcal L<3.84$, respectively~\cite{pdg}. 
The size of these confidence regions quantifies the sensitivity of each observable in a given channel to anomalies induced by a single EFT operator.

For each channel, the likelihood function is obtained from the distribution of the variables listed in table~\ref{tab:kinSel}. 
The best variable is identified by the sensitivity bounds at 68\% confidence level. 
Figure~\ref{fig:Combined_LL_Profiles_1D} shows the $-2\Delta\log \mathcal{L}$ fits as a function of the Wilson coefficient values for each operator considered. 
The black line shapes correspond to the combination of all the channels sensitive to each operator. In the plots, the horizontal dashed lines correspond to the values of $-2\Delta\text{log}\mathcal L=1(3.84)$ defining the 68\%(95\%) confidence levels. 
Table~\ref{tab:1DConstr} summarizes the results of the one-dimensional scan. Each operator is added individually to the SM Lagrangian. The sensitivity estimates for the semi-leptonic channels, VZZ and VZ$\g$, include the effects of the QCD-induced background processes. 
The bosonic EFT operators induce anomalous effects in the vast majority of the tribosonic channels. The main exceptions are Q$_W$ and Q$_{HB}$, which do not affect the ZZ$\g$ production process and the VZZ production (nor the corresponding main background QCD-ZZjj), respectively. 
\renewcommand{\arraystretch}{1.5}
\begin{table}[t!]\footnotesize
\begin{center}
\begin{tabular}{c|c||c|c|c|c|c}
\rotatebox[origin=c]{90}{$\Lsh$} Processes & Operators $\rightarrow$      & Q$_W$                                           & Q$_{HB}$                                          & Q$_{HW}$                                        & Q$_{HWB}$                                       & Q$_{HD}$                                        \\
\hline
  & Best var. & $p_T^{l1}$                 & $p_{T(Z)}^{\g}$             & $p_{T(Z)}^{\g}$           & $p_{T(Z)}^{\g}$                                   & $p_{T(WZ)}^{l1}$                                   \\
  & 68\% C.L.     & {[}-0.20,0.21{]}   & {[}-0.41,0.41{]}   & {[}-0.44,0.44{]} & {[}-0.50,0.73{]}                         & {[}-1.36,1.79{]}                          \\
\multirow{-3}{*}{WZ$\g$}     & 95\% C.L.     & {[}-0.31,0.32{]}   & {[}-0.60,0.60{]}   & {[}-0.65,0.65{]} & {[}-0.79,1.04{]}                         & {[}-2.50,11.2{]} \\
\hline
  & Best var. &   & $p_{T(Z_1)}^{\g}$                                     & $m_{\mu\mu}$                                      & $m_{\mu\mu}$                                      & $p_T^{e^+\mu^+}$                                   \\
  & 68\% C.L.     & No diagrams & {[}-0.62,0.61{]}                           & {[}-0.68,0.68{]}                         & {[}-0.81,1.06{]}                         & {[}-1.91,4.55{]}                          \\
\multirow{-3}{*}{ZZ$\g$}     & 95\% C.L.     &   & {[}-0.90,0.90{]}                           & {[}-0.98,0.99{]}                         & {[}-1.23,1.49{]}                         & {[}-3.27,6.53{]}                          \\
\hline
  & Best var. & $p_{T}^{l1}$                                         & $m_{jj}$                                          & $m_{jj}$                                        & $p_{T(\g)}^{l1}$           & $p_{T(\g)}^{l2}$            \\
  & 68\% C.L.     & {[}-0.26,0.26{]}                           & {[}-0.55,0.54{]}                           & {[}-0.60,0.60{]}                         & {[}-0.11,0.11{]} & {[}-0.17,0.17{]}  \\
\multirow{-3}{*}{VZ$\g$}     & 95\% C.L.     & {[}-0.37,0.37{]}                           & {[}-0.77,0.76{]}                           & {[}-0.84,0.84{]}                         & {[}-0.22,0.23{]} & {[}-0.33,0.34{]}  \\
\hline
  & Best var. & $p_{T}^{l1}$                                         &   & $p_T^{V}$                                        & $m_{\mu\mu}$                                        & $p_T^{e^+\mu^+}$                                   \\
  & 68\% C.L.     & {[}-0.63,0.63{]}                           & Negligible & {[}-4.78,4.08{]}                         & {[}-0.80,0.65{]}                         & {[}-2.73,1.82{]}                          \\
\multirow{-3}{*}{VZZ}         & 95\% C.L.     & {[}-0.97,0.97{]}                           &   & {[}-6.91,6.17{]} & {[}-2.22,1.20{]}  
& {[}-3.78,2.82{]}                          \\
\hline\hline
& 68\% C.L.     & {[}-0.18,0.19{]}   & {[}-0.37,0.37{]}   & {[}-0.40,0.40{]} & {[}-0.11,0.11{]} & {[}-0.17,0.17{]}  \\
\multirow{-2}{*}{Combination} 
& 95\% C.L.     & {[}-0.27,0.28{]}   & {[}-0.53,0.53{]}   & {[}-0.57,0.57{]} & {[}-0.21,0.21{]} & {[}-0.33,0.33{]} \\
\hline\hline
&&&&&\\
\multirow{-2}{*}{VBS} & \multirow{-2}{*}{95\% C.L.}     & \multirow{-2}{*}{{[}-0.19,0.18{]}}   & \multirow{-2}{*}{-}   & \multirow{-2}{*}{{[}-1.02,1.08{]}} & \multirow{-2}{*}{{[}-1.34,0.96{]}} & \multirow{-2}{*}{{[}-1.98,1.74{]}} \\
\hline
\end{tabular}
\end{center}
  \caption{ Confidence intervals on the estimates of the Wilson coefficients, relative to the subset of operators considered, extracted from the respective likelihood scan for the most sensitive variable, for each channel studied. 
  This table reports the confidence intervals obtained combining all the triboson channels involving diagrams induced by operators individually included in the SM Lagrangian. 
  The last line reports the results obtained in the study of Ref.~\cite{bellan2021sensitivity} combining many VBS channels. 
  All the reported results are obtained considering an integrated luminosity of 300~fb$^{-1}$. \\    \label{tab:1DConstr}} 
\end{table}

For all the channels, the leading lepton transverse momentum $p_T^{l1}$ was found to be the most Q$_W$-sensitive variable. The EFT quadratic term becomes increasingly dominant with higher transverse momentum, while the SM background spectrum decays rapidly. 
The main background for the semi-leptonic channels, induced by QCD vertices, also exhibits a continuously decaying $p_T^{l1}$ spectrum (see figure~\ref{subfig:VZZ_ptl1_cW}-\ref{subfig:VZG_cW}). 

The fully leptonic channels, WZ$\g$ and ZZ$\g$, display a similar sensitivity to the operators Q$_{HB}$ and Q$_{HW}$ as they affect the same vertices. 
For both operators and both channels, the $p_T^{\g}$ relative to the (leptonically decaying) Z-boson best candidate 
is very sensitive to any modification caused by the anomalous HZ$\gamma$ coupling since this variable is strongly related to the Z boson and photon kinematics (see figures~\ref{subfig:WZG_cHB}-\ref{subfig:ZZG_cHW}). 
For the semi-leptonic VZ$\g$ channel, the best observable is instead the dijet invariant mass (see figure~\ref{subfig:VZG_cHW}) 
due to the dominance of the EFT contribution in a well localized region around the nominal W and Z masses, over the flat QCD-induced background. 
The Q$_{HW}$-sensitivity of the VZZ channel is strongly suppressed by the inclusion of the dominant QCD-ZZjj background and by the lack of sensitivity to the HZ$\g$/H$\g\g$ anomalous couplings, compared to the VZ$\g$ production processes. 
 
For the WZ$\g$ (VZ$\g$) channels, the (sub-)leading lepton $p_T$ relative to the direction of the WZ diboson system (photon) is the most Q$_{HWB}$ sensitive variable. 
In the VZ$\g$ channel, the destructive interference of the QCD-Z$\g$jj diagrams enhances the sensitivity to both Q$_{HWB}$ and Q$_{HD}$ operators, leading to the most stringent single-channel limits in this study. 
For the channels with two Z bosons decaying leptonically, the same-flavor opposite-sign (SFOS) dilepton invariant masses are competitive variables, especially in the case of Q$_{HWB}$ and Q$_{HW}$. The VZ$\g$ channel was found to be the most sensitive to the anomalies induced by Q$_{HD}$. For ZZ$\g$ and VZZ, the total $p_T$ of the same-sign charged leptons (e.g. figure~\ref{subfig:ZZG_cHDD}) is able to better discriminate the Q$_{HD}$-induced anomalies, although the sensitivity is an order of magnitude lower than in the VZ$\g$ channel. Similar results are obtained for the WZ$\g$ channel with the leading lepton $p_T$ relative to the direction of the WZ diboson system (see figure~\ref{subfig:WZG_cHDD}). 
\\
It is interesting to note that the semi-leptonic VZ$\g$ channel is the most sensitive to anomalous effects due to the large contribution of processes with QCD-induced jets in the final state. In particular, the operators Q$_{HWB}$ and Q$_{HD}$ induce anomalous diagrams that are involved in the QCD-Z$\g$jj process itself. Therefore the deviations in the distribution of figure~\ref{subfig:VZG_cHWB} are further increased, leading to tighter constraints. 

A subset of the likelihood shapes shown in figure~\ref{fig:Combined_LL_Profiles_1D} feature a second local minimum in a region far from zero. 
This is caused by the interference of the diagrams induced by these operators with the SM component. The effect of this interference term is incorporated into the linear component. In the case of destructive interference, linear and quadratic components have opposite signs and cancel each other within a certain range of positive Wilson coefficients. This is the case of the anomalies induced by Q$_{HD}$ in the fully leptonic WZ$\g$ channel. 
When the interference is constructive, a similar behavior can be observed in a negative Wilson coefficient region, as in the case of Q$_{HWB}$ affecting the semi-leptonic VZZ channel. 
In most other cases the likelihood scans are quite symmetric due to the dominance of the quadratic contribution over the linear interference term. 

The combination of all the channels sensitive to the operators considered allows us to obtain more stringent constraints for all the operators studied. 
These can be compared with the results obtained for the VBS channels. In particular, the last line of table~\ref{tab:1DConstr} reports the 95\% C.L. exclusion intervals obtained in the VBS channel combination, at 300~fb$^{-1}$ of integrated luminosity, in the parton-level study of diboson channels of Ref.~\cite{bellan2021sensitivity}. 
The combination of triboson channels shows a higher sensitivity with respect to VBS to the anomalies induced by the operators Q$_{HW}$, Q$_{HWB}$, and Q$_{HD}$. 

For the Q$_W$ operator, the sensitivity of the combined triboson channels is found to be highly competitive with that of the VBS channels. In particular, the most sensitive one is found to be the fully leptonic WZ$\g$ channel, which allows probing in a unique way the anomalies in the quartic gauge coupling WWZ$\g$. In contrast, the VBS channels studied in Ref.~\cite{bellan2021sensitivity} involve only intermediate states with pairs of massive gauge bosons, hence they are not sensitive to this coupling.
The VZZ channel is also sensitive to the Q$_W$ operator through the WWZZ quartic gauge coupling, but the constraints are less stringent due to the significantly lower expected yield in this channel. 
A distinct feature of all the triboson channels is that the most Q$_W$-sensitive variable is the transverse momentum of the leading lepton. The quadratic term induces deviations in the high-$p_T^{l1}$ region, as shown in figures~\ref{subfig:VZZ_ptl1_cW}-\ref{subfig:VZG_cW}.

For the operator Q$_{HW}$, both the fully and semi-leptonic channels of VZ$\gamma$ triboson production exhibit larger sensitivity than the VBS processes involving only massive bosons. This is attributed to the tree-level HZ$\gamma$ coupling induced by the Q$_{HB}$ and Q$_{HW}$ operators, which results in a dominant quadratic term and a negligible linear interference, as shown in figures~\ref{subfig:WZG_cHB}-\ref{subfig:WZG_cHW}-\ref{subfig:ZZG_cHB}-\ref{subfig:ZZG_cHW}. However, this does not hold for the channels that do not contain a Z boson and a photon, such as VZZ and VBS channels of Ref.~\cite{bellan2021sensitivity}. 
Among the observables found to be the most Q$_{HW}$-sensitive for the VBS channels of Ref.~\cite{bellan2021sensitivity} there are the invariant masses of the di-jets and the di-leptons stemming from the vector boson candidates. This is consistent with the findings in the ZZ$\g$ channel and the inclusive semi-leptonic VZ$\g$ channel for the Q$_{HW}$ and Q$_{HB}$ operators. The fully leptonic WZ$\g$ channel emerges as the most sensitive one, where the most relevant variable is found to be the $p_T$ relative to non-standard longitudinal directions $p^{part.}_{T(any\;dir.)}$.

\begin{figure}[b!]
  \centering 
    \includegraphics[width=1.0
\textwidth]{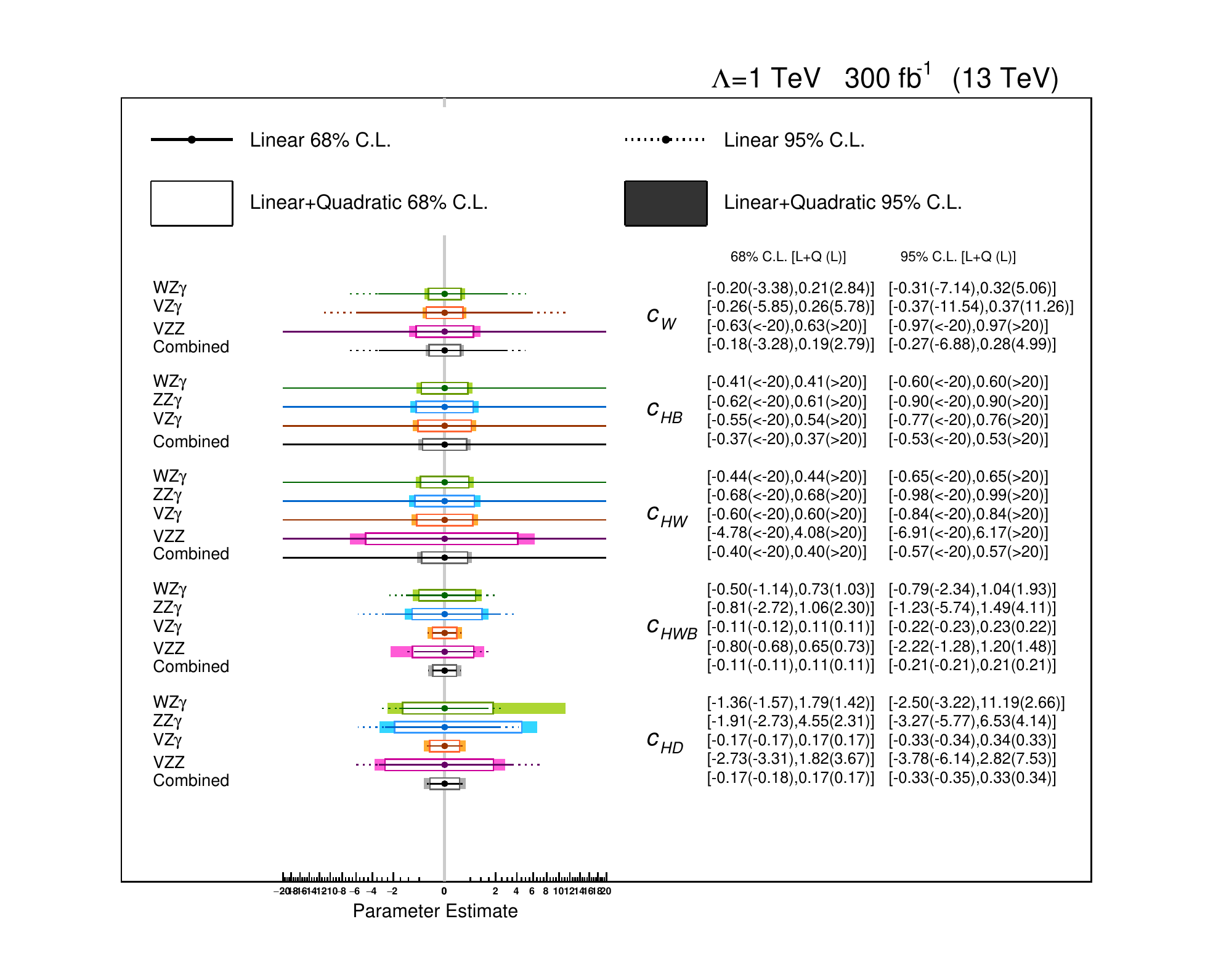}
  \caption{ Individual expected constraints on Wilson coefficients from the leptonic channels
WZ$\gamma$ (green) and ZZ$\gamma$ (light blue), the semi-leptonic VZ$\gamma$ (orange) and VZZ (purple) channels, and their combination (black). The solid points denote the SM
expectation. Solid (dashed) lines indicate the 68\% (95\%) confidence intervals obtained including
only terms linear in the Wilson coefficients in the signal predictions. Open (filled) boxes indicate
68\% (95\%) confidence intervals obtained including both linear and quadratic EFT components.
\label{Fig:LvsL+Q}}  
\end{figure}

The QCD-induced background processes are affected by the Q$_{HWB}$ and Q$_{HD}$ operators. As a consequence, the semi-leptonic VZ$\g$ channel appears to be significantly more sensitive than the other triboson channels.

The study of the $p^{part.}_{T(any\;dir.)}$ variables is crucial for the WZ$\g$ and VZ$\g$ channels. In the case of Q$_{HWB}$, for the channels with four charged leptons in the final state (VZZ and ZZ$\gamma$), the most sensitive variable is the invariant mass of the di-lepton system, as observed in several VBS channels~\cite{bellan2021sensitivity}. 
For the same channels, in the case of Q$_{HD}$, the most sensitive variable is the total $p_T$ of the same-charge lepton pairs in the final state, as also observed in the VBS ZZ channel~\cite{bellan2021sensitivity}. 
Overall, the results discussed highlight the importance of triboson channels,  
alongside with other multi-boson channels in the context of SM-EFT analyses.

\paragraph{Impact of one-dimensional quadratic terms on the sensitivity}
The effect of the quadratic terms on the one-dimensional constraints is examined. For each operator, the confidence intervals from table~\ref{tab:1DConstr} are compared with those extracted without the quadratic components in figure~\ref{Fig:LvsL+Q}. 

The constraints highlight the large impact of the quadratic terms on the sensitivity of the triboson analyses to the operators Q$_{HB}$, Q$_{HW}$, and Q$_{W}$. 
In these cases, the linear terms are negligible compared to the quadratic contribution. 

On the other hand, for the operators Q$_{HWB}$ and Q$_{HD}$, the interference with the SM diagrams leads to a significant contribution.
For the fully leptonic channels, WZ$\g$ and ZZ$\g$, in the case of Q$_{HD}$, as well as for the semi-leptonic VZZ in the case of Q$_{HWB}$, the linear and quadratic terms partially cancel each other for certain values of the Wilson coefficients (e.g. figures~\ref{subfig:WZG_cHDD}-\ref{subfig:ZZG_cHDD}), leading to asymmetric constraints when quadratic terms are included, as shown in Fig~\ref{fig:Combined_LL_Profiles_1D}. 

On the contrary, for the semi-leptonic VZ$\g$ channel, the linear contributions induced by Q$_{HWB}$ and Q$_{HD}$ are dominant over the respective quadratic components. This is visible in figure~\ref{subfig:VZG_cHWB}, which illustrates the absence of a significant effect of the quadratic terms on the one-dimensional constraints. 
%
%
%
%
%
%
%
%

\begin{figure}[b!]
  \centering 
  \includegraphics[width=.4\textwidth]{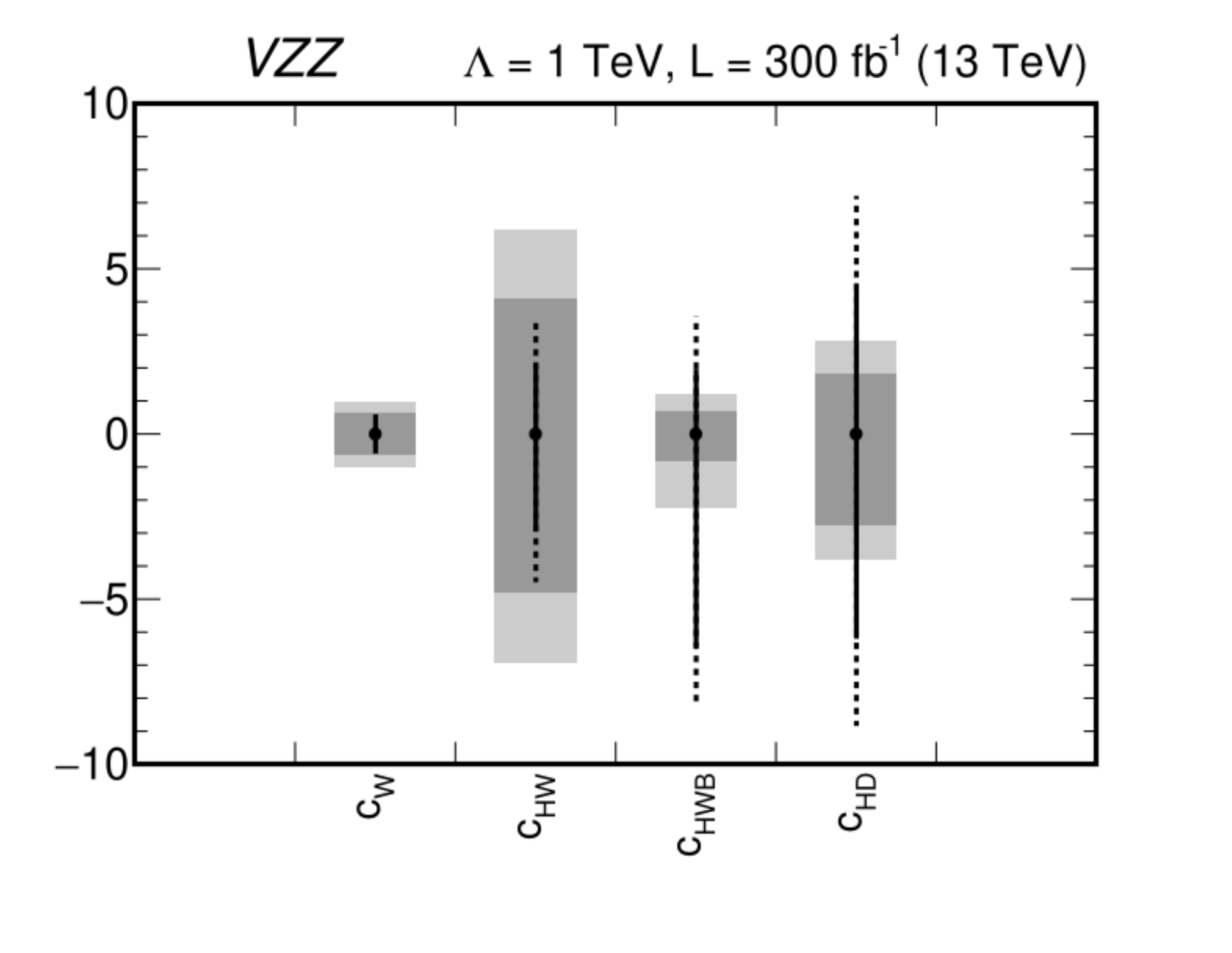}
  \includegraphics[width=.4\textwidth]{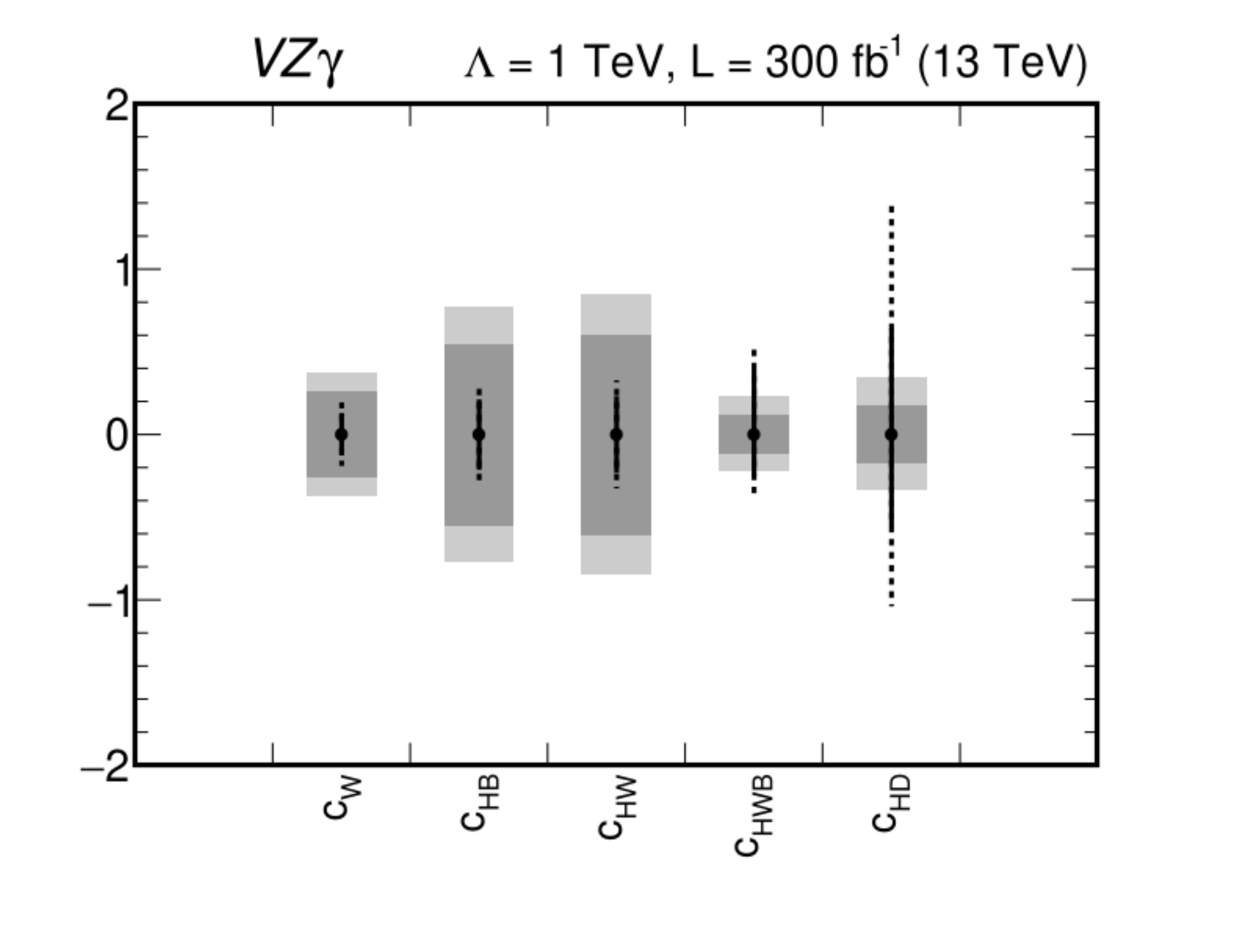}   
  \includegraphics[width=.18 \textwidth]{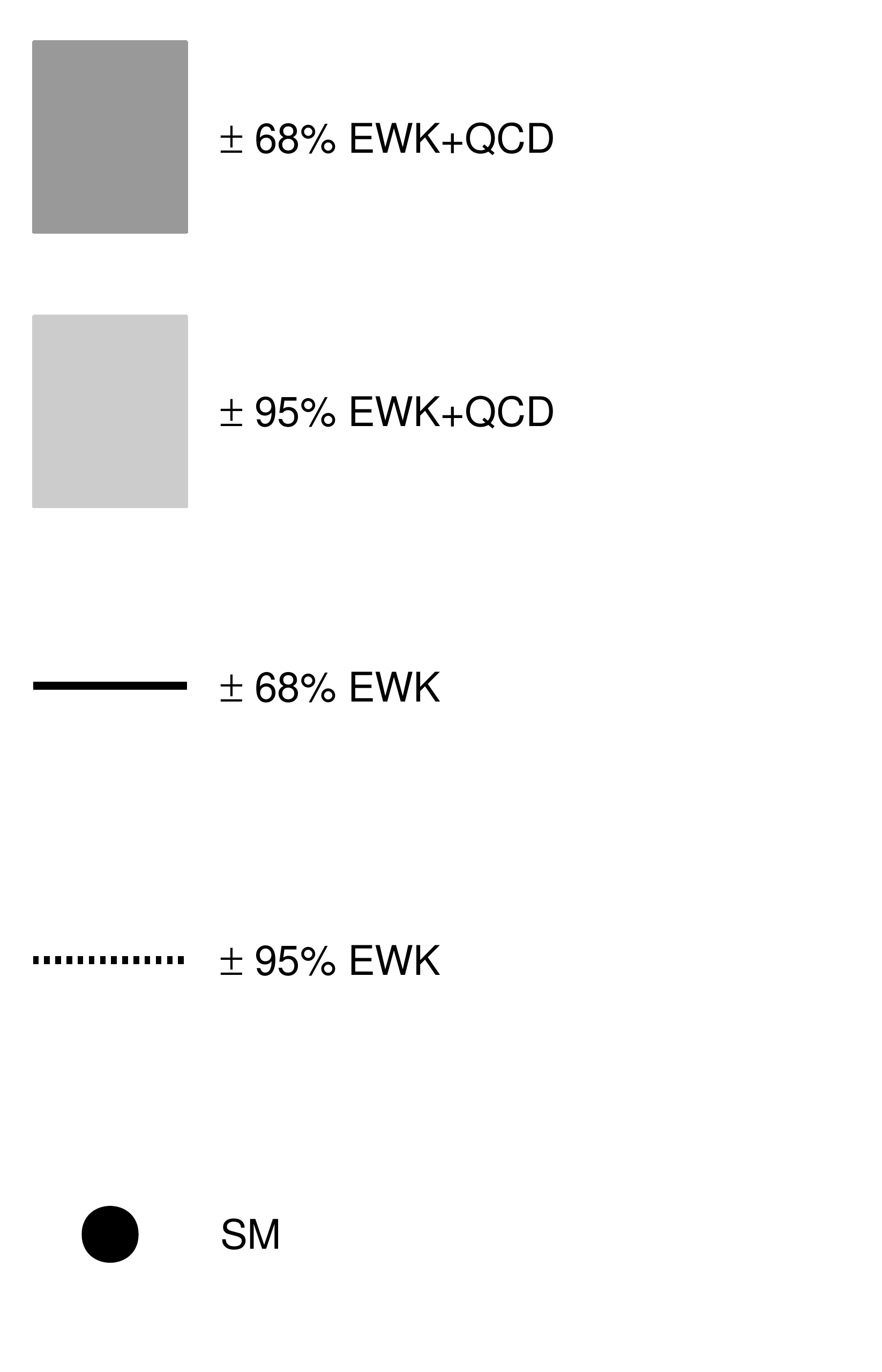}
  \caption{ Confidence intervals at 68\% and 95\% confidence level for the pure electroweak case (lines) and including processes with QCD vertices (gray boxes) extracted from the likelihood scans.}
  \label{fig:QCDimpact}
\end{figure}

\paragraph{Impact of processes with QCD-induced jets on the sensitivity}\label{qcd_effect}
The main background processes in the semi-leptonic VZZ and VZ$\g$ channels, labeled as QCD-ZZjj and QCD-Z$\g$jj, respectively, feature a jet pair induced by QCD vertices in the final state. Most of the operators under study do not affect them. 
Nevertheless, Q$_{HWB}$ and Q$_{HD}$ induce anomalous electroweak vertices in the QCD-ZZjj(-Z$\g$jj) diagrams leading to potential deviations other than the ones expected in the case of pure electroweak processes. 
Such anomalies are to be treated as BSM signal as well. 
Therefore, in the case of these two operators, the inclusion of processes with QCD-induced jets leads to two competing effects: a higher background component accompanied by a larger EFT signal contribution. 
\\
To determine the overall effect of the QCD-induced background on the single-operator-sensitivity of the VZZ and VZ$\g$ analyses, the impact on the estimated Wilson coefficient bounds is assessed. 
\\
Figure~\ref{fig:QCDimpact} shows a comparison of the confidence intervals for the pure electroweak cases with those obtained including background processes with QCD vertices. 
As expected, 
for the operators that do not affect the QCD-ZZjj(-Z$\g$jj) diagrams, i.e. Q$_W$, Q$_{HB}$, and Q$_{HW}$, the processes with QCD-induced jets constitute only a background source, leading to less stringent constraints. 
On the contrary, 
the confidence intervals on the coefficients c$_{HWB}$ and c$_{HD}$ improve considerably when  the QCD-induced events are taken into account.
%
%
%
%
\\
\\
\\

\begin{figure}[htbp]
  \flushright
  \includegraphics[width=.32\textwidth]{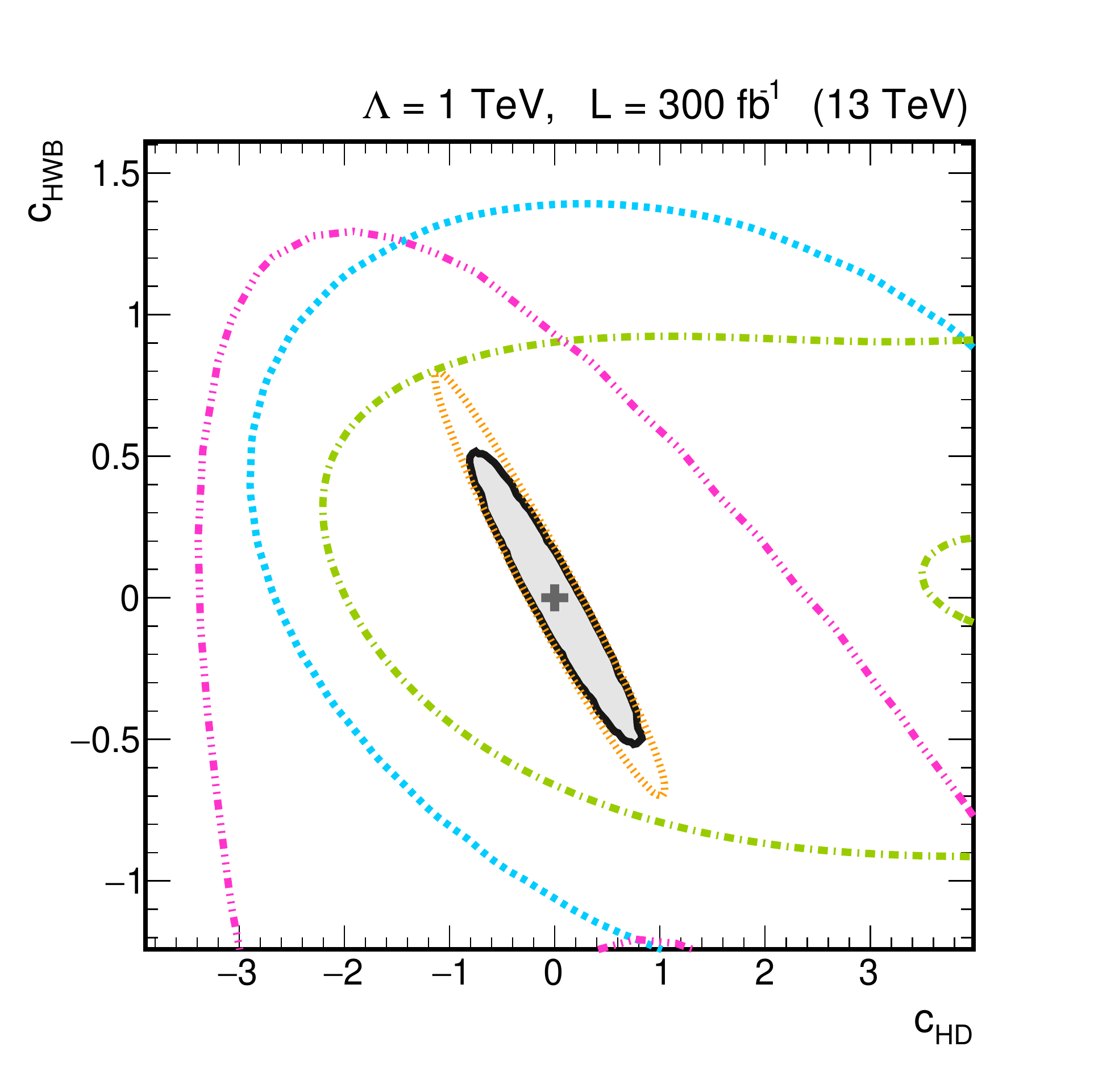}
  \includegraphics[width=.32\textwidth]{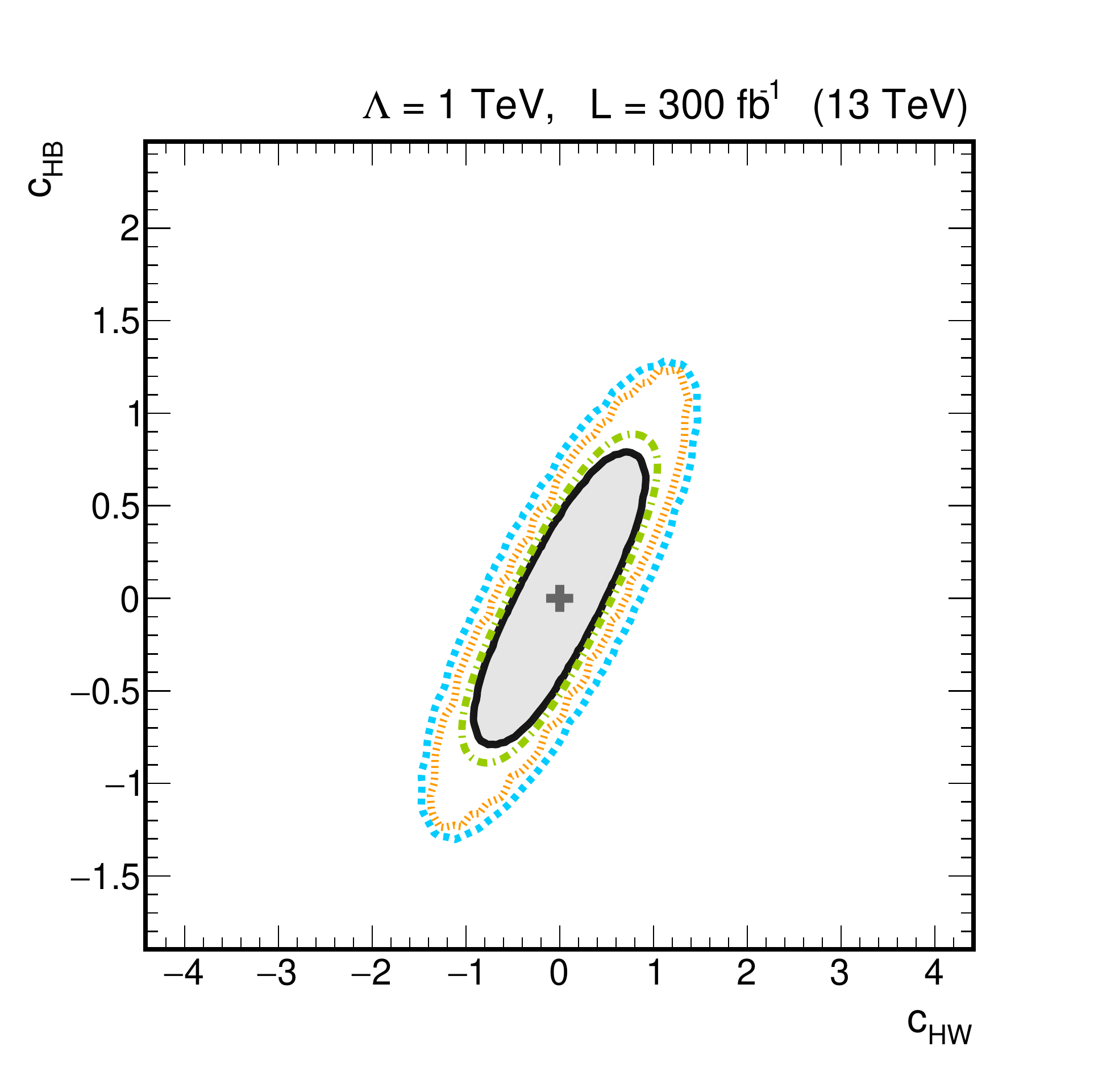}
  \\  
  \includegraphics[width=.32\textwidth]{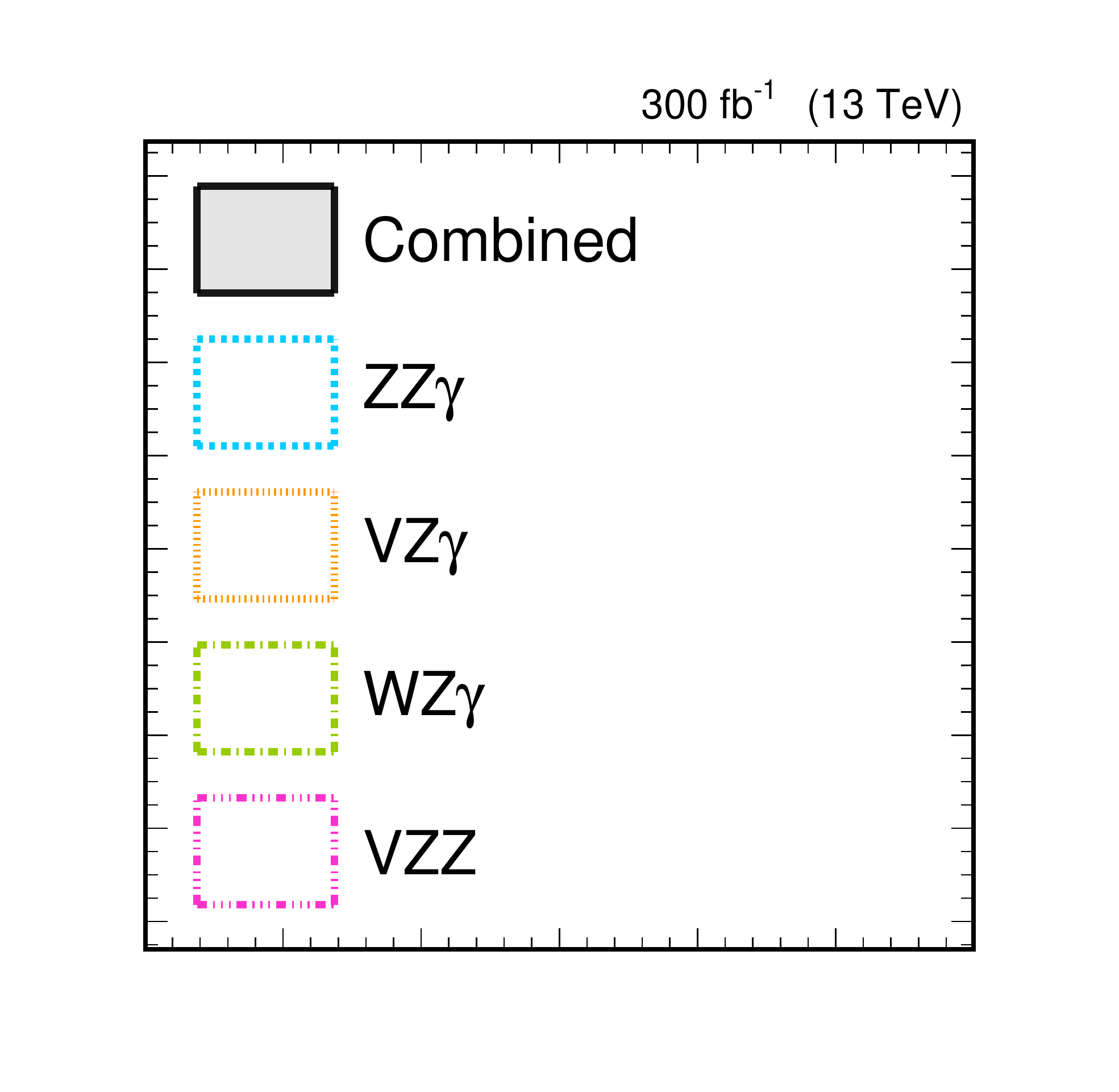}
  \includegraphics[width=.32\textwidth]{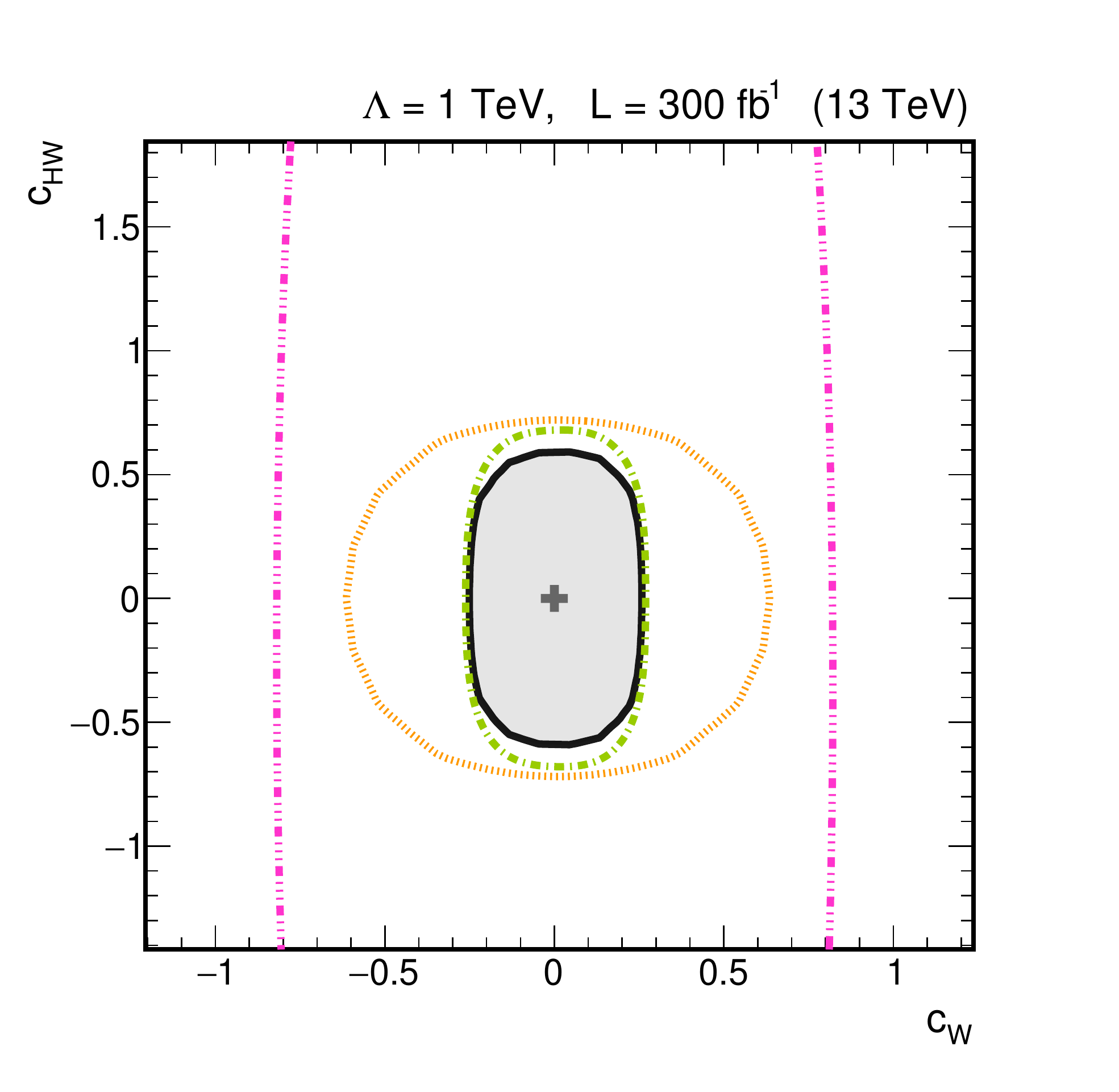}
  \includegraphics[width=.32\textwidth]{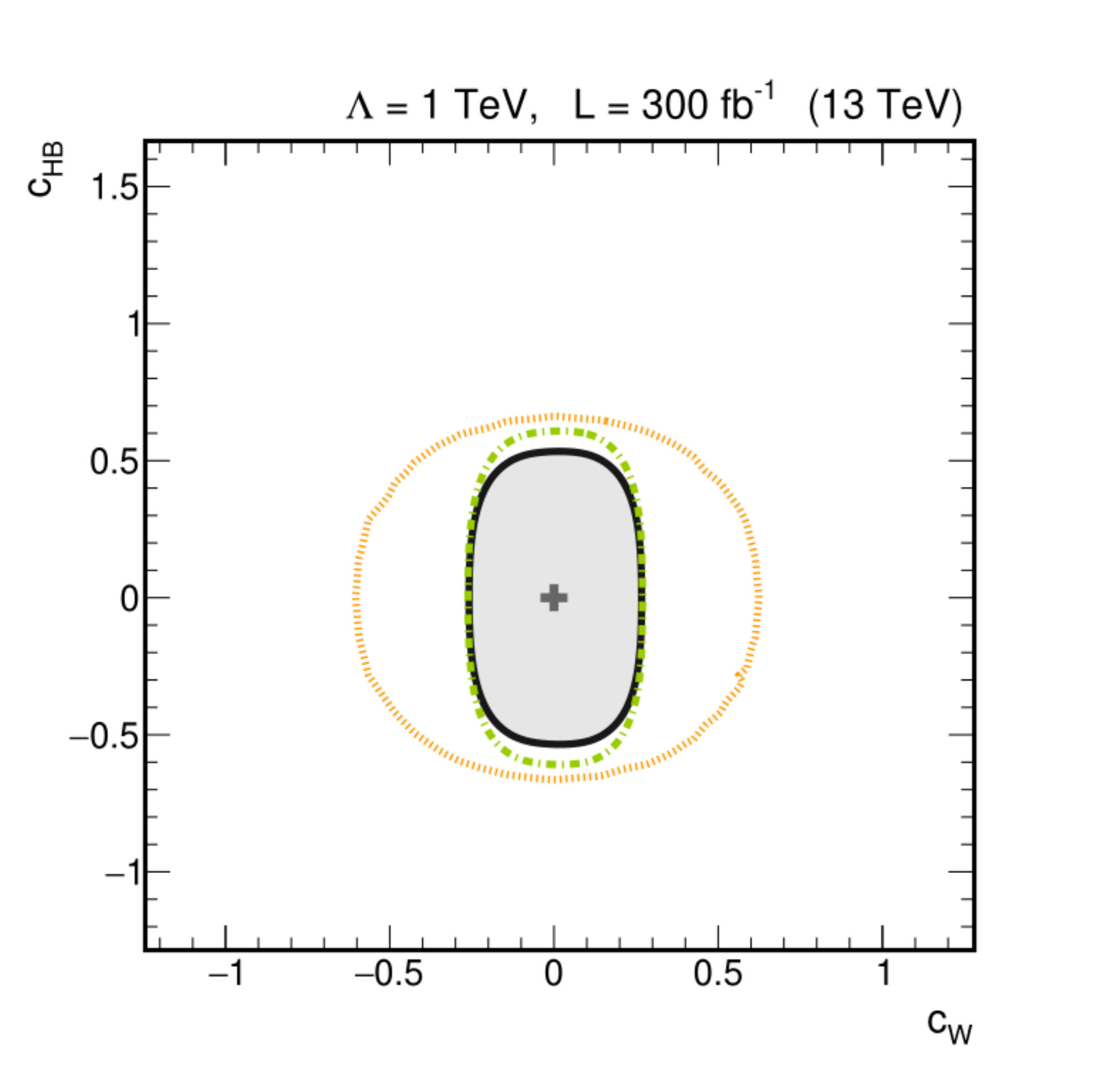} \\
  \includegraphics[width=.32\textwidth]{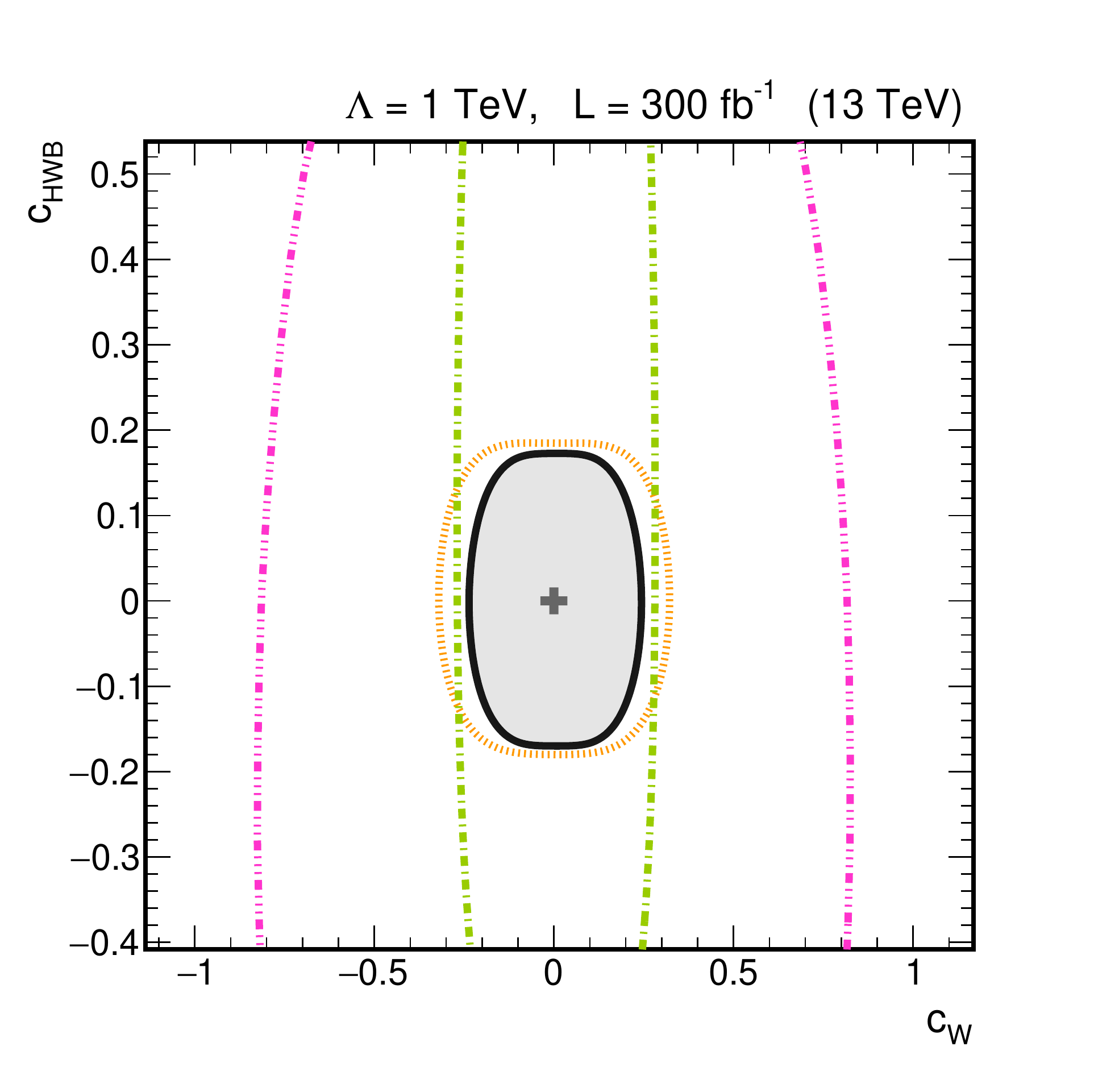} \includegraphics[width=.32\textwidth]{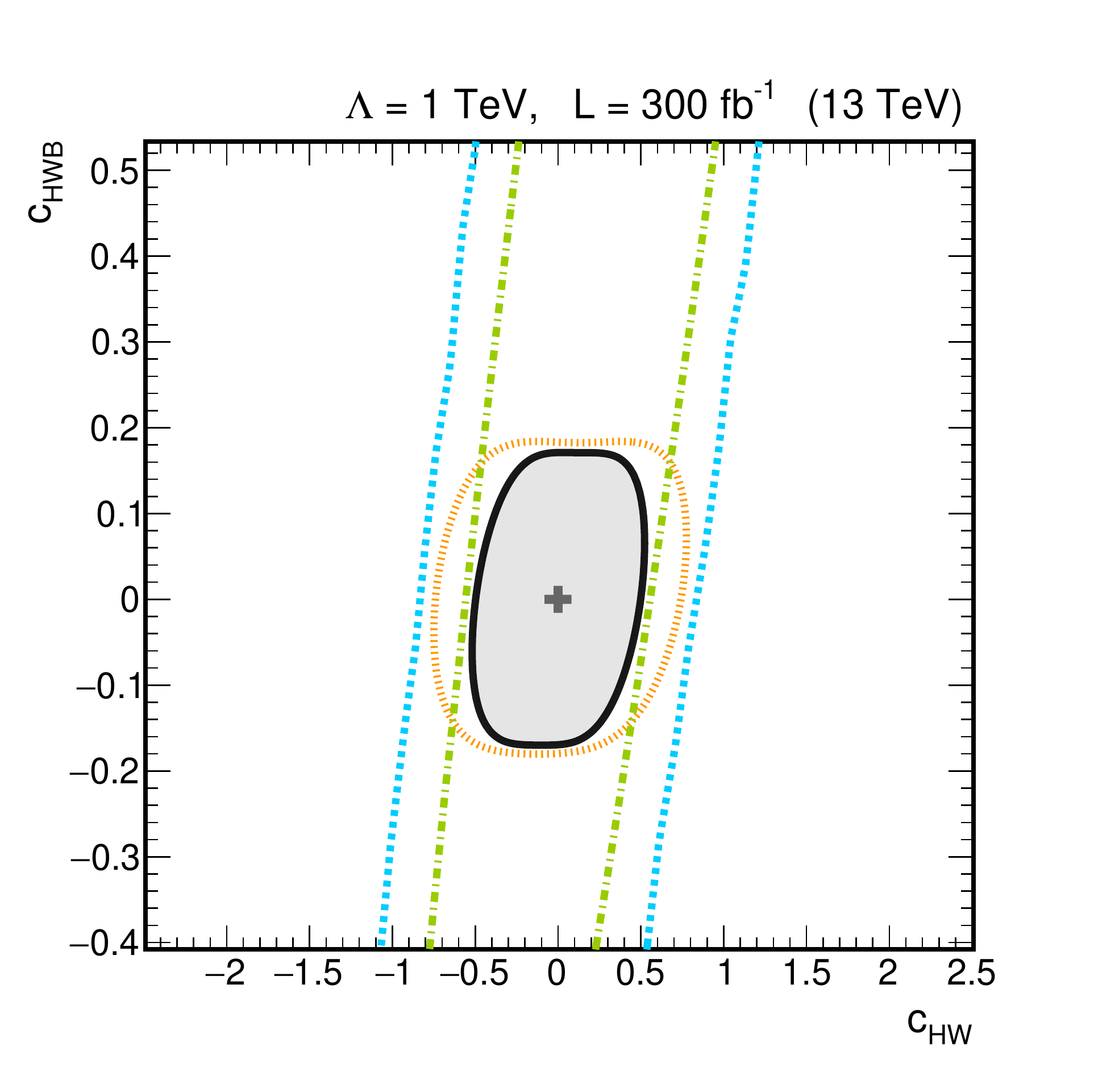}
  \includegraphics[width=.32\textwidth]{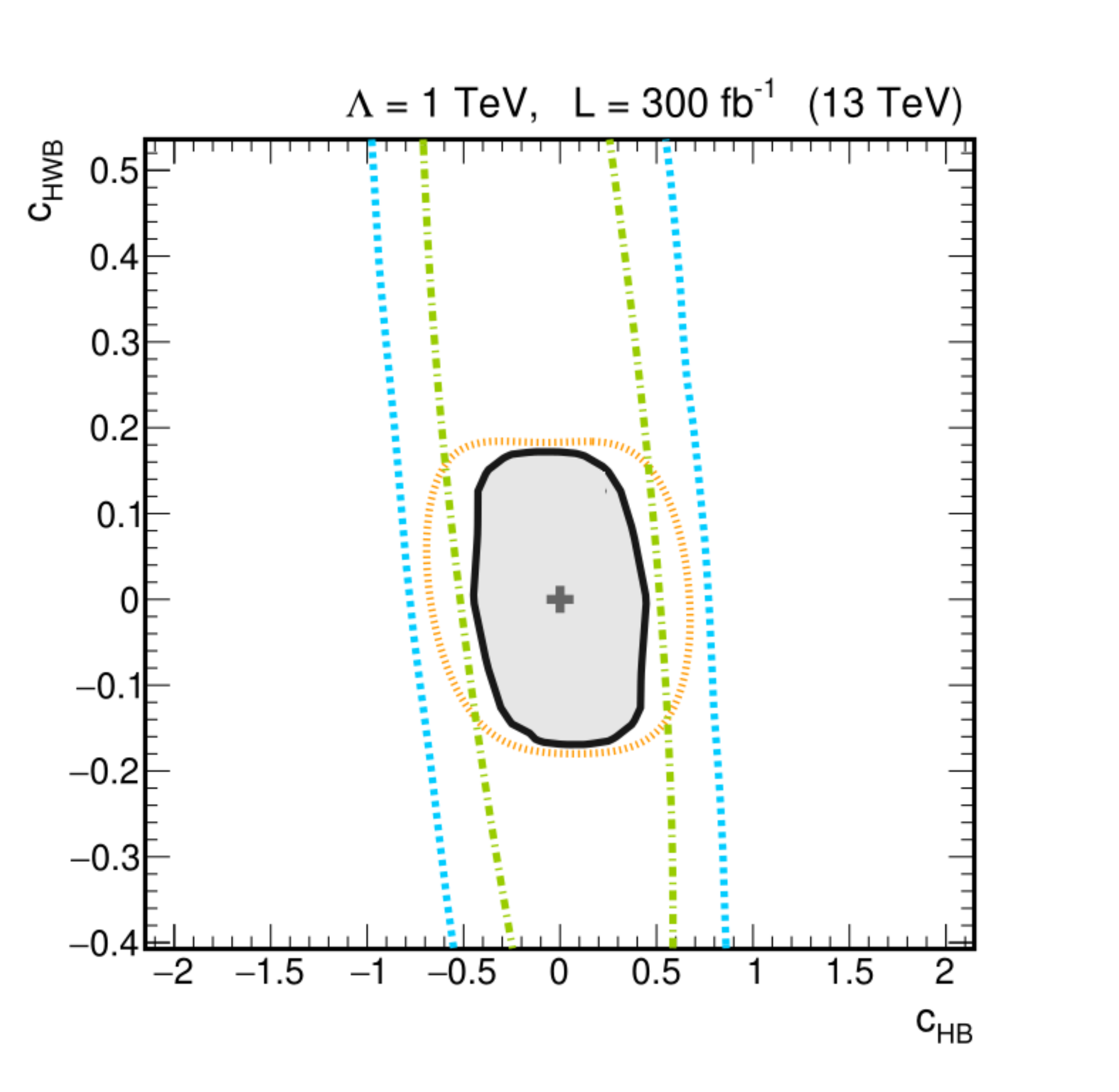}
  \\
  \includegraphics[width=.32\textwidth]{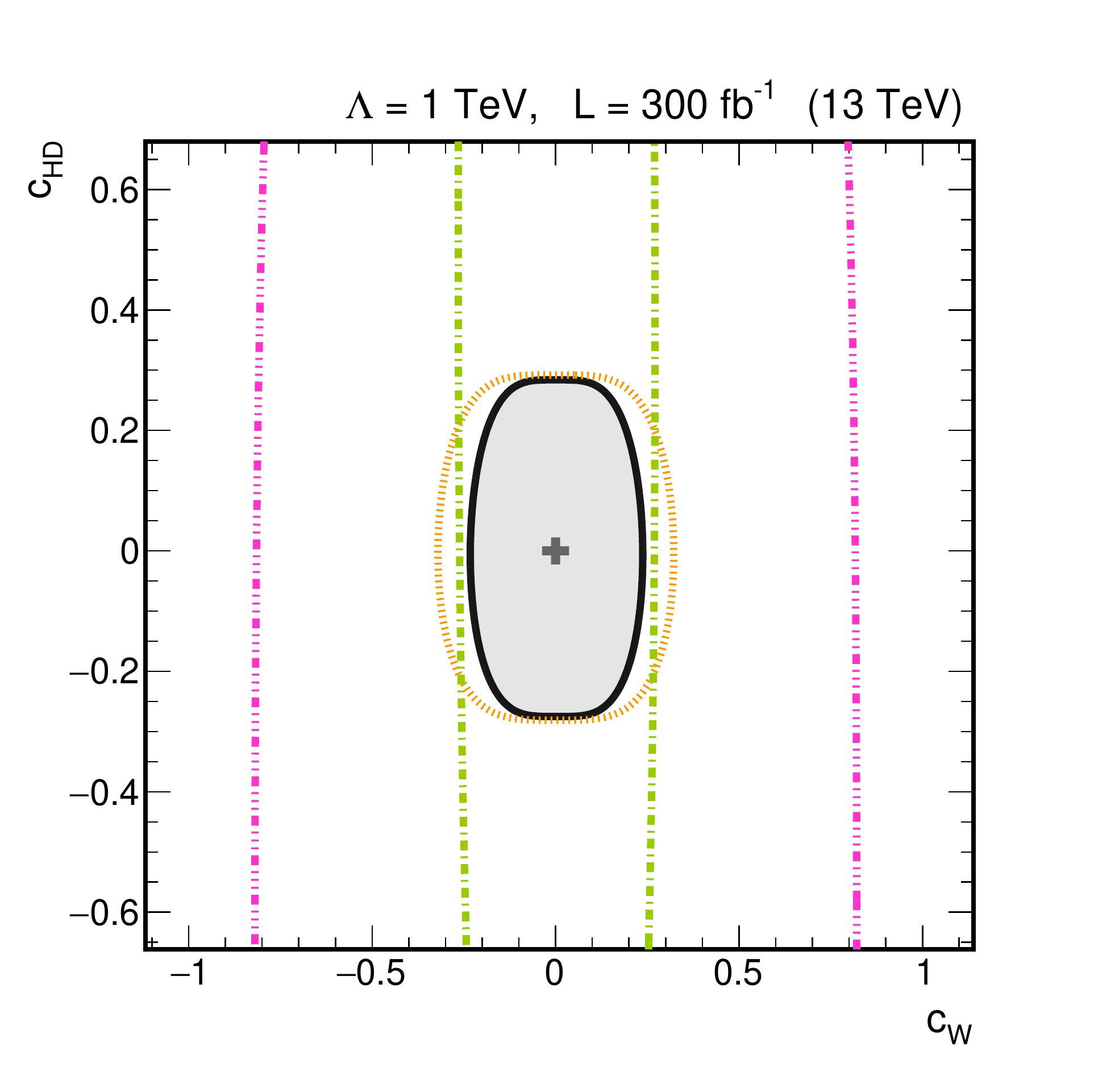}
  \includegraphics[width=.32\textwidth]{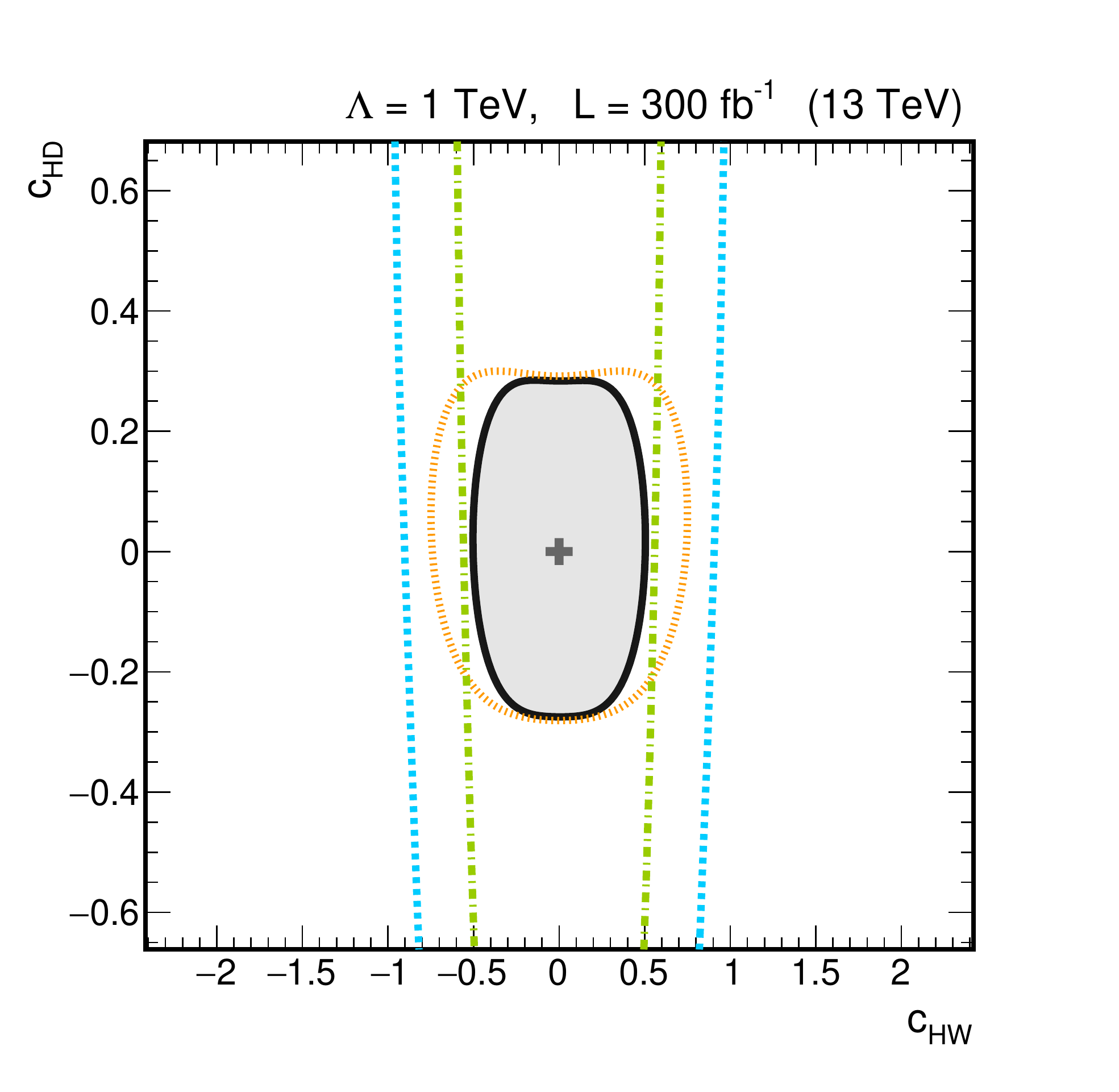}
  \includegraphics[width=.32\textwidth]{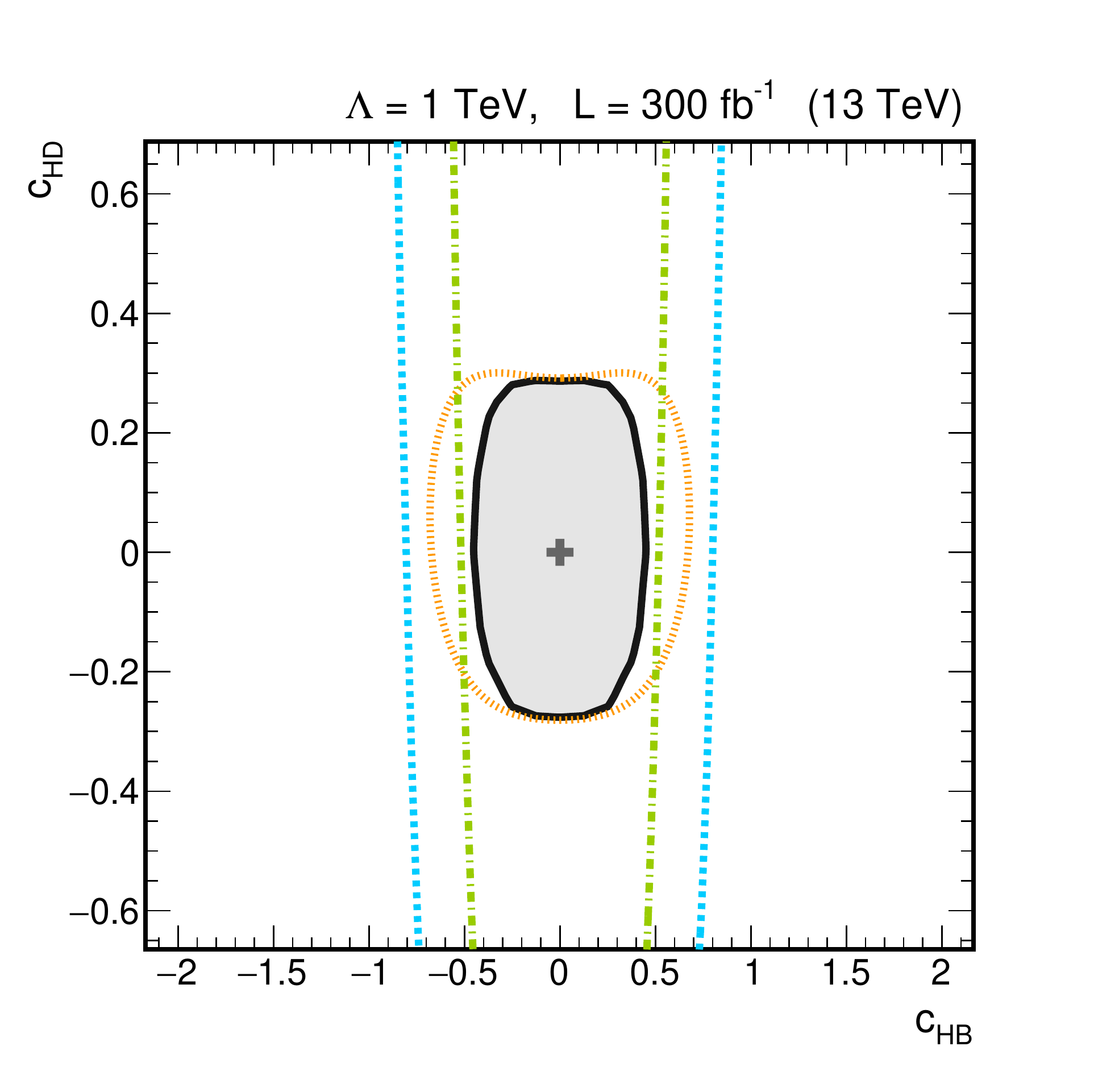}

  \caption{ Bi-dimensional likelihood contour levels at $-2\Delta \log \mathcal{L} = 2.3$  from the triboson channels sensitive for each pair of operators. In the case of the semi-leptonic channels VZZ and VZ$\g$, the QCD-induced backgrounds and the corresponding EFT contributions are included into the fit.}  
  \label{fig:Combined_LL_Profiles_2D}
\end{figure}

\paragraph{Two-dimensional constraints}
For each operator pair $\{\text Q_i,\text Q_j\}$, the effect of the mixed quadratic interference term on the sensitivity to anomalies is investigated by extending Eq.~\ref{LL} to the case of two non-zero Wilson coefficients. 
For each channel, the confidence regions are extracted from the likelihood scans using the variable most sensitive to the operator pair. 
The statistical combination of the likelihood is then performed for all the channels and the corresponding 68\% C.L. area is shown in figure~\ref{fig:Combined_LL_Profiles_2D}. 
The upper central plot in figure~\ref{fig:Combined_LL_Profiles_2D} displays the contour of the confidence region of the $\{\text c_{HWB}, \text c_{HD}\}$ pair. 
The corresponding operators are the only bosonic ones affecting the QCD-ZZjj(-Z$\gamma$jj) background processes. 
For the studied semi-leptonic channels, the contribution of the mutual interference between the QCD-induced anomalous tree-level diagrams is taken into account, as well as the interference of the electroweak VZZ(VZ$\g$) diagrams. 
This interference term in the inclusive Z$\g$jj production leads to a narrow elliptical shape of the contour, highlighting an anti-linear correlation between the estimates of $\text c_{HWB}$ and $\text c_{HD}$. 
In the region where both coefficients are positive, the mutual interference term partially cancels the linear interference terms.  

The upper right plot shows the combination of the operators $\{\text Q_{HB}, \text Q_{HW}\}$ inducing the same anomalous couplings between the Higgs boson and the neutral gauge bosons. 
For all the considered channels, a linear correlation between the estimates of these Wilson coefficients is observed. The latter is attributed to a large mutual interference term, since the linear interference terms with the SM are negligible.  

The second row of figure~\ref{fig:Combined_LL_Profiles_2D} illustrates the combination of the operator Q$_W$ with Q$_{HB}$ (left) and Q$_{HW}$ (right), respectively. The potential anomalies induced by the Q$_W$ operator in these channels can be regarded as uncorrelated with those caused by the other operators. In fact, Q$_W$ is the only operator modifying the gauge structure of the vector boson self-interactions.

The third row shows the pairwise combination of the operator Q$_{HWB}$ with Q$_W$ (left), Q$_{HB}$ (central), and Q$_{HW}$ (right), respectively. 
In the $\{\text c_{W}, \text c_{HWB} \}$ pair, the contours are symmetric with respect to the coefficient axes, as expected in the case of a negligible mutual interference term. 
For all the channels, a linear correlation is observed between the estimates of c$_{HB}$ and c$_{HWB}$, while it is anti-linear for the $\{\text c_{HW},$ $\text c_{HWB}\}$ pair. 
The main deviations from SM are due to the quadratic component of the Q$_{HB}$ and Q$_{HW}$ operators. 
On the contrary, the kinematic anomalies generated by Q$_{HWB}$ are dominated by the destructive interference with the SM diagrams.  
The confidence regions highlight the presence of a destructive interference term in the two cases of \{Q$_{HWB}$, Q$_{HB}$\} and \{Q$_{HWB}$, Q$_{HW}$\}, as opposed to the case of the \{Q$_{HB}$, Q$_{HW}$\} pair where the interference is constructive. 

The plots in the last row show the combination of the operator Q$_{HD}$ with Q$_W$ (left), Q$_{HB}$ (middle), and Q$_{HW}$ (right), respectively. 
In each case, the contours are quite symmetric with respect to the coefficient axes, due to the low contribution of both the mutual interference terms and of the individual linear terms associated with the Q$_W$, Q$_{HB}$, and Q$_{HW}$ operators. 
For the semi-leptonic channels, the Q$_{HD}$ interference term with the SM dominates when the QCD-induced backgrounds are included into the fit.

For all the operator pairs considered, the combination of all the channels leads to significantly more stringent confidence regions than the single channel with the highest sensitivity. 
\\
\\
\\

\paragraph{Profiled constraints} \label{subsec:profiledLL}
To preserve the model-independence of the EFT interpretation a global picture is needed. 
In this section all operators are left freely floating in the likelihood maximization.  Limits on a single coefficient are derived by profiling all parameters except for the one of interest. 
Floating parameters are treated as unconstrained nuisances with a flat prior in the maximum interval (-20,20).
A profiled fit allows a broader range of effects to be considered simultaneously, exploring the limitations of the EFT approach. 

\begin{figure}[t!]
  \centering 
    \includegraphics[width=.98\textwidth]{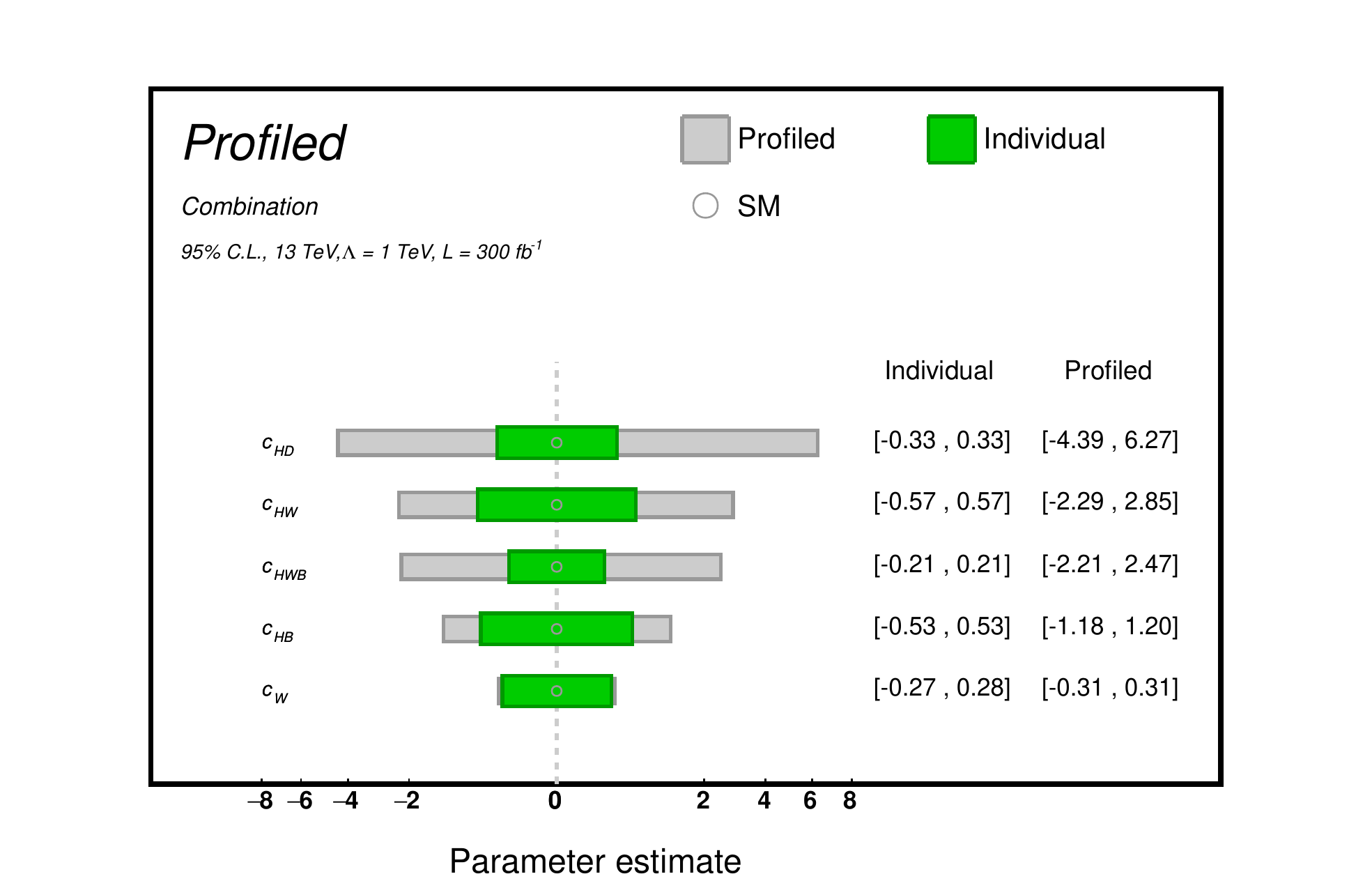}
  \caption{ Comparison between the profiled and the individual expected constraints (quadratic scale) on the Wilson coefficients from the combination of the leptonic channels WZ$\gamma$, ZZ$\gamma$, and the semi-leptonic channels VZ$\gamma$, VZZ. The solid dots denote the SM expectation. Gray-filled boxes indicate the 95\% C.L. intervals obtained by estimating the corresponding Wilson coefficient as the parameter of interest, with all the other coefficients left floating in the maximum interval (-20,20). Green-filled boxes indicate the 95\% C.L. individual constraints. \label{Prof_vs_Ind}}  
\end{figure}

The comparison between the profiled and individual constraints at 95\%~C.L. for each operator is shown in figure~\ref{Prof_vs_Ind}. The confidence intervals are extracted from the likelihood scans derived with the best variables from the individual constraints. 
The results corresponding to the combination of all the triboson channels illustrate the expected decrease in the sensitivity of the profiled fit with respect to the individual constraints. 
This effect is more pronounced for the EFT operators affecting the couplings between Higgs and gauge bosons, since they induce anomalies to the same vertices, leading to flat directions and a large mutual interference. 
The anomalous effects induced by Q$_W$ are instead uncorrelated with the other operators, as observed in the two-dimensional contours $\{\text c_W,\text c_i\}$ shown in figure~\ref{fig:Combined_LL_Profiles_2D}. 

Overall, the effect of the mutual interference contributions is analogous to the behavior observed in the global fit in the case of diboson processes~\cite{bellan2021sensitivity}. 
The sensitivity decrease highlighted by the profiled constraints is not negligible, but it does not compromise the constraining power of these analyses to the anomalies induced by dimension-6 operators. 
The impact of higher-dimension operators on the studied triboson channels is not assessed in this work, but it would potentially affect the global fit. The high interest in the combined anomalous effects of dimension-6 and -8 operators, particularly the mutual interference terms, requires the development of suitable, comprehensive theoretical models that remains open for future research. 

\newpage
\paragraph{Prospects for a fully leptonic WWW analysis \label{subsec:WWW}}
A benchmark sensitivity study is also performed for the WWW channel in the $3l+3\nu$ final state with three charged leptons without any same-flavor opposite-sign lepton pairs. The three W bosons decay leptonically respectively via W$^+\rightarrow\mu^+\nu_\mu$, 
W$^+\rightarrow~$e$^+\nu_e$, and W$^-\rightarrow\tau^-\Bar{\nu}_\tau$. The kinematic selection is taken from the search for anomalous quartic gauge couplings 
in the W$^\pm$W$^\pm$W$^\mp$ production at $\sqrt{s} =$ 13 TeV with CMS Run 2 data \cite{CMS:2019mpq}. In the SM, the WWW diagrams depend at the tree level on the triple WWW and quartic WWWW gauge couplings. In the context of dimension-6 EFT operators, this channel is expected to be most sensitive to the anomalies induced by the Q$_W$ operator on the W boson self-couplings. In this channel, it was found \cite{CMS:2019mpq} that the best variable to probe anomalous W boson self-interactions is $s_T$, defined as the sum of the transverse momenta of the particles in the final state. In this study, this variable is also used to assess the sensitivity to the kinematic anomalies induced by the bosonic dimension-6 EFT operators.  
The sensitivity of the tri-lepton final state to the EFT operators is derived by performing a cut-and-count analysis after requiring $s_T>1500$~GeV. This threshold is determined from the SM spectrum, which decreases sharply at this value of $s_T$. 

For an integrated luminosity of 300~fb$^{-1}$ ($\Lambda=1$~TeV) the exclusion interval at 68\%(95\%)  C.L. on c$_W$ is [-0.13(-0.22),~0.13(0.22)]. 
This channel alone is about a factor of two more sensitive to the anomalous effects induced by the Q$_W$ operator than the combination of the other triboson processes studied. This is expected since the dependence on the WWWW gauge couplings is enhanced with respect to the other channels. This result is also close to the Q$_W$-sensitivity in the diboson WW channel in the e$\mu$ final state \cite{bellan2021sensitivity}.

The multiplicity of the final states produced from the decay of WWW processes strongly advocates the need for a dedicated study, which is left for future work. 
\section{Summary and conclusions}\label{sec:conclusions}
The first benchmark study of the sensitivity of triboson measurements to dimension-6 SM-EFT operators has been presented.  
This study includes triboson channels with one photon in the final state, namely the fully leptonic WZ$\g$ and ZZ$\g$ and the semi-leptonic inclusive VZ$\g$, where V=\{W,Z\}. 
The semi-leptonic channel VZZ including both the WZZ and the ZZZ cases was also examined. 
In addition, we performed a feasibility study of the fully leptonic WWW analysis in a single tri-lepton final state. 

The expected sensitivity to anomalies arising from dimension-6 bosonic operators has been studied to cover the main class of corrections affecting triple and quartic, as well as Higgs-gauge couplings, which enter multiple triboson channels. 
The subset of operators in the Warsaw basis that satisfy CP symmetry has been used with $\{m_W,m_Z,G_F\}$ as the input parameter scheme. 
For all the channels and the studied dimension-6 operators, the EFT components were generated at the parton level up to the $\Lambda^{-4}$ order in the EFT expansion.

We investigated the effect of the quadratic terms (pure beyond-the-Standard-Model components) on the sensitivity to EFT dimension-6 bosonic operators. 
Their contribution increases the sensitivity of the triboson analyses to the SM deviation, especially when the anomalous couplings induced by the EFT bosonic operators are forbidden by the SM at the tree level and, as a consequence, the linear interference terms are negligible. 
 
The sensitivity constraints are computed for a LHC measurement with proton-proton collisions at a center-of-mass energy of 13 TeV and for a projected total luminosity of $\unit[300]{fb^{-1}}$.  
The triboson production processes are modeled inclusively as full $2\to 6$ fermions and $2\to 4$ fermions + $\g$ scattering at the leading order with the inclusion of the non-resonant diagrams. 
  
A template analysis of several differential distributions was performed to test the observables to be used in future LHC analyses targeting triboson production modes. 
Two classes of kinematic variables, namely the transverse momentum relative to non-standard longitudinal directions and the Fox-Wolfram moments, are found to be effective in isolating regions of phase space with pronounced anomalous effects. 
The impact on the shape analysis of the main backgrounds, which are QCD-induced processes, has been evaluated in semi-leptonic channels. The EFT corrections to the QCD vertices have been calculated at the leading order. 
Interestingly, the inclusion of EFT effects in the QCD-induced processes significantly increases the sensitivity to individual operators affecting them. 

Overall, we find that the processes most sensitive to the bosonic operators are those involved in the semi-leptonic VZ$\g$ channel (and its main background QCD-Z$\g$jj). 
The leading lepton transverse momentum provides the strongest constraints on the anomalous triple and quartic gauge boson couplings. 
It is interesting to make a comparison with the bounds on the dimension-6 Wilson coefficients from the combination of the vector boson scattering analyses in the literature. 
In the present study, the sensitivity reach to the operators affecting triple and quartic gauge couplings is competitive with the vector boson scattering combination, and it is larger by a factor six in the case of the operators affecting the Higgs-gauge bosons couplings. 
This unique EFT sensitivity of triboson analyses with a photon in the final state can be explained in light of the anomalous diagrams involving the  HZ$\g$ and H$\g\g$ couplings, which are instead excluded by the Standard Model at the tree level.

The results of the two-dimensional scan of operators illustrate the complementarity and interplay between different measurements, by identifying orthogonal directions in sensitivity for different pairs of operators. In most cases, the combination of all channels significantly increases the sensitivity to EFT effects, which is often dominated by a single measurement. The effect of the mutual interference term in the fit has also been investigated and found to determine the presence of linear (anti-)correlation in the sensitivity to pairs of operators. 

Constraints on the individual Wilson coefficients are extracted by profiling all the other coefficients in the set of interest. The latter are treated as nuisance parameters within the maximum interval (-20,20) in the likelihood fit. The results show a decrease in sensitivity for the profiled fit with respect to the individual constraints, which is more pronounced for the couplings between the Higgs and the gauge bosons, and found to be consistent with the results of the two-dimensional fit.

In addition, the first benchmark results on the sensitivity of a fully leptonic WWW analysis to the anomalies induced on W-boson self-couplings by the corresponding dimension-6 operator are derived and compared with the other triboson channels in this study.

The present study provides a first estimate of the potential constraining power of triboson LHC measurements on dimension-6 EFT effects. Possible improvements include refining the analysis to account for detector reconstruction effects, and performing a more detailed treatment of reducible background processes. A further development that we leave for future work would be to perform a global fit of vector boson scattering and multi-boson production, including both diboson \cite{bellan2021sensitivity} and other triboson processes with three massive boson decaying leptonically. 
In future studies, the scope of the investigation can be further expanded through the development of new models considering the next terms in the amplitude of the scattering matrix, namely the linear interference term of dimension-8 operators with the SM diagrams and the mixed interference between dimension-6 and dimension-8 operators. 

\acknowledgments

We express our gratitude to Dr. Ilaria Brivio for her invaluable assistance in setting up and using the SMEFTSim simulation package, as well as proofreading this manuscript. We extend our sincere appreciation to numerous colleagues for their engaging discussions and valuable communications. In particular, we would like to thank Prof. Roberto Covarelli and Dr. Giacomo Ortona for their contributions to this research.  Furthermore, we would like to express our appreciation to Ms. Magnaci and the “Fondazione CRT” (grant n. 2020.453) for the financial support provided to Dr. Antonio Vagnerini throughout this research.

\bibliographystyle{JHEP}
\bibliography{biblio.bib}

\end{document}